\documentclass[superscriptaddress,twocolumn,10pt,prx,aps,showpacs,longbibliography, floatfix]{revtex4-2}
\usepackage{float}
\usepackage{array}
\usepackage{graphicx}
\usepackage{amsbsy}
\usepackage[utf8x]{inputenc}
\usepackage{epstopdf}
\usepackage{amsmath,amssymb,amsfonts}
\usepackage{dsfont}
\usepackage{cprotect}
\usepackage{mathtools}
\usepackage[normalem]{ulem}
\usepackage{fancyhdr}

\usepackage[dvipsnames]{xcolor}

\newcommand{\bra}[1]{\langle #1|}
\newcommand{\ket}[1]{|#1\rangle}

\newcommand{\ketbra}[2]{\left| #1 \rangle \langle #2 \right|}

\newcommand{\expec}[1]{\left\langle #1 \right\rangle}

\newcommand{\figref}[1]{\mbox{Fig.~\ref{#1}}}

\renewcommand{\eqref}[1]{\mbox{Eq.~(\ref{#1})}}

\renewcommand{\ell}{l}

\newcommand{\be}{\begin{equation}}
\newcommand{\ee}{\end{equation}}
\newcommand{\bea}{\begin{eqnarray}}
\newcommand{\eea}{\end{eqnarray}}

\usepackage{multirow}
\usepackage{verbatim}

\usepackage{url}
\usepackage[colorlinks]{hyperref}
\hypersetup{
 plainpages=true,
 breaklinks=true,
 hypertexnames=false,
 pageanchor=true,
 colorlinks=true,
 linkcolor={blue},
 citecolor={red},
 urlcolor={blue},
 anchorcolor={black}
}

\usepackage{tabularx}

\newcommand{\LL}{\mathcal{L}}
\newcommand{\DD}{\mathcal{D}}
\newcommand{\rhot}{\hat{\rho}(t)}

\makeatletter
\newcommand*\bigcdot{\mathpalette\bigcdot@{.5}}
\newcommand*\bigcdot@[2]{\mathbin{\vcenter{\hbox{\scalebox{#2}{$\m@th#1\bullet$}}}}}
\makeatother

\usepackage{enumerate}

\usepackage{physics}

\makeatletter
\def\maketitle{
\@author@finish
\title@column\titleblock@produce
\suppressfloats[t]}
\makeatother

\begin{document}

\author{David S. Schlegel}
\affiliation{Institute of Physics, Ecole Polytechnique Fédérale de Lausanne (EPFL), CH-1015 Lausanne, Switzerland}
 \affiliation{Center for Quantum Science and Engineering, Ecole Polytechnique Fédérale de Lausanne (EPFL), CH-1015 Lausanne, Switzerland}
\author{Fabrizio Minganti}
\affiliation{Institute of Physics, Ecole Polytechnique Fédérale de Lausanne (EPFL), CH-1015 Lausanne, Switzerland}
\affiliation{Center for Quantum Science and Engineering, Ecole Polytechnique Fédérale de Lausanne (EPFL), CH-1015 Lausanne, Switzerland}
\author{Vincenzo Savona}
\email{vincenzo.savona@epfl.ch}
\affiliation{Institute of Physics, Ecole Polytechnique Fédérale de Lausanne (EPFL), CH-1015 Lausanne, Switzerland}
\affiliation{Center for Quantum Science and Engineering, Ecole Polytechnique Fédérale de Lausanne (EPFL), CH-1015 Lausanne, Switzerland}

\title{Quantum error correction using squeezed Schrödinger cat states}

\begin{abstract}
Bosonic quantum codes redundantly encode quantum information in the states of a quantum harmonic oscillator, making it possible to detect and correct errors. 
Schrödinger cat codes -- based on the superposition of two coherent states with opposite displacements -- can correct phase-flip errors induced by dephasing, but they are vulnerable to bit-flip errors induced by particle loss. 
Here, we develop a bosonic quantum code relying on squeezed cat states, i.e. cat states made of a linear superposition of displaced-squeezed states. Squeezed cat states allow to partially correct errors caused by particle loss, while at the same time improving the protection against dephasing.
We present a comprehensive analysis of the squeezed cat code, including protocols for code generation and elementary quantum gates. We characterize the effect of both particle loss and dephasing and develop an optimal recovery protocol that is suitable to be implemented on currently available quantum hardware. We show that with moderate squeezing, and using typical parameters of state-of-the-art quantum hardware platforms, the squeezed cat code has a resilience to particle loss errors that significantly outperforms that of the conventional cat code.
\end{abstract}

\date{\today}

\maketitle
\section{Introduction}

Quantum computers will leverage the phenomena of quantum superposition and entanglement to tackle computational tasks currently beyond the reach of conventional computers~\cite{PreskillNISQ2018,AruteNat19}.

Near-term quantum hardware is affected by noise originating from interaction with the surrounding environment. Noise induces errors which eventually lead to the loss of quantum information. To deal with errors in the short term, hybrid quantum-classical algorithms~\cite{Cerezo2021, EndoJPSJ2021} and error mitigation strategies~\cite{EndoJPSJ2021, EndoPRX2018} are being developed. To achieve fault tolerance and express the full potential of quantum computing however, quantum error correction (QEC) strategies~\cite{Shor1996,NielsenChuangBIBLE2010}, making a quantum computer resilient to a given set of errors, must eventually be developed and implemented.

QEC codes rely on encoding quantum information redundantly. The established QEC paradigm introduces redundancy by encoding $k$ \textit{logical quantum bits} (or qubits) onto a subspace -- hereafter called \textit{code space} -- of the state space of $n>k$ \textit{physical qubits}, in ways that allow detecting and correcting errors without disturbing the stored quantum information. While significant progress has recently been made in the experimental realization of many-qubit QEC codes~\cite{krinner2021, AndersenNatPhys20, Marques2021, Ryananderson2021, Chen2021}, this approach requires many interconnected physical qubits to efficiently correct errors on even a small number of logical qubits, resulting in a significant device footprint and a fundamental limitation to the design of scalable quantum devices. A promising route to improve the error rates on logical qubits, without the hardware overhead of conventional QEC, is to encode a single logical qubit into a bosonic mode. Within this approach, QEC can be performed on information that is redundantly encoded in multiple levels of the bosonic mode. The underlying general principle is then the same as in conventional QEC: the larger Hilbert space of a bosonic system allows detecting and correcting errors. Electromagnetic resonators -- both optical and in the microwave range -- coupled to nonlinear elements, represent an ideal device setting for the implementation of this QEC strategy. The resulting Bosonic Quantum Codes (BQCs) have been intensively investigated \cite{AlbertPRA2018,TerhalQuantumScienceandTechnology2020, JoshiQST21}, leading to proofs of principle of quantum gate operations and efficient error detection and correction.

Photons stored in a resonator are subject to two main dissipation mechanisms, photon loss and pure dephasing \cite{HouckNatPhys12,Carusotto_RMP_2013_quantum_fluids_light,AspelmeyerRMP2014,Gu2017,JoshiQST21}, which ultimately generate errors in BQCs. The main challenge in the design and realization of a BQC is therefore the ability to efficiently detect and correct errors resulting from photon loss and dephasing. An additional challenge is to realize \textit{autonomous} QEC~\cite{Lebreuilly2021}, i.e. to design a code which autonomously recovers from an error through an engineered dissipation scheme, without relying on \textit{error syndrome} measurement and conditional recovery operations.

Early proposals for BQCs -- aimed at implementing universal quantum computing onto a linear photonic platform -- were the single- and dual-rail codes, where a qubit is encoded onto single- or few-photon states of one or more optical modes~\cite{Knill2001, OBrien2003, LeungPRA1997, ChuangPRA1995}. At the same time, BQCs based on many-photon continuous-variable states were introduced, and the Gottesman-Kitaev-Preskill (GKP) and the Schrödinger cat codes emerged as the main paradigms for BQCs.

In the GKP code~\cite{GottesmanPRA01,GrimsmoPRXQuantum2021}, a qubit is logically encoded in two states that are invariant under discrete translations along the position and momentum variables, thereby forming a periodic lattice in the phase space of the harmonic oscillator. Errors associated to small displacements in phase space (i.e. smaller than the period of the lattice) can therefore be detected and corrected. GKP states are very effective for the detection and correction of particle loss~\footnote{From now on, we will use the terminology ``particle loss'' instead of ``photon loss'', to refer to more general physical realizations such as, for example, phonons in optomechanical systems.} errors, which entail small shifts of the state in the phase space \cite{JoshiQST21}. The GKP code is however not robust against dephasing errors, which can induce arbitrarily large phase-space shifts. The GKP code was recently demonstrated in experiments both on the trapped-ion~\cite{FluhmannNature2019} and superconducting-circuit platforms~\cite{Campagne2020}.

Schrödinger cat codes were the first example of an autonomous QEC code~\cite{RalphPRA2003, OfekNat16, BergmannPRA2016, LescanneNatPhys2020, GrimmNature2020, PurinpjQI17, PuriPRX2019, JeongPRA02, LeghtasPRA2013, LeghtasPRL2013, MirrahimiNJP14, LeghtasScience15, VlastakisScience2013, Monroe1996, DelegliseNat08, Hofheinz2009, WangScience2016}. In the most basic cat code (here dubbed 2-cat), the code space is spanned by the even and odd combinations of two coherent states with opposite displacement fields. The two codewords are therefore eigenstates of the parity operator, with $\pm1$ eigenvalues. When the 2-cat code is generated and stabilized through parametric two-photon drive and engineered two-photon dissipation, then in the limit of large displacement the encoded quantum information is autonomously protected against dephasing. A single particle loss process, on the other hand, induces a bit flip -- namely a transformation fully contained within the code space. This results in a non correctable error, as the error can only be detected by carrying out a measurement within the code space, which would unavoidably destroy the quantum information stored in the qubit. Proposals to detect and correct single-particle loss errors in cat codes typically rely on concatenation of multiple 2-cat codes~\cite{GuillaudPRX2019, PuriScienceAdvances2020, Chamberland2022, GuillaudPRA2021}, on biased-noise engineered dissipation~\cite{JoshiQST21}, on multicomponent cat codes \cite{LeghtasPRL2013,MirrahimiNJP14,LiPRL2017,BergmannPRA2016,GertlerNature2021}, or on cat codes in multiple resonators \cite{AlbertIOP2019}. 

The principle underlying BQCs can be generalized to address multiple errors. This has recently led to the development of binomial codes~\cite{MariosPRX16, HuNaturePhys2019}, rotation-symmetric codes~\cite{GrimsmoPRX2020}, and more general classes of BQCs~\cite{LiPRA2021,Wang2021}. While these codes are designed to protect quantum information against multiple particle loss and dephasing errors, their practical realization and stabilization in state-of-the-art quantum hardware platform still constitutes a major technological challenge. It has been recently suggested that squeezing can be introduced as a generalization of both the Schrödinger cat~\cite{GrimsmoPRX2020} and the GKP~\cite{ShuklaPRA2021} codes. The basic properties of squeezed cat states in particular have been investigated~\cite{NicacioPRA2010, ChaoAnnalenderPhysik2019, TehPRR2020, LiuPRA2005}, and the first experimental realizations have been recently demonstrated~\cite{Ourjoumtsev2007,Lo2015,JeannicPRL2018}.
In particular, it has been demonstrated that squeezed cats display promising properties that make them interesting states for quantum technologies. 
For instance, they show an enhanced resistance to dephasing with respect to standard 2-cats \cite{Lo2015}, and their quantumness (encoded in the negativity of the Wigner function) decays at a slower rate than standard 2-cats \cite{JeannicPRL2018}.

Here, we present a comprehensive analysis of the Squeezed Cat (SC) code -- a two-component cat code which relies on squeezing to mitigate particle loss errors, as compared to the 2-cat code, while simultaneously improving the resilience to dephasing errors.
Our main motivation lies in the fact that, squeezing being a Gaussian operation, the generation and manipulation of the SC code can be achieved by use of quadratic operations, without the need to introduce higher-order processes.
Differently from the 2-cat code, in the SC code a particle loss process transforms the logically encoded state onto a space with a nonzero component orthogonal to the original code space. As a result, the component outside the code space can in principle be \textit{detected} with an appropriate error syndrome measurement, and \textit{corrected} with an appropriate recovery operation. The detection and correction only succeed with a probability $p<1$, as the state after an error retains a component in the code space, which may be detected by the projective syndrome measurement, thus erroneously signaling that no error has occurred. The efficiency of the error mitigation therefore depends on the amplitude of the component outside code space, which is larger for larger squeezing parameter.

In a 2-cat code, the suppression of the dephasing error is due to the small overlap between the two coherent-state components of cat states. As a consequence, dephasing errors are suppressed exponentially with the amplitude of the displacement field. Increasing the displacement field leads however to an increase of the particle loss error rate which depends linearly on the number of particles, i.e. on the squared displacement field. In the SC code, the dephasing error rate is further suppressed with respect to a 2-cat code with the same displacement field~\cite{JeannicPRL2018}. This is achieved by selecting the squeezing angle in phase space to be orthogonal to the axis defining the cat state, leading to a smaller overlap between the two squeezed components.

After defining SC states, we investigate the properties of the SC code, by characterizing the code space and the error spaces generated respectively by the action of loss and dephasing errors on the codewords. We present a comprehensive, experimentally viable scheme for SC code generation and for the execution of elementary single-qubit gates. Errors are described by means of a Markovian model of dissipation, in terms of the (Gorini–Kossakowski–Sudarshan–)Lindblad master equation~\cite{GoriniJMP76, LindbladCMP76, BreuerBookOpen, NielsenChuangBIBLE2010}. The effect of errors is studied both using the Knill-Laflamme conditions~\cite{NielsenChuangBIBLE2010, Lebreuilly2021} and in the more general terms of channel fidelity \cite{AlbertPRA2018}. We describe a semi-autonomous QEC protocol based on the unconditional periodic application of an optimized recovery operation, and we detail a simple procedure to determine the optimal recovery operation for given displacement and squeezing parameters. In this way, we show that the SC code significantly outperforms the 2-cat code in its resilience to single particle loss errors, already at experimentally accessible values of the squeezing parameters~\cite{DassonnevillePRX2021, Gu2017, Eickbusch2021, Lo2015, Han:21, XiePRA2020}, while simultaneously improving the resilience to dephasing errors. Furthermore, we show that the correction of loss and dephasing errors in the SC code can naturally be extended to other types of errors, e.g. particle gain or higher-order error-processes.

The article is structured as follows. In Sec.~\ref{Sec:BQCs}, we briefly review error correction for BQCs, in particular by recalling the Knill-Laflamme conditions and notion of channel fidelity. In Sec.~\ref{Sec:Cats}, we recall SC states and describe how particle loss errors can be detected and corrected. In Sec.~\ref{Sec.SCC_channel} we assess the performance of the SC code by studying the channel fidelity. Sec.~\ref{Sec:Conclusion} contains the conclusions and outlook.

\section{Errors and their correction in bosonic codes}
\label{Sec:BQCs}

We describe the bosonic mode as a harmonic oscillator and the environment as a zero-temperature, Markovian reservoir. Thus, the system can be modeled using the (Gorini–Kossakowski–Sudarshan–)Lindblad master equation~\cite{GoriniJMP76, LindbladCMP76, BreuerBookOpen, NielsenChuangBIBLE2010}. 
In a frame rotating at the oscillator frequency, the reduced density matrix of $\rhot$ of the system is governed by the equations
\begin{equation}\label{Eq:Lindblad}
\begin{split}
\partial_t \rhot & =\LL \rhot = \kappa_1 \DD\left[ \hat{a} \right]\rhot + \kappa_2 \DD\left[ \hat{a}^\dagger \hat{a} \right] \rhot , \\
\DD[ \hat{J} ] \rhot & = \hat{J} \rhot \hat{J}^\dagger - \frac{ \hat{J}^\dagger \hat{J} \rhot + \rhot \hat{J}^\dagger \hat{J} }{2},
\end{split}
\end{equation}
where $\hat{a}$ and $\hat{a}^\dagger$ are the annihilation and creation operators, obeying $[\hat{a},\hat{a}^\dagger]=1$.
The dissipator superoperators $\DD[ \hat{a}]$ and $\DD[ \hat{a}^\dagger \hat{a}]$ describe particle loss and dephasing processes occurring with rates $\kappa_{1}$ and $\kappa_2$, respectively.
$\LL$ is the Liouvillian superoperator, generating the dynamics of the system. 
The evolution of an initial state $\hat{\rho}(0)$ under \eqref{Eq:Lindblad} for a time $\tau$ defines the \emph{noise channel}
\begin{equation} \label{Eq:Loss_channel_tau}
\begin{split}
 \mathcal{N}_{\kappa_1, \kappa_2} \left(\hat{\rho}(0) \right) &= e^{\mathcal{L} \tau} \hat{\rho}(0)\\ &= \exp\left[\tau\left(\kappa_1\mathcal{D}\left[{\hat{a}}\right] + \kappa_2\mathcal{D}\left[{\hat{a}^\dagger\hat{a}}\right]\right)\right] \hat{\rho}(0) \\ 
 &=\sum\limits_{j=0}^\infty \hat{K}_j \hat{\rho}(0) \hat{K}_j^\dagger,
 \end{split}
\end{equation}
where the map $\mathcal{N}_{\kappa_1, \kappa_2}$ is decomposed in terms of Kraus operators $\hat{K}_j$ ~\cite{BreuerBookOpen, KrausAnnalsofPhysics1971}.
To leading order in $\kappa_{1,2} \tau \ll 1$, the Kraus operators of $\mathcal{N}_{\kappa_1, \kappa_2}$ are
\begin{equation}\label{eq:Kraus_op_small_dt}
 \begin{gathered}
 \hat{K}_0 = \hat{\mathds{1}} - \frac{\kappa_1 \tau}{2} \hat{a}^\dagger \hat{a} - \frac{\kappa_2 \tau}{2} \left(\hat{a}^\dagger \hat{a}\right)^2, \\ 
 \hat{K}_1 = \sqrt{\kappa_1 \tau} \, \hat{a}, \qquad \hat{K}_2 = \sqrt{\kappa_2 \tau} \, \hat{a}^\dagger \hat{a}. 
 \end{gathered}
\end{equation}

In a BQC, the \textit{code space} is the two-dimensional subspace of the full Hilbert space, spanned by the two codeword states which encode the logical $\ket{0}_{\rm L}$ and $\ket{1}_{\rm L}$ states. The code space is ideally defined in such a way that, upon the action of distinct errors, each error results in a rotation onto an error subspace which is orthogonal both to the code space, and to the error subspaces associated to the other errors. Upon the occurrence of an error therefore, quantum information is not lost, but rather encoded in the error subspace. This allows to detect, and thus correct, errors without degrading the quantum information stored in the code.

The correction capability of a BQC for a finite set of errors can be characterized by the Knill–Laflamme (KL) conditions \cite{BennetPRA1996,KnillPRA97,NielsenChuangBIBLE2010}.
Given two distinct errors $\hat{E}_\ell$ and $\hat{E}_{\ell'}$, belonging to the set of errors $\{\hat{E}_k\}$, and codewords $\{\ket{\psi_{0}}, \ket{\psi_{1}}\}$, the KL conditions state that errors $\hat{E}_\ell$ and $\hat{E}_{\ell'}$ can be distinguished and corrected if and only if
\begin{equation}\label{eq:KL}
 \bra*{\psi_i}\hat{E}_\ell^\dagger \hat{E}_{\ell'}\ket*{\psi_j} = \delta_{ij} \alpha_{\ell \ell'}\,,
\end{equation} 
where $\alpha_{\ell \ell'}$ is a Hermitian matrix that is independent of $i$ and $j$.
To leading order in $\kappa_{1,2}\tau$, the KL conditions for the noise channel given in \eqref{Eq:Loss_channel_tau} are expressed in terms of the operators $\hat{K}_{0,1,2}$ defined in \eqref{eq:Kraus_op_small_dt}, or alternatively in terms of the set of errors $\{\hat{E}_k\}=\{\hat{\mathds{1}}, \hat{a}, \hat{a}^\dagger \hat{a}, \left(\hat{a}^\dagger \hat{a}\right)^2 \}$.

The KL conditions determine if a given set of errors can be \textit{exactly} corrected. 
Hereafter, however, we consider a BQC capable of \textit{approximately} correcting errors and assess how much a code is correctable with respect to the action of the noise channel $\mathcal{N}_{\kappa_1, \kappa_2}$ defined in \eqref{Eq:Loss_channel_tau}.
Conditions for approximate quantum error correction have been established starting from the KL conditions~\cite{LeungPRA1997, SchumacherQuInfoProcessing2002, NgPRA2010, BenyPRL2010, MandayamPRA2012}.

Hereafter, the KL conditions will be used as a guideline to assess the ideal performance of codes, in combination with a more quantitative complementary approach based on the notion of \textit{channel fidelity}.
Channel fidelity is suitable for assessing errors as modeled by \eqref{Eq:Lindblad}, is not restricted to short times~\cite{AlbertPRA2018}, and allows quantifying the error correction performance of a given code even if the KL conditions are not exactly fulfilled. The channel fidelity has the additional advantage of providing an explicit form of the recovery operation, needed for actual hardware implementations.

We assume to measure and correct the system periodically at time intervals $\tau$~\cite{NielsenChuangBIBLE2010,AlbertPRA2018}. Thus, over one period $\tau$ the system is subject to (i) the action of the Lindblad master equation for a time $\tau$ through the noise channel $\mathcal{N}_{\kappa_1, \kappa_2} $ [c.f. \eqref{Eq:Loss_channel_tau}]; (ii) the detection of the error syndrome $\mathcal{S}$, i.e. a projective measurement onto the orthogonal code and error subspaces, which we assume to take place instantaneously~\cite{AlbertPRA2018}; (iii) the error correction $\mathcal{C}$, i.e., an operation which projects the system back onto the code space and is also assumed instantaneous. The syndrome measurement $\mathcal{S}$ and correction $\mathcal{C}$ are expressed as a combined \textit{recovery} map, $\mathcal{R}=\mathcal{C}\circ\mathcal{S}$.
The combination of these three operations is then described by the map $\mathcal{E}$ as
\begin{equation}\label{eq:quantum_channel}
 \hat{\rho}(t+\tau) = \mathcal{E} \left(\hat{\rho}(t) \right)= \mathcal{R} \, \left( e^{\LL \tau} \rhot \right) = \mathcal{R} \circ \mathcal{N}_{\kappa_1, \kappa_2} (\rhot),
\end{equation}
The channel $\mathcal{E}$ depends on the dimensionless time $\kappa_\ell\tau$, [c.f. \eqref{Eq:Loss_channel_tau}], implying that a shorter period $\tau$ is equivalent to reducing both dissipation rates.

The channel $\mathcal{E}$ is expressed in terms of Kraus-operators $\hat{S}_\ell$ as 
\begin{equation}\label{Eq:Definition_recovery}
\mathcal{E}\left(\hat{\rho}\right) = \sum\limits_{\ell = 0}^\infty \hat{S}_\ell \hat{\rho} \hat{S}_\ell^\dagger = \sum\limits_{r,j=0}^{\infty}\hat{R}_r\hat{K}_j \hat{\rho} \hat{K}_j^\dagger \hat{R}_r^\dagger,
\end{equation}
where $\hat{K}_j$ are the Kraus operators associated with the noise channel $\mathcal{N}_{\kappa_1, \kappa_2}$ defined in \eqref{Eq:Loss_channel_tau}, and $\hat{R}_r$ are the Kraus operators associated with the recovery $\mathcal{R}$.
We define the \textit{average channel fidelity} \cite{SchuhmacherPRA1996, AlbertPRA2018} as \footnote{As discussed in~\cite{AlbertPRA2018}, the average channel fidelity of a single logical qubit, with codewords $\ket*{0}_{\rm L}$ and $\ket*{1}_{\rm L}$, can be equivalently expressed in terms of the Pauli operators in the code space $\{\hat{\mathds{1}}, \hat{\sigma}_x, \hat{\sigma}_y, \hat{\sigma}_z \}$ as
\begin{equation}
\mathcal{F} = \frac{1}{4}\sum_i{\mathcal{E}_{ii}}\,,
\end{equation}
where $\mathcal{E}_{ij} = \tfrac{1}{2}\Tr{\sigma_i \mathcal{E}(\sigma_j)}$}
\begin{equation}\label{eq:channel_fidelity}
 \mathcal{F}= \frac{1}{4} \sum\limits_{\ell = 0 }^\infty \left| \text{Tr}\left\{\hat{S}_\ell\right\} \right|^2.
\end{equation}
The average channel fidelity quantifies the average performance of a certain recovery operation $\mathcal{R}$ given the noise channel $\mathcal{N}_{\kappa_1, \kappa_2}$ acting on a specific BQC.

An explicit form of the recovery operation $\mathcal{R}$ should be specified, and different choices of $\mathcal{R}$ will, in general, result in different channel fidelities.
As a consequence, for a given BQC subject to the noise $\mathcal{N}_{\kappa_1, \kappa_2}$, an optimal recovery operation $\mathcal{R}$ that maximizes the corresponding channel fidelity can be found. 
The problem of finding the optimal recovery operation $\mathcal{R}$ can be expressed as a convex optimization problem, and solved using semi-definite programming~\cite{YamamotoPRA2005, FletcherPRA2007, KosutQuantum2009}. 
The optimal recovery map $\mathcal{R}$ must be determined for each given value of the dimensionless times $\kappa_1\tau$ and $\kappa_2\tau$~\cite{GrimsmoPRXQuantum2021}.
In Appendix \ref{App:Convex}, we introduce a simple parametrization of the recovery map $\mathcal{R}$ and provide details of the corresponding optimization procedure.

\section{Cat states and squeezing}
\label{Sec:Cats}

\subsection{2-cat states}
\label{Subsec:2-Cats}

Let $\ket*{\alpha}$ be a coherent state, defined via the relation $\hat{a}\ket*{\alpha}=\alpha\ket*{\alpha}$, where $\alpha$ is a complex-valued displacement field. 
Two-component cat states are defined as the even and odd superposition of coherent states with opposite displacement, i.e.
\begin{equation}
 \ket*{\mathcal{C}_\alpha^\pm} = \frac{1}{N_\alpha^\pm}\left(\ket*{\alpha} \pm \ket*{-\alpha} \right),
\end{equation}
where $N_\alpha^\pm$ is a normalization constant.
The states $\ket*{\mathcal{C}_{\alpha}^+}$ and $\ket*{\mathcal{C}_\alpha^-}$ are a superposition of only even and odd number states, respectively, and are therefore orthogonal.
They are eigenstates of the parity operator $\hat{\Pi} = \exp(i \pi \hat{a}^\dagger \hat{a})$ with eigenvalues $\pm 1$. 

Quantum information can be stored in the 2-cat BQC, where the cat states $\ket*{\mathcal{C}_\alpha^\pm}$ encode the basis states $\ket{0}_{\rm L}$ and $\ket{1}_{\rm L}$ of the logical qubit. Cat states suppress the dephasing error, as seen from the KL conditions \eqref{eq:KL}.
In particular, for large $\alpha$, the KL conditions for the set of errors $\left\{\hat{\mathds{1}}, \hat{a}^\dagger\hat{a}\right\}$ are (see also Appendix~\ref{appendixsubsec:catstateproperties})
\begin{equation}\label{Eq:CATKL}
\begin{split}
 &\bra*{\mathcal{C}_\alpha^\pm} \hat{a}^\dagger \hat{a} \ket*{\mathcal{C}_\alpha^\mp} =0 , \\
 & \bra*{\mathcal{C}_\alpha^-} \hat{a}^\dagger \hat{a}\ket*{\mathcal{C}_\alpha^-} - \bra*{\mathcal{C}_\alpha^+} \hat{a}^\dagger \hat{a}\ket*{\mathcal{C}_\alpha^+} \\ 
 & \qquad \quad = 2|\alpha|^2 \csch\left(2|\alpha|^2 \right) \xrightarrow[|\alpha| \gg 1]{} 2|\alpha|^2 e^{-2|\alpha|^2}. 
\end{split}
\end{equation}
As seen from the latter equation, the violation of the KL conditions [c.f. \eqref{eq:KL}] for dephasing errors $\hat{a}^\dagger\hat{a}$ is therefore exponentially small in the cat size $|\alpha|^2$.
 
The 2-cat code cannot correct the error caused by particle loss $\hat{a}$. 
particle loss errors cause a bit flip in the code space, thereby inducing a maximal violation in the KL condition. 
Indeed, particle loss induces a bit-flip in the code space by transforming the even cat state into the odd cat state ($\hat{a}\ket*{\mathcal{C}_{\alpha}^+} \propto \ket*{\mathcal{C}_{\alpha}^-}$) and vice versa and the KL condition for loss reads:
\begin{equation}\label{eq:coherent_cat_particle_loss}
\bra*{\mathcal{C}_\alpha^\mp}\hat{a}\ket*{\mathcal{C}_\alpha^\pm} = \alpha \frac{N_\alpha^{\mp}}{N_\alpha^{\pm}}.
\end{equation}
In a resonator, the particle loss rate increases linearly with $|\alpha|^2$~\cite{Haroche_BOOK_Quantum, GrimmNature2020}. Therefore, when increasing $|\alpha|^2$ in a 2-cat code, there is a trade-off between the (exponentially suppressed) phase-flip error rate and the (linearly increased) bit-flip error rate. 
Overall, the 2-cat code is effective in encoding a \textit{biased-noise} qubit on platforms where the particle loss rate is much smaller than the dephasing rate~\cite{JoshiQST21}. Within the biased-noise paradigm, the 2-cat code can be autonomously stabilized and controlled in a dissipation-engineered environment~\cite{MirrahimiNJP14, PurinpjQI17, PuriPRX2019, AlbertIOP2019, GrimmNature2020, PuriScienceAdvances2020, GertlerNature2021,MingantiSciRep16,BartoloPRA16,WangScience2016}.

\subsection{Squeezed cat states}
\label{Subsec:SCCs}

\begin{figure*}[ht]
 \includegraphics[width=\textwidth]{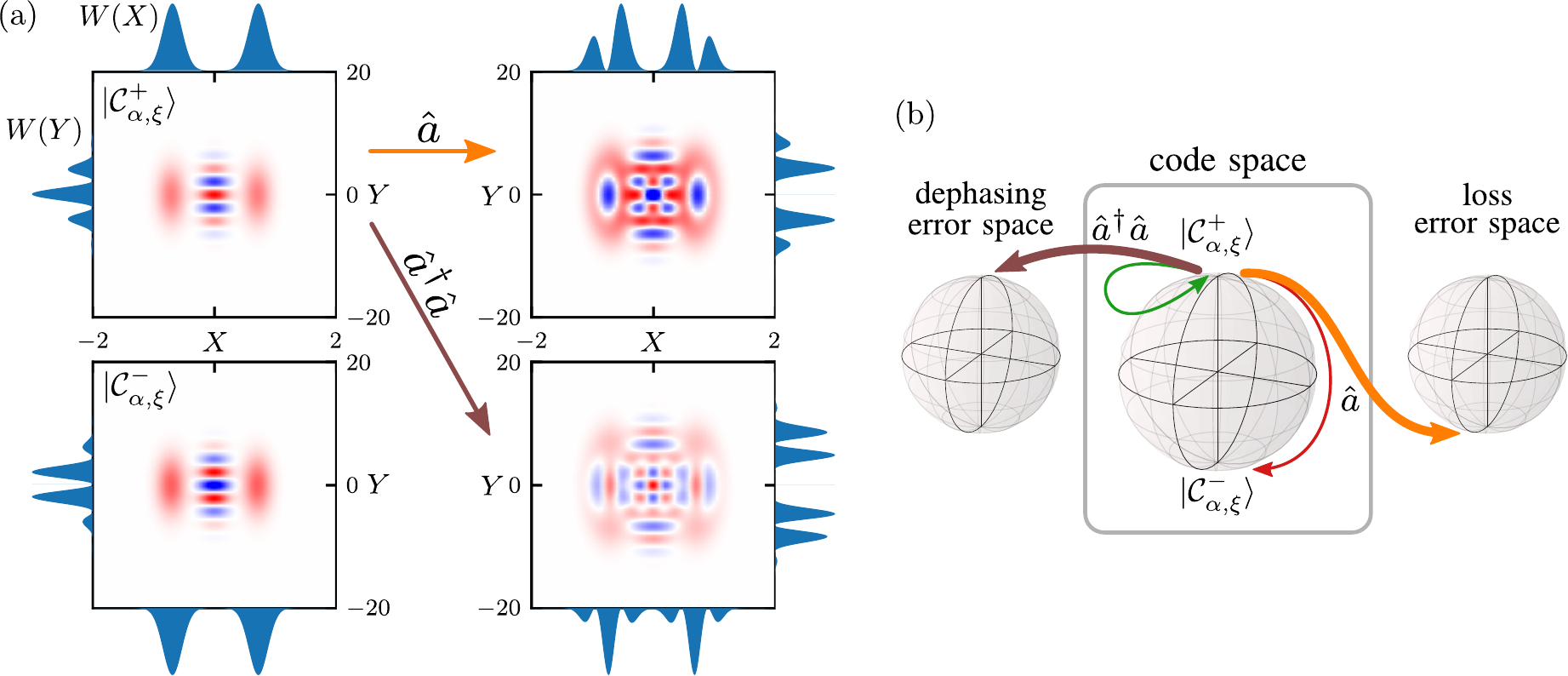}
 \caption{{(a) Wigner quasi-probability distribution function $W(X, Y)$ of the codewords \eqref{Eq:SCC} (left) and of two states generated by errors (right), for the SC code. $\alpha=0.5$, $\xi=1.5$ for all plots. On the sides of each panel, the integrated probabilities $W(X)$ and $W(Y)$, with $X = \tfrac{1}{2}\expval{\hat{a} + \hat{a}^\dagger}$ and $Y = \tfrac{i}{2}\expval{\hat{a}^\dagger - \hat{a}}$, are plotted. The scales of the quadratures $X$ and $Y$ differ by one order of magnitude, highlighting the squeezed nature of the code. Upon the action of particle loss $\hat{a}$ or dephasing $\hat{a}^\dagger\hat{a}$, as depicted for the state $\ket*{\mathcal{C}_{\alpha, \xi}^+}$, the codeword deforms in phase-space. (b) Schematic illustration of the effect of loss and dephasing errors on the state $\ket*{\mathcal{C}_{\alpha, \xi}^+}$. Codewords $\ket*{\mathcal{C}_{\alpha, \xi}^\pm}$ lie respectively at the north and south poles of Bloch sphere of the logical qubit. For finite squeezing, the state resulting from either particle loss or dephasing acting on $\ket*{\mathcal{C}_{\alpha, \xi}^+}$ has finite components both on the code space and on the respective error subspaces, as indicated by the arrows.}}
 \label{fig:action_a_n_on_cat}
\end{figure*}

Here, we introduce the squeezed-cat code, namely a two-component cat code based on squeezed-displaced states. A squeezed-displaced state is defined in terms of its displacement $\alpha$ and squeezing parameter $\xi$ as:
\begin{equation}\label{eq:squeezed-displaced-state}
\ket*{\alpha, \xi} = \hat{D}(\alpha)\hat{S}(\xi)\ket*{0},
\end{equation}
where $\hat{D}(\alpha)$ and $\hat{S}(\xi)$ are the displacement and squeezing operator, respectively, defined by
\begin{align}\label{eq:displacement-squeezing}
\hat{D}(\alpha) &= \exp\left[\alpha \hat{a}^\dagger - \alpha^* \hat{a} \right], \\
\hat{S}(\xi) &= \exp\left[\tfrac{1}{2}\left(\xi^* \hat{a}^2 - \xi {{}\hat{a}^\dagger}^2\right)\right].
\end{align}
In what follows, squeezing is always assumed to be orthogonal to the direction of displacement, by setting $\xi=|\xi|e^{i\theta/2}$, $\alpha= |\alpha| e^{i\theta}$. Without loss of generality, we set $\theta =0$ in the analysis that follows. This implies in particular that $\alpha$ is a positive real-valued quantity. Similarly to coherent states with opposite displacements, squeezed-displaced states are quasi-orthogonal for large displacement and squeezing parameters, as 
\begin{equation}\label{Eq:SCorth}
 \expec{\alpha, \xi| -\alpha, \xi} = \exp(-2 |\alpha|^2 e^{2 |\xi|}) .
\end{equation}
The codewords in the SC code are defined as
\begin{equation}\label{Eq:SCC}
 \ket*{\mathcal{C}_{\alpha,\xi}^\pm} = \frac{1}{N_{\alpha,\xi}^\pm} \left(\ket*{\alpha, \xi} \pm \ket*{-\alpha, \xi}\right),
\end{equation}
where $N_{\alpha, \xi}^\pm$ is a normalization constant.
The Wigner function of $\ket*{\mathcal{C}_{\alpha,\xi}^+}$ and $\ket*{\mathcal{C}_{\alpha,\xi}^-}$ are displayed in Fig.~\ref{fig:action_a_n_on_cat}(a).

\subsubsection{Asymptotic discrete translation invariance}
In the limit of large squeezing, the SC codewords are invariant under phase-space translation orthogonal the displacement direction $\alpha$. For any finite displacement $\eta$ orthogonal to $\alpha$, we have
\begin{equation}\label{eq:scc_translation_invariance}
\begin{split}
 \expval*{\hat{D}(\eta)}{\mathcal{C}_{\alpha, \xi}^+} &= \cos(|\alpha \eta|)\exp\left(-\frac{1}{2}e^{-2|\xi|}|\eta|^2\right) \\
 &\xrightarrow{\xi \rightarrow \infty } \cos(|\alpha \eta|).
 \end{split}
\end{equation}
A similar result can be proven for the odd SC state $\ket*{\mathcal{C}_{\alpha, \xi}^-}$. Hence, in the limit of large squeezing, the SC code is invariant under translations along one direction in phase space, by $|\eta| = n \frac{2\pi}{|\alpha|}$, with $n \in \mathds{Z}$. 
This property bears close analogy with the translation invariance of the the squeezed comb state~\cite{ShuklaPRA2021} and the GKP code~\cite{GottesmanPRA01}, although the latter is invariant under discrete translations along two directions in phase space.
As discussed in Refs.~\cite{JoshiQST21,AlbertPRA2018, GrimsmoPRXQuantum2021}, the translational invariance in the GKP code entails its capability to approximately detect and correct particle loss errors. In the SC code, particle loss results in a phase-space change to the codewords similar to the one occurring in the GKP code, as can be inferred in particular by the integrated probability distribution function $W(X)$ shown in \figref{fig:action_a_n_on_cat}(a). Under this light, the SC code enables approximate correction of both particle loss and dephasing errors by combining the error-correction capabilities of the 2-cat and GKP codes. 

\subsubsection{Code Generation and Gates}

\label{sec:Gates}

Using the relations
\begin{equation}\label{Eq:fundamental_commutation}
\begin{gathered}
 \hat{D}(\alpha)\hat{S}(\xi) = \hat{S}(\xi)\hat{D}(\zeta), \\
 \zeta = \alpha\cosh{(|\chi|)} - \alpha^* e^{i\theta} \sinh{(|\chi|)}, \quad \chi/|\chi| =e^{i \theta},
\end{gathered}
\end{equation}
any linear combination of squeezed-displaced states can be expressed as
\begin{equation}\label{Eq:SCC_as_squeezed_2-cats}
 c_1 \ket{\alpha, \xi} + c_2 \ket{-\alpha, \xi} = S(\xi) \left(c_1 \ket{\zeta} + c_2\ket{-\zeta} \right).
\end{equation}
In other words, any SC can be obtained by squeezing a 2-cat code (with a different amplitude).
Therefore, the SC code can potentially be implemented on all platforms where squeezed-displaced states can be generated, including superconducting circuits~\cite{Yurke:87, GopplJAP2008}, 3D superconducting cavities~\cite{ReagorAPL2013}, cavity quantum optomechanics~\cite{AspelmeyerRMP2014}, and photonics~\cite{JeannicPRL2018, MyattNature2000, OurjoumtsevScience06, Neergaard-NielsenPRL2006, TakahashiPRL2008, Ourjoumtsev2007, EtessePRL2015}.
For instance, an SC state can be generated by squeezing a 2-cat state generated through standard two-photon processes \cite{GillesPRA94,LeghtasScience15}.

A different scheme to generate an arbitrary squeezed-coherent state $\ket*{\alpha, \xi}$ is based on a series of squeezing and displacements of lower magnitude, namely
\begin{equation}
 \hat{D}(\alpha)\hat{S}(\xi) = \prod_n \hat{D}(\alpha_{n})\hat{S}(\xi_{n}),
\end{equation} 
The equality holds for an appropriate choice of $\alpha_{n}$ and $\xi_{n}$.
Starting again from \eqref{Eq:fundamental_commutation}, it can be shown that
\begin{equation}
\begin{split}
 \alpha_{0} &= \frac{\alpha}{N}, \qquad \xi_{n} = \xi/N
 \\
 \alpha_{m} &= \alpha_{m-1} \cosh(r) - \alpha_{m-1}^* \sinh(r) e^{i\theta}.
 \end{split} 
\end{equation}
Finally, by means of a parity measurement, or through beam-splitter operations, the codewords of the SC code can be prepared from the squeezed-displaced states.

Gates can be implemented similarly to those for 2-cat codes \cite{MirrahimiNJP14}.
Indeed, as it stems from \eqref{Eq:SCC_as_squeezed_2-cats}, any arbitrary gate $\hat{G}_\text{SC}(\alpha, \xi)$ on the SC code can be obtained in a straightforward way by applying a squeezing operator on a gate designed for the 2-cat code, according to the relation 
\begin{equation}\label{eq:scc_cat_gates}
\hat{G}_\text{SC}(\alpha, \xi) = \hat{S}(\xi)\hat{G}_\text{cat}(\zeta)\hat{S}^\dagger(\xi).
\end{equation}

Finally, we remark that the SC code can be stabilized via engineered dissipation.
A 2-cat is stabilized by a dissipator $ \mathcal{D}\left[\hat{a}^2 - \alpha^2 \right]$, or equivalently by a two-photon dissipator and a coherent two-photon drive~\cite{MirrahimiNJP14}.
Similarly, the stabilization of squeezed cat states requires an engineering of squeezed two-photon dissipation processes, such that
\begin{equation}
 \mathcal{D}\left[\hat{b}^2 - \zeta^2\right] \ketbra*{\mathcal{C}_{\alpha, \xi}^\pm}{\mathcal{C}_{\alpha, \xi}^\pm} = 0,
\end{equation}
where $\zeta$ has been given in \eqref{Eq:fundamental_commutation}, and $\hat{b}$ is the squeezed (Bogoliubov) mode, reading
\begin{equation}
 \hat{b} = \hat{S}(\xi)\hat{a}\hat{S}^\dagger(\xi) = \cosh(r) \hat{a} + \sinh(r) e^{i \theta} \hat{a}^\dagger.
\end{equation}
Similar dissipative stabilization of squeezed modes have been proposed in the context of, e.g., squeezed lasing \cite{Munozarxiv20}, by means of resonantly coupling the squeezed mode to an additional reservoir, possibly adiabatically eliminated.

\subsection{Error correction conditions for the SC code}
\label{Subsubsec:KL}

Here, we want to determine if, in some ideal case, the SC code can perform error correction of particle loss and dephasing errors. This first step, based on KL conditions, is the starting point to determine the error approximate error correction capability of the code in realistic scenarios.

Let us first consider a 2-cat state $\ket*{\mathcal{C}_{\alpha}^\pm}$, subject to a particle loss event.
Since $\hat{a}\ket*{\mathcal{C}_{\alpha}^\pm} \propto \ket*{\mathcal{C}_{\alpha}^\mp}$, no syndrome will be able to detect if such an error occurred.
%
%
%
%
%
Consider now the SC codewords $\ket*{\mathcal{C}_{\alpha, \xi}^\pm}$.
Here, the action of the error $\hat{a}$ maps any state in the SC code space into a state that lies partially outside of it,
\begin{equation}
 \hat{a} \ket*{\mathcal{C}_{\alpha, \xi}^\pm} = c \ket*{\mathcal{C}_{\alpha, \xi}^\mp} + d \ket*{\tilde{\mathcal{C}}_{\alpha, \xi}^\mp},
\end{equation}
where the states $\ket*{\tilde{\mathcal{C}}_{\alpha, \xi}^\mp}$ span an error space orthogonal to the code space.
For instance, $\ket*{\mathcal{C}_{\alpha, \xi}^+}$ is an even state, thus $\hat{a}\ket*{\mathcal{C}_{\alpha, \xi}^+}$ is a state with odd particle number parity (and vice versa for $\ket*{\mathcal{C}_{\alpha, \xi}^-}$ and $\hat{a}\ket*{\mathcal{C}_{\alpha, \xi}^-}$).
Therefore, $\hat{a}\ket*{\mathcal{C}_{\alpha, \xi}^+}$ can be decomposed into a contribution within the code space and into
 $\ket*{\tilde{\mathcal{C}}_{\alpha, \xi}^\pm}$, representing its orthogonal complement.
This is due to the fact that squeezed-coherent states are no longer eigenstates of $\hat{a}$.
Thus, states $\ket*{\tilde{\mathcal{C}}_{\alpha, \xi}^\pm}$ carry information about the occurrence of the error.
Since, by construction, $\ket*{\mathcal{C}_{\alpha, \xi}^\mp}$ and $\ket*{\tilde{\mathcal{C}}_{\alpha, \xi}^\mp}$ are orthogonal, one could identify a syndrome that checks whether the state is in the \textit{code space} or in the \textit{error space}.


This picture illustrates the mechanism to detect errors in the SC, and can be generalized to multiple error spaces,
as shown in \figref{fig:action_a_n_on_cat}(a) and illustrated in \figref{fig:action_a_n_on_cat}(b), where we also consider dephasing processes.
In this case, the loss ($\hat{a}$) and dephasing ($\hat{a}^\dagger\hat{a}$) errors transform the SC codewords into states with finite components outside the code space. 
To quantify the error correction capability, here we characterize error correction in terms of the Knill-Laflamme conditions. Technical details of this analysis are provided in Appendix \ref{appendixsubsec:SCCproperties} and in the Supplementary Material.

\subsubsection{KL condition for the SC}

Let us consider the set of errors $\{\hat{\mathds{1}}, \hat{a}, \hat{a}^\dagger \hat{a}\}$.
For the action of $\hat{a}$ on the SC codewords, in the limit of large squeezing (see also Appendix~\ref{App:KL}), we have 
\begin{equation}\label{KL:loss_SC}
 \bra*{\mathcal{C}_{\alpha, \xi}^\mp}\hat{a}\ket*{\mathcal{C}_{\alpha,\xi}^\pm} \to \alpha.
\end{equation}
Therefore, in this limit, the loss error $\hat{a}$ is increasingly correctable by reducing the amount of displacement $\alpha$.

The SC codewords are eigenstates of the particle number parity $\hat{\Pi}$ and thus $\bra*{\mathcal{C}_{\alpha, \xi}^\pm} \hat{a}^\dagger \hat{a} \ket*{\mathcal{C}_{\alpha, \xi}^\mp}=0$.
Hence, the KL conditions require
\begin{equation}
 \bra*{\mathcal{C}_{\alpha, \xi}^+} \hat{a}^\dagger \hat{a} \ket*{\mathcal{C}_{\alpha, \xi}^+} = \bra*{\mathcal{C}_{\alpha, \xi}^-} \hat{a}^\dagger \hat{a} \ket*{\mathcal{C}_{\alpha, \xi}^-}.
\end{equation}
This condition is satisfied by the SC code in the limit of large displacement or, at finite displacement, in the limit of large squeezing [see details in Appendix \ref{appendixsubsec:SCCproperties}, Eqs.~(\ref{Eq:KL_SC_dephasing1})~and~(\ref{Eq:KL_SC_dephasing2})].
In this limit, one can show that
\begin{equation}
 \bra*{\mathcal{C}_{\alpha, \xi}^+} \hat{a}^\dagger \hat{a} \ket*{\mathcal{C}_{\alpha, \xi}^+} - \bra*{\mathcal{C}_{\alpha, \xi}^-} \hat{a}^\dagger \hat{a} \ket*{\mathcal{C}_{\alpha, \xi}^-} \simeq 2 |\alpha| ^2 e^{{-2 |\alpha| ^2 \exp({2 |\xi|})}}
\end{equation}
The improved suppression of dephasing errors is intuitively explained as a consequence of the factor $e^{2 |\xi|}$ appearing in \eqref{Eq:SCorth}, which makes the value of the right-hand-side smaller than in the $\xi=0$ case [c.f. \eqref{Eq:CATKL}]. 

The KL coefficients for loss for the SC in \eqref{KL:loss_SC} and for the 2-cat in \eqref{eq:coherent_cat_particle_loss} have approximately the same magnitude for the same value of the displacement $\alpha$. Therefore, in the limit of large displacement, the two codes are equally vulnerable to particle loss. 
Notably, increasing the squeezing $\xi$ does not increase the KL coefficient in \eqref{KL:loss_SC}, and SC states with very large squeezing $\xi$ become correctable from the action of $\hat{a}$ for small $\alpha$.
The main difference with respect to 2-cats is that, despite the small $\alpha$, the SC is also correctable with respect to $\hat{a}^\dagger \hat{a}$.

Summarizing, an ideal SC code allows correction of particle loss errors, while also providing a correction from dephasing errors~\cite{JeannicPRL2018}.
The intuition is that we can increase the correctability of particle loss by setting a lower displacement $\alpha$, and compensate for the reduced performance against dephasing by increasing the squeezing parameter $\xi$.
Notice that this is not the case in the GKP code, where dephasing errors become increasingly less correctable when increasing the amount of squeezing (see Appendix~\ref{Appendix:GKP_vs_SC}).

In any realistic application, squeezing is a limited resource. 
Nonetheless, for finite squeezing $\xi$, these results hint that the SC code enables approximate detection and correction of particle loss and dephasing errors.
In this non-ideal case, optimal values of displacement and squeezing must be determined as a function of the physical error rates. This optimization procedure is detailed in the Appendix~\ref{sec:Convex recovery optimization}, and it underlies the channel fidelity study in Sec.~\ref{Sec.SCC_channel}.

\subsubsection{Correction of multiple errors}

In the following, to demonstrate the SC code performance with respect to the standard 2-cat code, we will focus on dephasing and loss errors to leading order in $\tau$ [c.f. \eqref{eq:Kraus_op_small_dt}].
Nevertheless, the SC code can also correct for other errors, such as particle gain or higher-order dephasing or particle-loss processes. 
This follows from the fact that, in the limit of large displacement, or finite displacement and large squeezing, the action of $\hat{a}^\dagger$ and $\hat{a}$ on the code become similar (up to a phase). 
As an illustrative example, in this limit the KL conditions for $\hat{a}^\dagger \hat{a}$ and $\hat{a}^2$ are given by
\begin{equation}
 \expval*{\hat{a}^\dagger \hat{a}}{\mathcal{C}_{\alpha, \xi}^\pm} = \frac{e^{2|\xi|}}{4} = -\expval*{\hat{a}^2}{\mathcal{C}_{\alpha, \xi}^\pm}.
\end{equation}
Hence the KL conditions for $\hat{a}^\dagger \hat{a}$ and $\hat{a}^2$ become identical (up to a sign in this limit). 
Since, as previously demonstrated, $\hat{a}^\dagger\hat{a}$ is a correctable error, so is $\hat{a}^2$. 
This is a fundamental difference between the SC and other bosonic codes, such as binomial codes~\cite{MariosPRX16}.
%
In this limit, the SC code thus constitutes a degenerate bosonic quantum code in which multiple errors act identically onto the code space. It is therefore possible to devise error correction schemes also for errors of higher order.

\section{Approximate Quantum Error Correction for the squeezed cat code}
\label{Sec.SCC_channel}

As we detailed above, assessing the performance of the code for finite values of squeezing requires to explicitly take into account both the displacement $\alpha$ and the squeezing $\xi$, resulting in a trade-off between the correction of dephasing and particle loss errors.
In this regard, to assess the performance of the code via the channel fidelity described in \eqref{eq:channel_fidelity}, we need to evaluate (i) the optimal encoding and (ii) the optimal recovery.

\subsection{Recovery and Encoding}
\label{Subsec:Rec_and_enc}

To construct the optimal encoding and recovery, we will consider a leading-order Kraus representation expansion of the Liouvillian in \eqref{eq:Kraus_op_small_dt}.
However, when gauging the performance of the code, we will consider the full Liouvillian noise channel in \eqref{Eq:Loss_channel_tau}, thus allowing all orders of errors to occur. 

To define a recovery operation $\mathcal{R}$ for the SC code, we first characterize the error subspaces [see also \figref{fig:action_a_n_on_cat}(b)]. 
To this purpose, we define the 8-dimensional subspace generated by applying the operators $\{\hat{\mathds{1}}, \hat{a}, \hat{a}^\dagger \hat{a}, \left(\hat{a}^\dagger \hat{a}\right)^2\}$ [i.e., the operators in \eqref{eq:Kraus_op_small_dt}] onto the codewords $\ket*{\mathcal{C}_{\alpha, \xi}^\pm}$. This subspace contains the code space.
A Gram-Schmidt orthonormalization procedure with respect to the code space then results in the three single-error subspaces, respectively generated by the operators $\{\hat{a}, \hat{a}^\dagger \hat{a}, \left(\hat{a}^\dagger \hat{a}\right)^2\}$.
The recovery operation is defined as the map from these error subspaces back onto the code space as
\begin{equation}
 \mathcal{R}\left(\hat{\rho} \right) = \sum\limits_{r=0}^3 \hat{R}_r \hat{\rho} \hat{R}_r^\dagger\,,
 \label{eq:recovery}
\end{equation}
where $r=0$ denotes the code space and $r=1,2,3$ the three single-error subspaces. Here, each operator $\hat{R}_r$ transforms the $r$-th subspace onto the code space.
\eqref{eq:recovery} is the most general form of a recovery operation for single errors, however it does not define the action of $\mathcal{R}$ onto states lying outside both code space and single-error subspace.
Here, we assume that the recovery acts as the identity outside these subspaces. Detail on the characterization of the error subspaces is given in Appendix \ref{appendixsubsec:errorsubspaces}.

For given values of the loss and dephasing rates $\kappa_{1, \,2}$ (or for different times $\tau$), the recovery operation must be optimized in order to maximize the channel fidelity. 
The optimal recovery operators $\hat{R}_r^\text{opt}$ are obtained through convex optimization, as detailed in Appendix \ref{sec:Convex recovery optimization}.
The recovery optimization yields an explicit form of the recovery. As the optimal recovery is directly obtained in an operator-sum representation constituting a quantum map [c.f. \eqref{eq:recovery}], actual hardware operations could be engineered to match the desired target operation using, e.g., pulse-engineering techniques
~\cite{KhanejaJournalMagneticResonance2005, GlaserEuropeanPhysJournalD2015}.
As an alternative recovery procedure, similarly to the gate operations discussed in Sec. \ref{sec:Gates}, the recovery operation could be carried out by (i) unsqueezing the state onto a 2-cat basis; (ii) applying the recovery gates; (iii) squeezing the state back to restore the SC.
In an autonomously stabilized implementation, such recovery operation would thus amount to inject a particle upon the occurrence of a particle loss event, in close resemblance to 4-cats \cite{MirrahimiNJP14}. 

Notice that the overall channel fidelity depends both on the displacement $\alpha$ and squeezing $\xi$, as the probabilities of single loss and dephasing events in a short time $\tau$ scale respectively as
\begin{equation}\label{eq:error_probabilities}
 \begin{split}
 p_1 &= \kappa_1 \tau \expval{\hat{a}^\dagger\hat{a}},\\
 p_2 &= \kappa_2 \tau \expval{(\hat{a}^\dagger \hat{a})^2}.
 \end{split}
\end{equation}
These quantities grow with both $\alpha$ and $\xi$, making the occurrence of error events more probable.
However, the correction capability of the code also increases.
For this reason, a nested optimization procedure is needed, whereby we perform convex optimization of the recovery operation, while also varying the parameters $\alpha$ and $\xi$ to find the best possible encoding and recovery. 

As a final remark, notice that in the procedure just summarized, the optimal recovery operation is uniquely defined for the code and is not conditional to the outcome of the syndrome measurement. In this sense, the present SC code is \textit{semi-autonomous}, in that it requires a periodic syndrome measurement, but only an unconditional recovery operation. In principle, a generalized procedure could be developed for the SC code, where the recovery operation is chosen conditionally to the outcome of the syndrome measurement. While the conditional approach may further improve the error correction efficiency of the code, its advantage should be weighted against its ease of implementation in each specific quantum hardware platform.

\subsection{Channel fidelity for dephasing and loss}

\begin{figure*}[ht]
 \includegraphics[width=\textwidth]{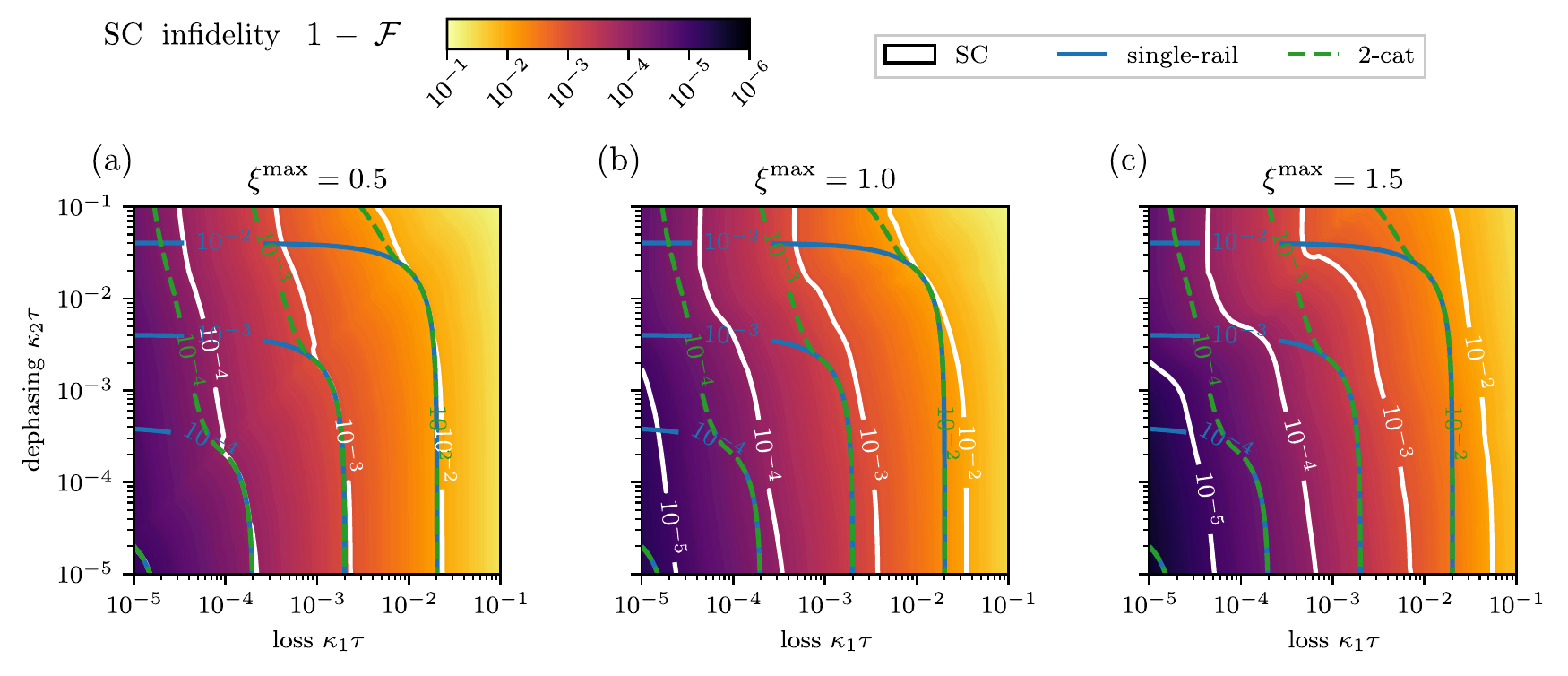}
 \caption{
 Color-map plot of the channel infidelity $1-\mathcal{F}$ [c.f. \eqref{eq:channel_fidelity}], as computed for the optimal SC code for various maximum squeezing parameters: $\xi^\mathrm{max}=$ $0.5$ (a), $1.0$ (b), and $1.5$ (c) as a function of the particle loss parameter $\kappa_1\tau$ and dephasing parameter $\kappa_2\tau$. 
 The white contour plot represents the channel-infidelity iso-lines for the SC code.
 The other iso-lines represent the channel infidelity computed for the single-rail code (blue) and the optimal 2-cat code (green).}
 \label{fig:channel_fidelity}
\end{figure*}

We compute the optimal channel fidelity $\mathcal{F}$ defined in \eqref{eq:channel_fidelity}, as a function of the dimensionless times $\kappa_1\tau$ and $\kappa_2\tau$, by applying the optimization procedure introduced in the previous Section (See also Appendix \ref{App:Convex}). 
Since $\tau$ is the time between two subsequent recoveries, a larger $\kappa_1\tau$ or $\kappa_2\tau$ implies either stronger dissipation or longer intervals between two recoveries.
The channel fidelity $\mathcal{F}$ provides a measurement of the code performance (for the best encoding and recovery) given the jump rates in \eqref{eq:error_probabilities}.
A higher population in the bosonic mode induces higher loss and dephasing error rates, but, according to the KL conditions, the performance of the code also increases with the amount of squeezing. The best possible encoding (i.e., the amount of squeezing and the displacement), is therefore determined by an optimization procedure which we carry out numerically.

As shown in Sec. \ref{Sec:Cats}, large squeezing and low displacement is favorable to simultaneous error correction of particle loss and dephasing.
However, to be consistent with squeezing achievable in state-of-the-art hardware platforms, the optimization procedure is carried out by setting an upper bound $\xi^\mathrm{max}$ to the allowed values of the squeezing parameter $\xi$.
In Fig. \ref{fig:channel_fidelity}(a-c), the channel infidelity $1-\mathcal{F}$ is plotted as a function of $\kappa_1\tau$ and $\kappa_2\tau$, for three values of the maximal squeezing parameter, $\xi^\mathrm{max} = 0.5$, $\xi^\mathrm{max} = 1$, and $\xi^\mathrm{max} = 1.5$.
The color-plot and the white iso-lines are the results for the optimal SC code.

We compare the performances of the SC to the \emph{optimal} 2-cat code (green iso-lines) and to the single-rail code (i.e., the encoding onto the number states $\ket{0}$ and $\ket{1}$, in blue). This choice is justified by the fact that the SC -- like the 2-cat code and the single rail code as its limiting case -- entails at most quadratic operations. These codes are therefore expected to require similar resources to be experimentally realized on a given hardware platform.
For the 2-cat code, only the coherent amplitude $\alpha$ is optimized, and the recovery simply transforms the state resulting from a dephasing error directly back onto the code space, while disregarding the non-correctable particle loss errors.
The single-rail code, instead, is only subject to the noise channel $\mathcal{N}_{\kappa_1, \kappa_2}$ without any recovery operation.
In Appendix \ref{Appendix:GKP_vs_SC} we also compare the performances of SC and GKP with same amount of squeezing via the KL conditions.
Our analysis highlights the similarities between the GKP and SC for particle loss correction, but shows that GKP cannot fully correct for dephasing errors, independently of the amount of squeezing.

The SC code asymptotically approaches the 2-cat code in the limit of $\xi \rightarrow 0$, and the 2-cat code approaches the single-rail code in the limit of $\alpha \rightarrow 0$. Hence, the relation $\mathcal{F}^\text{SC, opt} \geq \mathcal{F}^\text{cat, opt} \geq \mathcal{F}^\text{single-rail}$ holds in general.
In particular, for $\xi^\mathrm{max}=0.5$ in Fig. \ref{fig:channel_fidelity}(a), and for large one-particle loss parameter $\kappa_1\tau$, the channel fidelities of the three codes nearly coincide, indicating that their correction capability saturates.

For increasing squeezing parameter $\xi$, the channel fidelity of the SC code increasingly improves with respect to the 2-cat and single-rail codes. In the regime dominated by particle loss $\kappa_1 > \kappa_2$, the SC code outperforms the 2-cat code thanks to the optimal error detection and recovery procedure described above. The SC code outperforms the 2-cat code also in the regime where dephasing dominates, in accordance to the experimental results in Ref.~\cite{Lo2015}. This improved performance of the SC code originates from smaller overlap between squeezed states of opposite displacement, with respect to coherent states with the same displacement $\alpha$, as discussed above. The SC code also displays a significantly increased channel fidelity with respect to the 2-cat code in the region where $\kappa_1 \sim \kappa_2$, highlighting the potential of this code for mainstream quantum hardware platforms.

The requirement to maximize the channel fidelity under the combined action of loss and dephasing channel causes the upper bound on squeezing to also set an upper bound on the particle number.
We report in Appendix \ref{appendixsubsec:Characteristics_of_the_optimally-resistant_state} the average particle number, the squeezing, and the displacement, as evaluated for the optimal encoding shown in Figs.~\ref{fig:channel_fidelity}(a-c).

\begin{figure}[htb]
 \includegraphics[width=\linewidth]{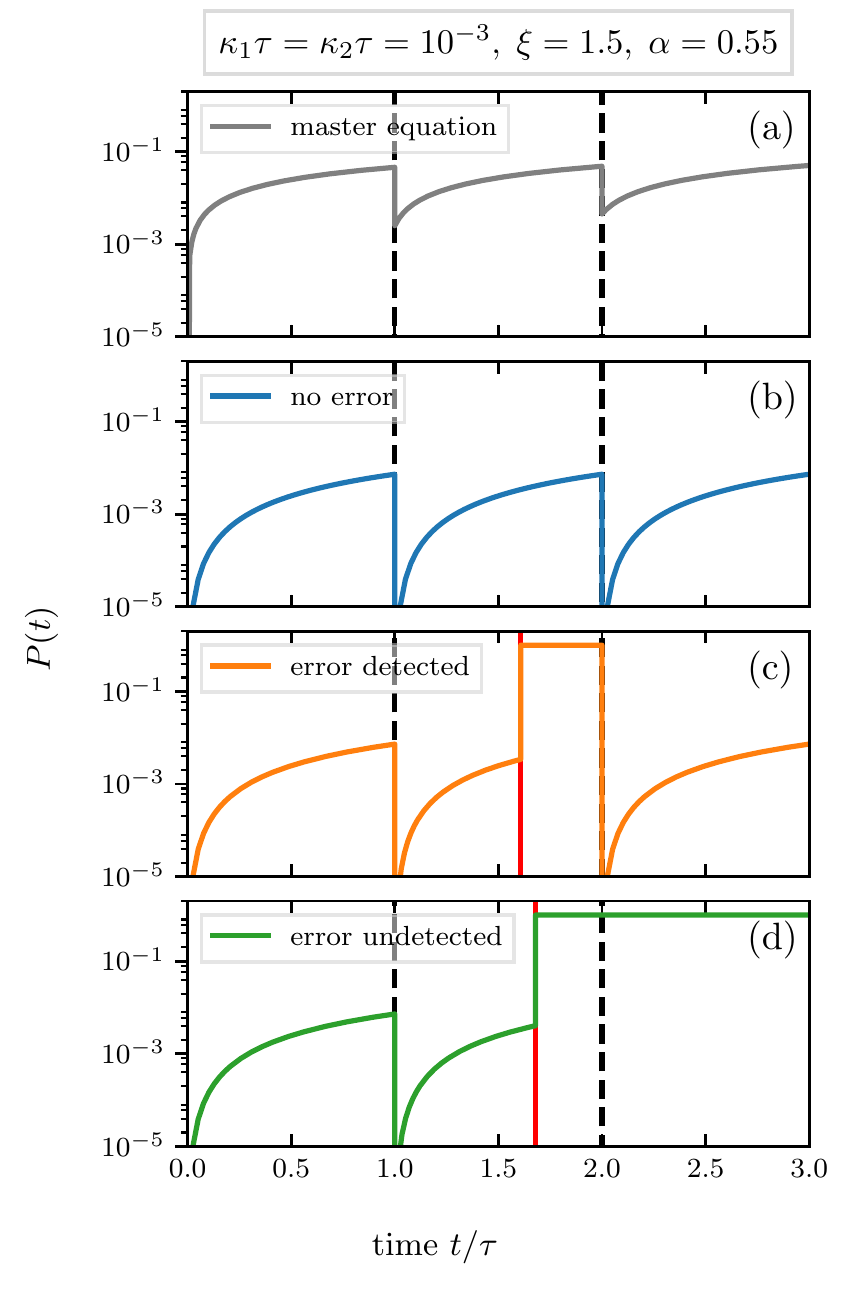}
 \caption{The quantity $P(t)$ defined in Eqs.~(\ref{Eq:P(t)_1})~and~(\ref{Eq:P(t)_2}) of a time-evolved quantum state initially prepared in an even SC state $\ket*{\mathcal{C}_{\alpha, \xi}^+}$ with squeezing $\xi=1.5$ and displacement $\alpha = 0.55$ for a loss channel with loss and dephasing parameters $\kappa_1\tau = \kappa_2\tau = 10^{-3}$ as a function of time $t/\tau$.
 (a) $P(t)$ computed from the Lindblad master equation [c.f. \eqref{Eq:P(t)_1}]. (b) $P(t)$ computed from a single quantum trajectory in the case where no error occurs [c.f. \eqref{Eq:P(t)_2}]. (c) $P(t)$ computed from a single quantum trajectory in the case where an error $\hat{a}$ (depicted in red) occurs and is correctly detected and corrected. (d) $P(t)$ computed from a single quantum trajectory in the case where an error $\hat{a}$ occurs that is undetected and therefore not corrected.
 In all panels, the dashed vertical line indicates the stroboscopic time at which the recovery operation is applied.
 In panels (c)~and~(d) the vertical red lines indicates the occurrence of a particle loss. 
 }
 \label{fig:sample_trajectories_and_effective_decay_rate}
\end{figure}

To illustrate the effect of particle loss and recovery on the SC code, in \figref{fig:sample_trajectories_and_effective_decay_rate} the behavior of the code under the action of the map $\mathcal{E}$, i.e., the combined action of $\mathcal{N}_{\kappa_1, \kappa_2}$ and $\mathcal{R}$, is displayed. 
More precisely, in \figref{fig:sample_trajectories_and_effective_decay_rate}(a) we show the quantity 
\begin{equation}\label{Eq:P(t)_1}
 P(t)=1-\bra*{\mathcal{C}_{\alpha, \xi}^+}\rhot\ket*{\mathcal{C}_{\alpha, \xi}^+},
\end{equation}
representing the difference in overlap between the initial state and the state of the system at time $t$.
The code at $t=0$ is set to the state $\hat{\rho}(0)=\ketbra*{{\mathcal{C}_{\alpha, \xi}^+}}$, for which $P(0)=0$, and the quantity $P(t)$ is plotted as a function of time.

In Figs.~\ref{fig:sample_trajectories_and_effective_decay_rate}(b-d) single quantum trajectories \cite{Molmer1993,DalibardPRL92,Gardiner_BOOK_Quantum,BreuerBookOpen} sampled from the map $\mathcal{E}$ are plotted. In a quantum trajectory, an initial state $\ket*{\psi(0)}$ evolves in time according to a non-Hermitian effective Hamiltonian
\begin{equation}\label{eq:Heff}
 \hat{H}_{\rm eff} = -i \frac{\kappa_1}{2} \hat{a}^\dagger \hat{a} -i \frac{\kappa_2}{2} \hat{a}^\dagger \hat{a}\hat{a}^\dagger \hat{a}
\end{equation}
until a quantum jump occurs after a random time interval.
The non-Hermitian effective Hamiltonian represents the back-action of the noise process, in which information is gained about the system when no jump occurs~\cite{BreuerBookOpen}.

At every time $\tau$, instead, the recovery operation acts.
The quantum trajectory is an \textit{unraveling} of the Lindblad master equation, which is recovered by averaging over the statistical ensemble of quantum trajectories.
In this case,
 \begin{equation}\label{Eq:P(t)_2}
 P(t)=1-|\bra*{\mathcal{C}_{\alpha, \xi}^+}\ket*{\psi(t)}|^2,
 \end{equation}
 where the code at $t=0$ is set to the state $\ket*{\psi(0)}=\ket*{{\mathcal{C}_{\alpha, \xi}^+}}$.
\figref{fig:sample_trajectories_and_effective_decay_rate}(b) shows a quantum trajectory in the case where no error occurred. In this case, the finite value of $P(t)$ is only due to the time evolution under the non-Hermitian Hamiltonian (\ref{eq:Heff}). The recovery operation in the absence of errors therefore undoes this non-Hermitian evolution by driving the system back onto the code space.
\figref{fig:sample_trajectories_and_effective_decay_rate}(c) displays a quantum trajectory in the case where a particle loss error occurs and is successfully corrected. In this case, the projective syndrome measurement included in the recovery protocol correctly detects the error by projecting onto the particle loss error subspace. Then, the remaining stage of the recovery corrects the error and transforms the system back onto the code space. Notice that, immediately after the error, the quantity $P(t)=1$. This is due to the fact that the loss of a particle transforms the state into one of opposite parity. Finally, in \figref{fig:sample_trajectories_and_effective_decay_rate}(d), a quantum trajectory is shown, where an error occurs, but the subsequent recovery operation doesn't succeed. This occurs because, at the time when the recovery operation takes place, the syndrome measurement projects the system onto the code space, thus erroneously signaling that no error has occurred. Then, the remaining stage of the recovery just drives back the system onto the code space, like in~\figref{fig:sample_trajectories_and_effective_decay_rate}(b), without correcting the wrong parity of the state. 
We recall that \figref{fig:sample_trajectories_and_effective_decay_rate}(a), showing the quantity $P(t)$ defined in \eqref{Eq:P(t)_1} and computed from the Lindblad master equation, can be obtained by the average over many single quantum trajectories. 
These results shown in \figref{fig:sample_trajectories_and_effective_decay_rate} highlights the effectiveness of the $\tau$-periodic protocol for error detection and correction.

\section{Conclusions}
\label{Sec:Conclusion}

Bosonic quantum codes are a very promising method to encode and process quantum information in a fault-tolerant way. Most BQCs to date impose a trade-off between ease of generation and stabilization, and the capability to correct both particle loss and dephasing errors effectively~\cite{JoshiQST21}. In particular, the 2-cat code can be autonomously generated and stabilized, but is vulnerable to particle loss processes, while the GKP code allows to partially correct particle loss error, while being vulnerable to dephasing and challenging in its preparation and stabilization. 

In this work, we have developed a BQC based on squeezed cat states, i.e. on Schrödinger cat states resulting from the superposition of squeezed-displaced states with opposite displacement. Thanks to squeezing, particle loss on the codewords leads to states that only partially overlap with the code space, thus making the error detectable with finite efficiency. At the same time, the SC code improves on the protection against dephasing compared to the 2-cat code with the same average number of particles. In the limit of large squeezing, we have shown that both single particle loss and dephasing errors can be detected and corrected exactly by the SC code. 
We have designed an error-detection and recovery protocol which is semi-autonomous -- in that it requires a periodic, unconditional application of a recovery operation. 
We have characterized the error-correction capability of the SC code both in terms of the Knill-Laflamme conditions and of the channel fidelity, and developed an optimization procedure for the recovery operation, which maximizes the channel fidelity for given values of the particle loss and dephasing rates. Our analysis shows that with a squeezing parameter $\xi \sim 1$ -- within reach of current superconducting-circuit~\cite{Eickbusch2021,DassonnevillePRX2021} and trapped-ion~\cite{Lo2015} architectures -- the channel fidelity of the SC code exceeds that of the 2-cat code by almost one order of magnitude. We have finally indicated protocols for code generation and gate operation.

The SC code allows to detect and correct multiple particle loss and dephasing errors, as these errors bring the codewords to subspaces that only partially overlap with the code space and with the single-error subspaces. The recovery optimization procedure that we developed could therefore be generalized to account for multiple-error subspaces. This generalization should lead to a further improved channel fidelity for the code. The question about how to specifically implement the recovery operation or an approximation thereof in each specific quantum hardware platform, however, remains open.

In the limit of large squeezing, the SC code acquires discrete translation invariance in the direction orthogonal to the displacement. 
The translational invariance in the SC code bears close analogy with that of the GKP code, and is at the origin of the capability of both codes to detect and correct particle loss errors. 
In this respect, the SC code bridges the properties of the 2-cat and GKP codes, by inheriting some error correction features of both codes. 

The error-correction strategy of the SC code can naturally be extended to other types of errors beyond the errors considered here, such as particle gain or higher-order error processes. 
This is an interesting perspective we plan to investigate in the future, in particular for systems operating in regime of moderate particle loss rate, or in the biased-noise limit~\cite{JoshiQST21}.

With the ease of generation and stabilization of two-component Schrödinger cat states, and the recent progress in the generation of squeezed states, we anticipate that the SC code will become an efficient method to enhance the lifetime of quantum information in a bosonic qubit, thus paving the way towards fault-tolerant and scalable quantum information processing.

\begin{acknowledgments}
We acknowledge discussions with Victor V. Albert, Luca Gravina, Alexander Grimm, and Carlos S\'anchez-Mu\~noz. 
This work was supported by the Swiss National Science Foundation through Project No. 200021\_162357 and 200020\_185015. This project was conducted with the financial support of the EPFL Science Seed Fund 2021.
\end{acknowledgments}

\appendix

\section{Error subspaces}
\label{appendixsubsec:errorsubspaces}
To leading order in $\kappa_{1,2}\tau$, the noise channel $\mathcal{N}_{\kappa_1, \kappa_2}$ is described by the Kraus-operators $\hat{K}_0$, $\hat{K}_1$, and $\hat{K}_2$, as defined in \eqref{eq:Kraus_op_small_dt}. $\hat{K}_0$ characterizes the non-Hermitian evolution, whereas the operators $\hat{K}_1$, and $\hat{K}_2$ describe jumps associated to particle loss and dephasing, respectively. To describe the action of the operators $\hat{K}_0$, $\hat{K}_1$, and $\hat{K}_2$ on the code space, we can equivalently consider the action of the operators $\{ \hat{\mathds{1}}, \hat{a}, \hat{a}^\dagger \hat{a}, (\hat{a}^\dagger \hat{a})^2\}$ that appear in \eqref{eq:Kraus_op_small_dt}. We thereby define the non-orthogonal subspaces spanned by the states $\ket*{\widetilde{\psi}_i^\pm}$ as
\begin{equation}
 \ket*{\widetilde{\psi}_i^\pm} = \hat{E}_i \ket*{\mathcal{C}_{\alpha, \xi}^\pm}\;,\quad \hat{E}_i \in \left\{ \hat{\mathds{1}}, \hat{a}, \hat{a}^\dagger \hat{a}, \left(\hat{a}^\dagger \hat{a}\right)^2\right\}.
\end{equation}
The states $\ket*{\widetilde{\psi}_i^\pm}$ are parity eigenstates (although the subspace generated by $\hat{a}\ket*{\mathcal{C}_{\alpha, \xi}^\pm}$ flips the parity).
We orthogonalize the states $\ket*{\widetilde{\psi}_i^\pm}$ using Gram-Schmidt orthogonalization, to obtain the final error subspaces spanned by the orthogonal states $\{\ket*{{\psi}_i^\pm},\; i=0\dots3\}$. Note that the states $\ket*{{\psi}_0^\pm}=\ket*{\mathcal{C}_{\alpha, \xi}^\pm}$ coincide with the codewords. The choice of an orthogonal basis is not unique and is defined up to a rotation within each error subspace. Notice also that the three original error subspaces $\ket*{\widetilde{\psi}_i^\pm}$ are not mutually orthogonal. This feature is one of the reasons underlying the approximate error-correction capability of the SC code, as can be also inferred from the KL relations derived in Appendix~\ref{App:KL}.

\section{Convex recovery optimization}\label{sec:Convex recovery optimization}
\label{App:Convex}

Our goal is to numerically find the best recovery operation, i.e., the $\mathcal{R}$ which maximizes the average channel fidelity.
The average channel fidelity of the channel $\mathcal{E} = \mathcal{R}\circ\mathcal{N}_{\kappa_1, \kappa_2}$ is given by [c.f. Eqs.~(\ref{eq:quantum_channel})~and~(\ref{eq:channel_fidelity})]
\begin{equation}\label{eq:appendix_channelfidelity}
\begin{split}
\mathcal{F}_\text{avg} &= \frac{1}{d^2}\sum\limits_\ell \lvert\text{Tr}\left\{ \hat{S}_\ell\right\} \rvert^2\\
&=\frac{1}{d^2}\sum\limits_{r, k} \lvert\text{Tr}\left\{ \hat{R}_r\hat{K}_k\right\} \rvert^2,
\end{split}
\end{equation}
where $\hat{S}_\ell$ are the Kraus operators of the entire channel $\mathcal{E}$, and $\hat{R}_r$, $\hat{K}_k$ are the Kraus operators of the recovery and error channel, respectively [c.f. \eqref{Eq:Definition_recovery}]. Here, $d$ denotes the qudit dimension ($d=2$ in our case).
To optimize the recovery operation $\mathcal{R}$, we can express the operators $\hat{R}_r$ in terms of a set of basis operators $\{\hat{B}_i\}$, such that 
\begin{equation}
 \hat{R}_r = \sum_i x_{r, i} \hat{B}_i.
\end{equation}
The basis operators $\hat{B}_i$ describe how states outside of the code space are projected back onto the code space.
The operators $\hat{B}_i$ are orthogonal, i.e. $\text{Tr}\{\hat{B}_i^\dagger \hat{B}_j\} = \delta_{ij}$.
The optimization procedure amounts to determine, for a given set $\{\hat{B}_i\}$, the coefficients $x_{r, i}$ which maximize the channel fidelity $\mathcal{F}_\text{avg} $.

Using the definition of the recovery in terms of the basis operators $\hat{B}_i$,
\eqref{eq:appendix_channelfidelity} can be rewritten as~\cite{KosutQuantum2009}
\begin{equation}\label{eq:fid_convexoptim}
\mathcal{F}_\text{avg} = \frac{1}{d^2}\sum\limits_{i,j} X_{ij} W_{ij},
\end{equation}
where the recovery matrix $\mathbf{X}$ and the process matrix $\mathbf{W}$ are given as
\begin{align}\label{eq:fid_process_matrix}
X_{ij} &= \sum\limits_r x_{r,i} x_{r, j}^*,\\
W_{ij} &= \sum\limits_k \text{Tr}\left\{\hat{B}_i\hat{K}_k\right\}\text{Tr}\left\{\hat{B}_j\hat{K}_k\right\}^*.
\end{align}
Using the property $\text{Tr}\{\hat{A}\}\text{Tr}\{\hat{B}\} = \text{Tr}\{\hat{A}\otimes \hat{B}\}$, and the Kraus-representation of the error channel $\sum_k \hat{K}_k\hat{\rho}\hat{K}_k^\dagger = e^{\mathcal{L}\tau}\hat{\rho}$, the process matrix can be expressed as
\begin{equation}
 W_{ij} = \text{Tr}\left\{\left(e^{\mathcal{L}\tau}\right) \left(\hat{B}_i\otimes \hat{B}_j^\dagger\right)\right\} = \text{Tr}\left\{\left(\mathcal{N}_{\kappa_1, \kappa_2}\right) \left(\hat{B}_i\otimes \hat{B}_j^\dagger\right)\right\},
\end{equation}
where $\text{Tr}$ in this case denotes the trace of the super-operators.
As a result, the process matrix $\mathbf{W}$ can be obtained by calculating the product of the superoperators $\hat{B}_i\otimes \hat{B}_j^\dagger$ with the noise channel $\mathcal{N}_{\kappa_1\kappa_2}(\tau)$ in its super-operator representation.
Since \eqref{eq:fid_convexoptim} is linear in $X_{ij}$, and hence convex, the equality $X_{ij} = \sum_r x_{r,i}x_{r,j}^*$ is a quadratic constraint and does not form a convex set. However, one can relax the constraint by allowing the matrix $\mathbf{X}$ to be positive semidefinite, where the problem becomes a convex \textit{semidefinite program}~\cite{KosutQuantum2009} (SDP):
\begin{align}\label{eq:SDP}
 \text{maximize} \quad \frac{1}{d^2}\sum\limits_{i,j} X_{ij}W_{ij} &= \frac{1}{d^2} \text{Tr}\{\mathbf{X} \mathbf{W}\}\\
 \text{subject to} \quad \sum\limits_{i,j} X_{ij} \hat{B}_i^\dagger \hat{B}_j &= \hat{\mathds{1}},\; \text{(trace preservation)}\nonumber\\
 \mathbf{X} &\succcurlyeq 0\; \text{(positive semidefinite)}. \nonumber
\end{align}

The convex optimization of the SC code via \eqref{eq:SDP} requires the matrix $\mathbf{X}$, and thus an explicit form of the basis operators [c.f. \eqref{eq:fid_process_matrix}].
By expanding $\mathcal{N}_{\kappa_1, \kappa_2}$ up to first order in $\kappa_{1,2}\tau$ in terms of the Kraus-operators given in \eqref{eq:Kraus_op_small_dt}, and using the definition of the error subspaces in Appendix~\ref{appendixsubsec:errorsubspaces},
we choose the following set of basis operators $\hat{B}_i$: For each error subspace, spanned by the basis states $\{\ket*{\psi_m^\pm}\}$ with $m=0,\ldots,\,3$, we define
\begin{align}
 \hat{P}_m^{(0)} &= \ketbra*{\mathcal{C}_{\alpha, \xi}^+}{\psi_m^+} + \ketbra*{\mathcal{C}_{\alpha, \xi}^-}{\psi_m^-},\\
 \hat{P}_m^{(1)} &= \ketbra*{\mathcal{C}_{\alpha, \xi}^+}{\psi_m^-} + \ketbra*{\mathcal{C}_{\alpha, \xi}^-}{\psi_m^+},\\
 \hat{P}_m^{(2)} &= i\ketbra*{\mathcal{C}_{\alpha, \xi}^+}{\psi_m^-} - i\ketbra*{\mathcal{C}_{\alpha, \xi}^-}{\psi_m^+},\\
 \hat{P}_m^{(3)} &= \ketbra*{\mathcal{C}_{\alpha, \xi}^+}{\psi_m^+} - \ketbra*{\mathcal{C}_{\alpha, \xi}^-}{\psi_m^-}.
\end{align}
The 16 operators $\hat{P}_m^{(n)}$ define a basis of the recovery operations, as they are mutually orthogonal and describe an arbitrary action of the recovery operation projecting $\{\ket*{\psi_m^\pm}\}$ back onto the code space.
Thus, we define $\{\hat{B}_i\} \equiv \{ \hat{P}_m^{(n)}, \, m\in [0,3], \, n\in [0,3]\}$.
Solving the SDP gives an optimal solution $\mathbf{X}^\text{opt}$ that maximizes the average channel fidelity.

There are various efficient solvers available to solve the above SDP.
We solve the SDP, defined in \eqref{eq:SDP}, in the Hilbert subspace spanned by $\{\ket*{\psi_i^\pm}\; i=0\dots3\}$ that characterize the error subspaces, including the code space.
An explicit form of the optimal recovery operations $\hat{R}_r^\text{opt}$ can be obtained through a singular value decomposition of the solution $\mathbf{X}^\text{opt}$~\cite{KosutQuantum2009},
\begin{align}
 \mathbf{X}^\text{opt} &= \mathbf{V}\mathbf{\Sigma}\mathbf{V}^\dagger,\\
 \hat{R}_r^\text{opt} &= \sqrt{\sigma_r} \sum\limits_i V_{ir}\hat{B}_i,
\end{align}
where the elements $\sigma_r$ of the diagonal matrix $\mathbf{\Sigma}$ are the singular values of $\mathbf{X}^\text{opt}$ and $\mathbf{V}$ is a unitary matrix.

To numerically solve the SDP, we resort to the \texttt{SDPA-GMP} solver~\cite{NakataIEEE2010} within the \texttt{Convex} framework~\cite{UdellSC142014} in the \texttt{julia} programming language~\cite{BezansonSIAM2017}. All numerical calculations use the \texttt{julia} framework for open quantum dynamics \texttt{QuantumOptics}~\cite{KramerComputerPhysicsCommunications2018}.

\ 

\section{KL condition}
\label{App:KL}

In this Appendix we report the KL conditions for the 2-cat and SC codes.
For the sake of simplicity, we will consider the set of errors $\{\hat{E}_k\}=\{\hat{\mathds{1}}, \hat{a}, \hat{a}^\dagger \hat{a}, \left(\hat{a}^\dagger \hat{a}\right)^2 \}$.
From these, we can always reconstruct the effect of the Kraus opeators in \eqref{eq:Kraus_op_small_dt}.

\subsection{Cat state properties}

As we detailed in the main text, the 2-cats are defined by
\begin{equation}
 \ket*{\mathcal{C}_\alpha^\pm} = \frac{1}{N_\alpha^\pm}\left(\ket*{\alpha} \pm \ket*{-\alpha} \right),
\end{equation}
where
\begin{equation}\label{eq:cat_normalization}
 N_{\alpha}^\pm = \sqrt{2\left(1\pm e^{-2|\alpha|^2}\right)}.
\end{equation}
Thus, the following relation between normalization constants of even and odd 2-cat holds
\begin{equation}
 \left(\frac{{{}N_{\alpha}^-}}{{{}N_{\alpha}^+}}\right)^2 = \tanh(|\alpha|^2).
\end{equation}

\renewcommand{\arraystretch}{2.2}

The KL conditions for the 2-cat code are reported in Tables \ref{tab:KLcat1} and \ref{tab:KLcat2}.

 \begin{table*}[ht]
 \begin{tabularx}{\textwidth}{| X | c | c | c | c |}
 \hline
 $\hat{E}_l^\dagger \, 	\backslash \, \hat{E}_{l'}$ & $\hat{\mathds{1}}$ & $\hat{a}$ & $\hat{a}^\dagger \hat{a}$& $(\hat{a}^\dagger \hat{a})^2$ \\
 \hline
 $\hat{\mathds{1}}$ & 1 & 0 & $|\alpha|^2 \frac{
 \left(N_{\alpha }^{\mp
 }\right){}^2}{\left(N_{\alpha
 }^{\pm }\right){}^2}$ & $|\alpha|^2 \left(|\alpha|^2+\frac{\left(N_{\alpha }^{\mp
 }\right){}^2}{\left(N_{\alpha
 }^{\pm }\right){}^2}\right)$ \\ \hline 
 $\hat{a}^\dagger$ & 0 & $ |\alpha|^2 \frac{\left(N_{\alpha }^{\mp
 }\right){}^2}{\left(N_{\alpha
 }^{\pm }\right){}^2}$ & 0 & 0 \\ \hline 
 $\hat{a}^\dagger \hat{a}$& $ |\alpha|^2 \frac{
 \left(N_{\alpha }^{\mp
 }\right){}^2}{\left(N_{\alpha
 }^{\pm }\right){}^2} $& 0 & $|\alpha|^2 \left(|\alpha|^2
 +\frac{\left(N_{\alpha }^{\mp
 }\right){}^2}{\left(N_{\alpha
 }^{\pm }\right){}^2}\right) $& $
 |\alpha|^2 \left(3 |\alpha|^2+\frac{\left(|\alpha|^4+1\right)
 \left(N_{\alpha }^{\mp
 }\right){}^2}{\left(N_{\alpha
 }^{\pm }\right){}^2}\right)$ \\ \hline
 $(\hat{a}^\dagger \hat{a})^2$ &
 $|\alpha|^2 \left(|\alpha|^2+\frac{\left(N_{\alpha
 }^{\mp
 }\right){}^2}{\left(N_{\alpha
 }^{\pm }\right){}^2}\right)$ & 0 & $
 |\alpha|^2 \left(3 |\alpha|^2+\frac{\left(|\alpha|^4+1\right)
 \left(N_{\alpha }^{\mp
 }\right){}^2}{\left(N_{\alpha
 }^{\pm }\right){}^2}\right)$ &$
 |\alpha|^2 \left(|\alpha|^2 \left(|\alpha|^4+7\right)+\frac{\left(6
 |\alpha|^4+1\right)
 \left(N_{\alpha }^{\mp
 }\right){}^2}{\left(N_{\alpha
 }^{\pm }\right){}^2}\right)$
 \\
 \hline
 \end{tabularx}
 \caption{Knill-Laflamme condition, defined in \eqref{eq:KL}, for $\bra*{\mathcal{C}_\alpha^\pm}\hat{E}_l^\dagger \hat{E}_{l'}\ket*{\mathcal{C}_\alpha^\pm}$.}
 \label{tab:KLcat1}
 \end{table*}

 \begin{table*}[ht]
 \begin{tabularx}{0.7\textwidth}{| X | c | c | c | c |}
 \hline
 $\hat{E}_l^\dagger \, 	\backslash \, \hat{E}_{l'}$ & $\hat{\mathds{1}}$ & $\hat{a}$ & $\hat{a}^\dagger \hat{a}$& $(\hat{a}^\dagger \hat{a})^2$ \\
 \hline
 $\hat{\mathds{1}}$ & 0 & $\alpha \frac{ N_{\alpha }^{\pm
 }}{N_{\alpha }^{\mp }}$ & 0 & 0 \\
 \hline
 $\hat{a}^\dagger$ & $ \alpha ^* \frac{ N_{\alpha }^{\mp
 }}{N_{\alpha }^{\pm }}$ & 0 &
 $|\alpha|^2 \frac{ \alpha
 ^* N_{\alpha }^{\pm
 }}{N_{\alpha }^{\mp }}$ &
 $|\alpha|^2 \frac{ \alpha
 ^* \left(|\alpha|^2
 \left(N_{\alpha }^{\mp
 }\right){}^2+\left(N_{\alpha }^{\pm
 }\right){}^2\right)}{N_{\alpha
 }^{\mp } N_{\alpha }^{\pm }}$ \\
 \hline
 $\hat{a}^\dagger \hat{a}$ & 0 & $|\alpha|^2 \frac{\alpha
 N_{\alpha }^{\mp }}{N_{\alpha
 }^{\pm }}$ & 0 & 0 \\ \hline
 $(\hat{a}^\dagger \hat{a})^2$ & 0 & $|\alpha|^2 \frac{\alpha
 \left(|\alpha|^2
 \left(N_{\alpha }^{\pm
 }\right){}^2+\left(N_{\alpha }^{\mp
 }\right){}^2\right)}{N_{\alpha
 }^{\mp } N_{\alpha }^{\pm }}$ & 0 &
 0 
 \\ \hline
 \end{tabularx}
 \caption{Knill-Laflamme condition, defined in \eqref{eq:KL}, for $\bra*{\mathcal{C}_\alpha^\pm}\hat{E}_l^\dagger \hat{E}_{l'}\ket*{\mathcal{C}_\alpha^\mp}$.}
 \label{tab:KLcat2} 
 \end{table*} 

\label{appendixsubsec:catstateproperties}

\subsection{Squeezed cat state properties}
\label{appendixsubsec:SCCproperties}
As defined in \eqref{Eq:SCC}, the squeezed cat states, defining the logical codewords of the bosonic code, are given by
\begin{equation}
 \ket*{\mathcal{C}_{\alpha, \xi}^\pm} = \frac{1}{N_{\alpha, \xi}^\pm} \left(\ket*{\alpha, \xi} \pm \ket*{-\alpha, \xi} \right),
\end{equation}
where $\ket{\pm\alpha, \xi}$ are opposite squeezed-displaced states [c.f \eqref{eq:squeezed-displaced-state}, \eqref{eq:displacement-squeezing}] with complex displacement $\alpha$ and complex squeezing $\xi = re^{i\theta}$.
The normalization constant $N_{\alpha, \xi}^\pm$ is given by
\begin{equation}\label{eq:scc_normalization}
 N_{\alpha, \xi}^\pm = \sqrt{2\left(1\pm e^{-2|\gamma|^2}\right)},
\end{equation}
where we define the squeezed displacement as
\begin{equation}\label{eq:squeezed-coherent-amplitude}
 \gamma = \alpha\cosh(r)+e^{i\theta}\alpha^*\sinh(r).
\end{equation}
In the basis of number states, the SC states are given by
\begin{equation}\label{eq:SC_fock_basis}
\begin{split}
\ket*{\mathcal{C}_{\alpha, \xi}^\pm} =& \frac{1}{\sqrt{2\cosh (r)\left(1\pm e^{-2|\gamma|^2}\right)}} \\
& \times \exp \left[-\frac{1}{2}|\alpha|^{2}-\frac{1}{2} \alpha^{* 2} e^{i \theta} \tanh (r)\right] \\
& \times \sum_{n=0}^{\infty} \Bigg[(1 \pm (-1)^n)\frac{\left[\frac{1}{2} e^{i \theta} \tanh (r)\right]^{n / 2}}{\sqrt{n !}}\times \\ 
& \qquad H_{n}\left[\gamma\left(e^{i \theta} \sinh (2 r)\right)^{-1 / 2}\right] \Bigg]\ket{n},
\end{split}
\end{equation}
where $H_n(x)$ is the Hermite polynomial of degree $n$.
From the above expression the even and odd parity structure of the states $\ket*{\mathcal{C}_{\alpha, \xi}^\pm}$ becomes directly evident, as $\ket*{\mathcal{C}_{\alpha, \xi}^+}$ is composed of only even number states and $\ket*{\mathcal{C}_{\alpha, \xi}^-}$ of only odd number states [due to the factor $(1 \pm (-1)^n)$ in \eqref{eq:SC_fock_basis}].

One can calculate the error correction properties of the SC code by expressing the codewords in terms of squeezing operators acting on a 2-cat code,
\begin{equation}
 \ket*{\mathcal{C}_{\alpha, \xi}^\pm} = \frac{1}{N_{\alpha,\xi}^\pm} \hat{S}(\xi)\left(\ket*{\gamma} \pm \ket*{-\gamma}\right) = \hat{S}(\xi) \ket*{\mathcal{C}_{\gamma}^\pm},
\end{equation}
where $\gamma$ is the squeezed displacement defined in \eqref{eq:squeezed-coherent-amplitude}.
Using this relation, and the fact that $\hat{S}^\dagger(\xi) \hat{a} \hat{S}(\xi) = \hat{a} \cosh{(r)} - \hat{a}^\dagger e^{i \theta} \sinh{(r)}$, where $\xi=r e^{i \theta}$, the KL condition for the SC can be directly derived.
For the sake of brevity, we report here just some of them, highlighting the capability of the SC code to correct particle loss and dephasing errors.
We refer the interested reader to the Supplementary Material, where the full set of expressions for the KL conditions can be found.

The capability to correct particle loss errors is assessed by considering $\hat{E}_l = \hat{\mathds{1}}$ and $\hat{E}_{l'} = \hat{a}$.
Since the SC code has the same parity structure as the 2-cat code, we have
\begin{equation}\label{Eq:KL_SC_loss2}
 \bra*{\mathcal{C}_{\alpha, \xi}^\pm} \hat{a} \ket*{\mathcal{C}_{\alpha, \xi}^\pm} =0.
\end{equation}
Just like in the 2-cat code, where the particle-loss KL condition scales linearly with the displacement $\alpha$, this behavior is recovered for the squeezed cat in the large squeezing limit, as
\begin{equation}\label{Eq:KL_SC_loss1}
\begin{split}
 \bra*{\mathcal{C}_{\alpha, \xi}^\pm} \hat{a} \ket*{\mathcal{C}_{\alpha, \xi}^\mp} = & \frac{\gamma \cosh (\xi ) N_{\alpha
 \xi }^{\mp }}{N_{\alpha \xi }^{\pm
 }}-\frac{\gamma \sinh (\xi )
 N_{\alpha \xi }^{\pm }}{N_{\alpha
 \xi }^{\mp }} \\
 & \quad \xrightarrow[|\xi| \gg 1]{} \alpha.
\end{split} 
\end{equation}
This clearly indicates that, by increasing the squeezing parameter, and in the limit in which $N_{\alpha
\xi }^{\pm } \simeq N_{\alpha
\xi }^{\mp }$, particle loss errors become increasingly correctable, \emph{upon lowering $\alpha$}. 

The key difference with respect to the 2-cat code is that, by lowering $\alpha$ and and at the same time increasing $\xi$, also dephasing errors are suppressed, thus making both errors correctable.
Indeed, if we consider $\hat{E}_l = \hat{\mathds{1}}$ and $\hat{E}_{l'} = \hat{a}^\dagger \hat{a}$, we have
\begin{equation}\label{Eq:KL_SC_dephasing1}
\begin{split}
\bra*{\mathcal{C}_{\alpha, \xi}^\pm} \hat{a}^\dagger \hat{a} \ket*{\mathcal{C}_{\alpha, \xi}^\pm}=& \sinh (\xi ) \left(\sinh (\xi )-2
 \gamma ^2 \cosh (\xi
 )\right) \\ &+\frac{\gamma ^2 \cosh (2
 \xi ) \left(N_{\alpha \xi }^{\mp
 }\right){}^2}{\left(N_{\alpha \xi
 }^{\pm }\right){}^2} \end{split}\,,
\end{equation}
while
\begin{equation}\label{Eq:KL_SC_dephasing2}
\bra*{\mathcal{C}_{\alpha, \xi}^\pm} \hat{a}^\dagger \hat{a} \ket*{\mathcal{C}_{\alpha, \xi}^\mp}=0.
\end{equation}
Again, in the limit where $ N_{\alpha
 \xi }^{\pm } \simeq N_{\alpha
 \xi }^{\mp }$, the KL conditions are fulfilled.
In particular, 
\begin{equation}\label{Eq:KL_SC_dephasing3}
\begin{split}
 |\bra*{\mathcal{C}_{\alpha, \xi}^+} \hat{a}^\dagger \hat{a} & \ket*{\mathcal{C}_{\alpha, \xi}^+} - \bra*{\mathcal{C}_{\alpha, \xi}^-} \hat{a}^\dagger \hat{a} \ket*{\mathcal{C}_{\alpha, \xi}^-}| \\ & \xrightarrow[|\xi| \gg 1]{} 2 |\alpha| ^2 \exp\left({4 r-2 |\alpha| ^2 e^{2
 r}}\right),
 \end{split}\,
\end{equation}
where $\xi = r e^{i \theta}$.
Notice that this KL condition becomes zero both when $\alpha$ is large, but also if $\xi$ is large and $\alpha \neq 0$, as required for a perfectly correctable error.

Similar relations hold for other choices of discrete errors $\hat{E}_l$ and $\hat{E}_{l'}$. Thus, particle loss and dephasing can be partially corrected (up to order $\kappa_{1,2}\tau$) using SC states. The correction capability of the code increases as squeezing is increased.

\eqref{Eq:KL_SC_loss1} and \eqref{Eq:KL_SC_dephasing3} highlight the mutual dependence of the code resilience against loss and dephasing errors. For finite amount of squeezing, decreasing $\alpha$ improves the resilience against loss but makes the code more vulnerable to dephasing. Hence, depending on the decay rates $\kappa_{1,\,2}$, an optimal choice of $\xi$ and $\alpha$ is determined, which maximizes the channel fidelity. It is thanks to this optimization procedure that the overall performance of the SC is generally better than that of the 2-cat code.

\section{Features of the optimal SC code}
\label{appendixsubsec:Characteristics_of_the_optimally-resistant_state}

\begin{figure*}
 \centering
 \includegraphics[width=0.8 \textwidth]{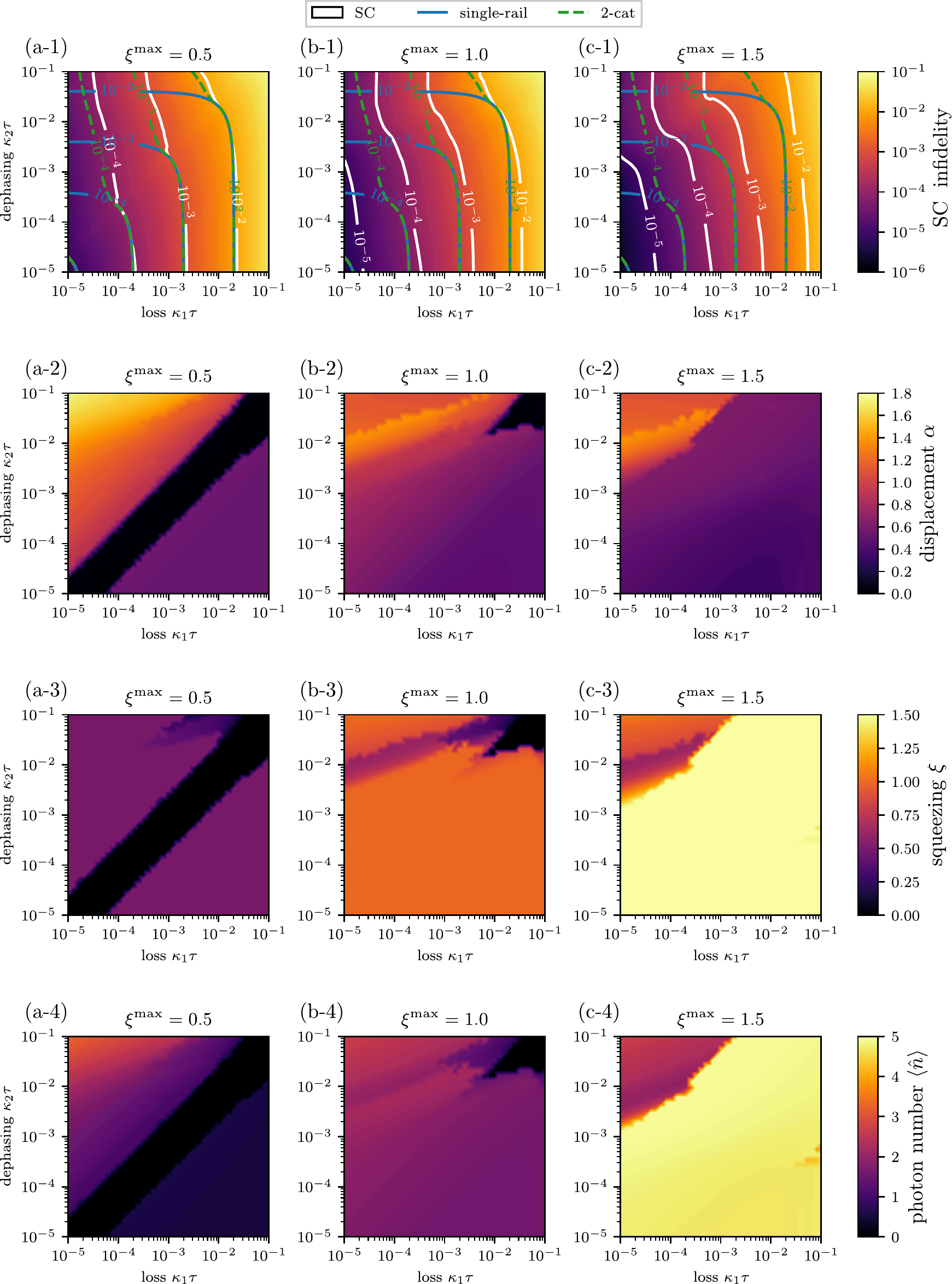}
 \caption{
 Color-map plot of quantities associated to the optimal SC code, computed as a function of the particle loss parameter $\kappa_1\tau$ and dephasing parameter $\kappa_2\tau$, and for three values of the maximum squeezing parameters [(a-1)-(a-4) $\xi^\mathrm{max}=0.5$, (b-1)-(b-4) $\xi^\mathrm{max}=1.0$, and (c-1)-(c-4) $\xi^\mathrm{max}=1.5$].
 (a-1), (b-1), and (c-1): the channel infidelity $1-\mathcal{F}$ [same as Fig.~\ref{fig:channel_fidelity}(a-c)].
 (a-2), (b-2), and (c-2): the optimal displacement $\alpha$.
 (a-3), (b-3), and (c-3): the optimal amount of squeezing $\xi$.
 (a-4), (b-4), and (c-4): the average particle number in the resonator.}
 \label{fig:encoding_structure}
\end{figure*}

In Fig.~\ref{fig:encoding_structure}, we plot quantities associated to the optimal SC code, as obtained by the study of the channel fidelity.
Figures~\ref{fig:encoding_structure} (a-1), (b-1), and (c-1) are identical to Fig.~\ref{fig:channel_fidelity}, and are reported here for ease of comparison.
Figures~\ref{fig:encoding_structure} (a-2), (a-3), and (a-4) show respectively the displacement, the amount of squeezing $\xi$, and the average particle number $\expec{\hat{n}}$, evaluated for the optimal SC code obtained by setting a maximal amount of squeezing $\xi = 0.5$. 
We identify two regimes: one of small displacement (loss dominated), and one of larger displacement (dephasing dominated).
The results confirm the conclusions drawn from the analysis of the KL conditions in Sec.~\ref{Subsubsec:KL}. Indeed, in the loss-dominated case, $\alpha$ is to be taken small enough to ensure the correction capability of the code.
In both small- and large-$\alpha$ regimes, maximal squeezing is almost always reached.
Interestingly, when the dephasing and loss rates are almost equal ($\kappa_1 = \kappa_2$), the resulting optimal code approaches the single-rail code.
In the whole parameter region considered, the particle number remains fairly small.

Increasing the maximal amount of squeezing [Figs.~\ref{fig:encoding_structure} (b-2), (b-3), and (b-4)] we see that the boundary between the dephasing and loss dominated regimes becomes less defined (although the separation is still visible for large-enough error rates). 
In addition, for $\xi^\mathrm{max}=1.0$ the optimal value of $\alpha$ is correspondingly smaller than for $\xi^\mathrm{max}=0.5$. This is to be understood as a consequence of the advantage brought by the larger squeezing, which enables correction of dephasing errors while keeping $\alpha$ -- and thus the code correctability against loss -- correspondingly smaller, in agreement with the KL conditions. In this case as well, the optimization procedure takes maximal advantage of the allowed amount of squeezing, and the average particle number remains moderate.

Finally, for $\xi^{\rm max} = 1.5$ in Figs.~\ref{fig:encoding_structure} (c-2), (c-3), and (c-4), displacement remains moderate in a large region of parameters. Notably, however, in the dephasing dominated regime, the amount of squeezing required to achieve optimal correction is significantly smaller than the maximal value $\xi^{\rm max}$.

These data are consistent with the analysis of the KL conditions in Sec.~\ref{Subsubsec:KL}, in that the optimization procedure in general results in a small value of $\alpha$ combined with a significant amount of squeezing to suppress dephasing.

\section{Comparison between the GKP and SC codes}
\label{Appendix:GKP_vs_SC}
We compare the error correction properties of the SC code with those of the \emph{approximate}, i.e. energy-constrained, GKP code.
In particular, we assume a Gaussian envelope of the form \cite{GrimsmoPRX2020} 
\begin{equation}
 \left|{\mu}(\Delta)\right\rangle \propto e^{-\Delta^{2} \hat{a}^{\dagger} \hat{a}}\left|\mu\right\rangle
\end{equation}
where $\mu \in \{0, 1\}$ labels the codewords, which in the ideal GKP limit are 
\begin{equation}
\left|\mu\right\rangle \propto \sum_{k, l=-\infty}^{\infty} e^{-i \pi(k l+\mu / 2)} \hat{D}\big((2 k+\mu) \alpha+l \beta \big) |0\rangle,
\end{equation}
where $\hat{D}(\beta)$ is the displacement operator introduced in \eqref{eq:displacement-squeezing}.

In order to explicitly show the similarity with the SC code, the approximate GKP codewords can be written in terms of the squeezing operator as
\begin{equation}
\begin{split}
 \left|\mu(\Delta)\right\rangle \propto \sum_{n \in \mathbb{Z}} & e^{-\frac{\pi}{2} \Delta^{2}(2 n+\mu)^{2}} \\ & \quad \hat{D}\left(\sqrt{\frac{\pi}{2}(2 n+\mu)}\right) 
 \hat{S} \left({-\ln \Delta}\right) \ket{0}.
\end{split} 
\end{equation}
Thus, we can define the squeezing parameter for the GKP as
\begin{equation}
 \xi = -\ln \Delta.
\end{equation}

Using this formalism, we can compare the performance of the GKP and SC codes for the same value of the squeezing parameter $\xi$.
As a metric, we adopt a KL cost function \cite{Reinhold_PhD_Thesis}. Given the KL tensor
\begin{equation}
 f_{i j l l'}=\left\langle\psi_{i}\left|\hat{E}_{l}^{\dagger} \hat{E}_{l'}\right| \psi_{j}\right\rangle
\end{equation}
we evaluate the cost function as
\begin{equation}\label{Eq:KL_cost}
 C_{\rm KL}(\{\hat{E}\})=\sum_{l, l'}\left|f_{00 l l'}-f_{11 l l'}\right|^{2}+\left|f_{01 l l'}\right|^{2} ,
\end{equation}
where $\{\hat{E}\}$ represents the set of errors upon which the sum in $l, \, l'$ is taken.
In the limit of a perfectly correctable code, one has $C_{\rm KL}=0$.
An increasing $C_{\rm KL}$, as a function of the amount of squeezing, indicates a less-correctable code.

We plot $C_{\rm KL}$ as a function of the squeezing parameter for the SC with different values of $\alpha$ and the GKP. The results for $\{\hat{E}\} = \{\hat{\mathds{1}}, \hat{a}\}$ and $\{\hat{E}\} = \{\hat{\mathds{1}}, \hat{a}^\dagger \hat{a}\}$, corresponding to loss and dephasing, are displayed with full and dashed lines respectively.
For the loss error, the KL cost function shows similar features in the GKP and SC cases, and saturates to a finite value, which depends on $\alpha$ in the case of the SC.
For the dephasing error on the other hand, the KL cost function rapidly approaches zero as $\alpha$ increases in the SC case, while it grows indefinitely for the GKP code.
This analysis clearly shows the difference between the GKP and SC codes, and illustrates how the latter takes advantage of squeezing as a resource for improved error correction.

\begin{figure}
 \centering
 \includegraphics[width = 0.49\textwidth]{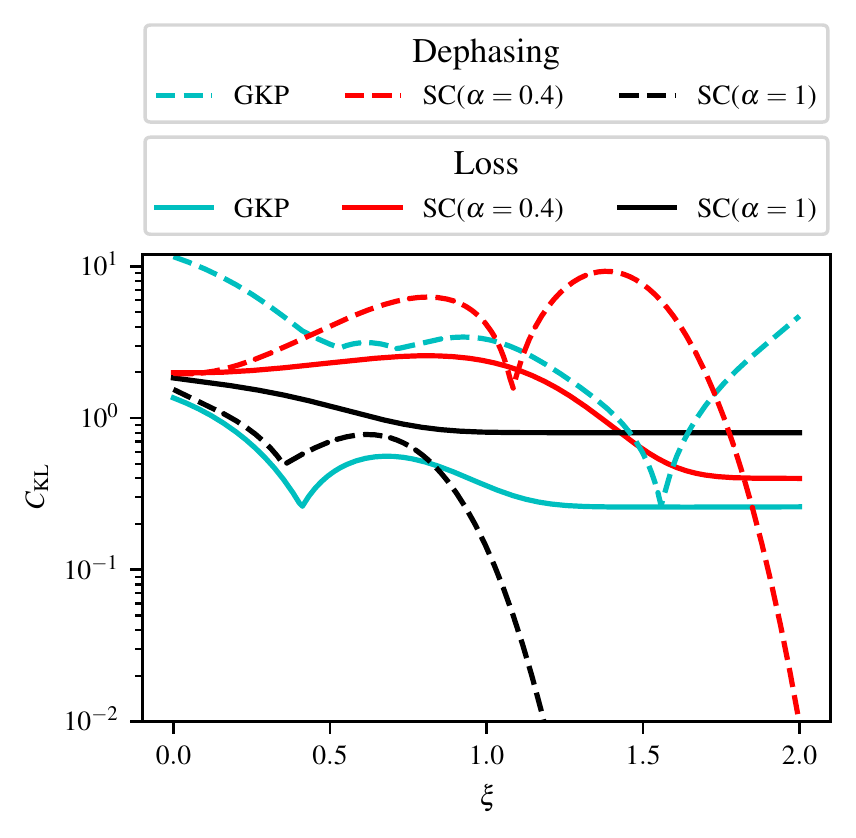}
 \caption{The KL cost function \eqref{Eq:KL_cost} plotted as a function of the squeezing $\xi$, for the GKP (light blue), the SC with displacement $\alpha=0.4$ (red), and the SC with displacement $\alpha=1$ (blue).}
 \label{fig:KL_GKP}
\end{figure}




\bibliography{biblio}

\begin{thebibliography}{108}%
\makeatletter
\providecommand \@ifxundefined [1]{%
 \@ifx{#1\undefined}
}%
\providecommand \@ifnum [1]{%
 \ifnum #1\expandafter \@firstoftwo
 \else \expandafter \@secondoftwo
 \fi
}%
\providecommand \@ifx [1]{%
 \ifx #1\expandafter \@firstoftwo
 \else \expandafter \@secondoftwo
 \fi
}%
\providecommand \natexlab [1]{#1}%
\providecommand \enquote  [1]{``#1''}%
\providecommand \bibnamefont  [1]{#1}%
\providecommand \bibfnamefont [1]{#1}%
\providecommand \citenamefont [1]{#1}%
\providecommand \href@noop [0]{\@secondoftwo}%
\providecommand \href [0]{\begingroup \@sanitize@url \@href}%
\providecommand \@href[1]{\@@startlink{#1}\@@href}%
\providecommand \@@href[1]{\endgroup#1\@@endlink}%
\providecommand \@sanitize@url [0]{\catcode `\\12\catcode `\$12\catcode
  `\&12\catcode `\#12\catcode `\^12\catcode `\_12\catcode `\%12\relax}%
\providecommand \@@startlink[1]{}%
\providecommand \@@endlink[0]{}%
\providecommand \url  [0]{\begingroup\@sanitize@url \@url }%
\providecommand \@url [1]{\endgroup\@href {#1}{\urlprefix }}%
\providecommand \urlprefix  [0]{URL }%
\providecommand \Eprint [0]{\href }%
\providecommand \doibase [0]{http://dx.doi.org/}%
\providecommand \selectlanguage [0]{\@gobble}%
\providecommand \bibinfo  [0]{\@secondoftwo}%
\providecommand \bibfield  [0]{\@secondoftwo}%
\providecommand \translation [1]{[#1]}%
\providecommand \BibitemOpen [0]{}%
\providecommand \bibitemStop [0]{}%
\providecommand \bibitemNoStop [0]{.\EOS\space}%
\providecommand \EOS [0]{\spacefactor3000\relax}%
\providecommand \BibitemShut  [1]{\csname bibitem#1\endcsname}%
\let\auto@bib@innerbib\@empty
\bibitem [{\citenamefont {Preskill}(2018)}]{PreskillNISQ2018}%
  \BibitemOpen
  \bibinfo {author} {J.~Preskill},\ \emph {\bibinfo {title} {Quantum
  {C}omputing in the {NISQ} era and beyond}},\ \href
  {\doibase10.22331/q-2018-08-06-79} {\bibfield  {journal} {\bibinfo  {journal}
  {{Quantum}}\ }\textbf {\bibinfo {volume} {2}},\ \bibinfo {pages} {79}
  (\bibinfo {year} {2018})}\BibitemShut {NoStop}%
\bibitem [{\citenamefont {Arute}\ \emph {et~al.}(2019)\citenamefont {Arute},
  \citenamefont {Arya}, \citenamefont {Babbush}, \citenamefont {Bacon},
  \citenamefont {Bardin}, \citenamefont {Barends}, \citenamefont {Biswas},
  \citenamefont {Boixo}, \citenamefont {Brandao}, \citenamefont {Buell},
  \citenamefont {Burkett}, \citenamefont {Chen}, \citenamefont {Chen},
  \citenamefont {Chiaro}, \citenamefont {Collins}, \citenamefont {Courtney},
  \citenamefont {Dunsworth}, \citenamefont {Farhi}, \citenamefont {Foxen},
  \citenamefont {Fowler}, \citenamefont {Gidney}, \citenamefont {Giustina},
  \citenamefont {Graff}, \citenamefont {Guerin}, \citenamefont {Habegger},
  \citenamefont {Harrigan}, \citenamefont {Hartmann}, \citenamefont {Ho},
  \citenamefont {Hoffmann}, \citenamefont {Huang}, \citenamefont {Humble},
  \citenamefont {Isakov}, \citenamefont {Jeffrey}, \citenamefont {Jiang},
  \citenamefont {Kafri}, \citenamefont {Kechedzhi}, \citenamefont {Kelly},
  \citenamefont {Klimov}, \citenamefont {Knysh}, \citenamefont {Korotkov},
  \citenamefont {Kostritsa}, \citenamefont {Landhuis}, \citenamefont
  {Lindmark}, \citenamefont {Lucero}, \citenamefont {Lyakh}, \citenamefont
  {Mandr{\`a}}, \citenamefont {McClean}, \citenamefont {McEwen}, \citenamefont
  {Megrant}, \citenamefont {Mi}, \citenamefont {Michielsen}, \citenamefont
  {Mohseni}, \citenamefont {Mutus}, \citenamefont {Naaman}, \citenamefont
  {Neeley}, \citenamefont {Neill}, \citenamefont {Niu}, \citenamefont {Ostby},
  \citenamefont {Petukhov}, \citenamefont {Platt}, \citenamefont {Quintana},
  \citenamefont {Rieffel}, \citenamefont {Roushan}, \citenamefont {Rubin},
  \citenamefont {Sank}, \citenamefont {Satzinger}, \citenamefont {Smelyanskiy},
  \citenamefont {Sung}, \citenamefont {Trevithick}, \citenamefont
  {Vainsencher}, \citenamefont {Villalonga}, \citenamefont {White},
  \citenamefont {Yao}, \citenamefont {Yeh}, \citenamefont {Zalcman},
  \citenamefont {Neven},\ and\ \citenamefont {Martinis}}]{AruteNat19}%
  \BibitemOpen
  \bibinfo {author} {F.~Arute}, \bibinfo {author} {K.~Arya}, \bibinfo {author}
  {R.~Babbush}, \bibinfo {author} {D.~Bacon}, \bibinfo {author} {J.~C. Bardin},
  \bibinfo {author} {R.~Barends}, \bibinfo {author} {R.~Biswas}, \bibinfo
  {author} {S.~Boixo}, \bibinfo {author} {F.~G. S.~L. Brandao}, \bibinfo
  {author} {D.~A. Buell} et~al.,\ \emph {\bibinfo {title} {Quantum supremacy
  using a programmable superconducting processor}},\ \href
  {\doibase10.1038/s41586-019-1666-5} {\bibfield  {journal} {\bibinfo
  {journal} {Nature}\ }\textbf {\bibinfo {volume} {574}},\ \bibinfo {pages}
  {505} (\bibinfo {year} {2019})}\BibitemShut {NoStop}%
\bibitem [{\citenamefont {Cerezo}\ \emph {et~al.}(2021)\citenamefont {Cerezo},
  \citenamefont {Arrasmith}, \citenamefont {Babbush}, \citenamefont {Benjamin},
  \citenamefont {Endo}, \citenamefont {Fujii}, \citenamefont {McClean},
  \citenamefont {Mitarai}, \citenamefont {Yuan}, \citenamefont {Cincio},\ and\
  \citenamefont {Coles}}]{Cerezo2021}%
  \BibitemOpen
  \bibinfo {author} {M.~Cerezo}, \bibinfo {author} {A.~Arrasmith}, \bibinfo
  {author} {R.~Babbush}, \bibinfo {author} {S.~C. Benjamin}, \bibinfo {author}
  {S.~Endo}, \bibinfo {author} {K.~Fujii}, \bibinfo {author} {J.~R. McClean},
  \bibinfo {author} {K.~Mitarai}, \bibinfo {author} {X.~Yuan}, \bibinfo
  {author} {L.~Cincio}\ and\ \bibinfo {author} {P.~J. Coles},\ \emph {\bibinfo
  {title} {Variational quantum algorithms}},\ \href
  {\doibase10.1038/s42254-021-00348-9} {\bibfield  {journal} {\bibinfo
  {journal} {Nature Reviews Physics 2021 3:9}\ }\textbf {\bibinfo {volume}
  {3}},\ \bibinfo {pages} {625} (\bibinfo {year} {2021})}\BibitemShut {NoStop}%
\bibitem [{\citenamefont {Endo}\ \emph {et~al.}(2021)\citenamefont {Endo},
  \citenamefont {Cai}, \citenamefont {Benjamin},\ and\ \citenamefont
  {Yuan}}]{EndoJPSJ2021}%
  \BibitemOpen
  \bibinfo {author} {S.~Endo}, \bibinfo {author} {Z.~Cai}, \bibinfo {author}
  {S.~C. Benjamin}\ and\ \bibinfo {author} {X.~Yuan},\ \emph {\bibinfo {title}
  {Hybrid Quantum-Classical Algorithms and Quantum Error Mitigation}},\ \href
  {\doibase10.7566/JPSJ.90.032001} {\bibfield  {journal} {\bibinfo  {journal}
  {Journal of the Physical Society of Japan}\ }\textbf {\bibinfo {volume}
  {90}},\ \bibinfo {pages} {032001} (\bibinfo {year} {2021})}\BibitemShut
  {NoStop}%
\bibitem [{\citenamefont {Endo}\ \emph {et~al.}(2018)\citenamefont {Endo},
  \citenamefont {Benjamin},\ and\ \citenamefont {Li}}]{EndoPRX2018}%
  \BibitemOpen
  \bibinfo {author} {S.~Endo}, \bibinfo {author} {S.~C. Benjamin}\ and\
  \bibinfo {author} {Y.~Li},\ \emph {\bibinfo {title} {Practical Quantum Error
  Mitigation for Near-Future Applications}},\ \href
  {\doibase10.1103/PhysRevX.8.031027} {\bibfield  {journal} {\bibinfo
  {journal} {Phys. Rev. X}\ }\textbf {\bibinfo {volume} {8}},\ \bibinfo {pages}
  {031027} (\bibinfo {year} {2018})}\BibitemShut {NoStop}%
\bibitem [{\citenamefont {Shor}(1996)}]{Shor1996}%
  \BibitemOpen
  \bibinfo {author} {P.~Shor},\ in\ \href {\doibase10.1109/SFCS.1996.548464}
  {\emph {\bibinfo {booktitle} {Proceedings of 37th Conference on Foundations
  of Computer Science}}}\ (\bibinfo {year} {1996})\ pp.\ \bibinfo {pages}
  {56--65}\BibitemShut {NoStop}%
\bibitem [{\citenamefont {Nielsen}\ and\ \citenamefont
  {Chuang}(2010)}]{NielsenChuangBIBLE2010}%
  \BibitemOpen
  \bibinfo {author} {M.~A. Nielsen}\ and\ \bibinfo {author} {I.~L. Chuang},\
  \href {\doibase10.1017/CBO9780511976667} {\emph {\bibinfo {title} {Quantum
  Computation and Quantum Information: 10th Anniversary Edition}}}\ (\bibinfo
  {publisher} {Cambridge University Press},\ \bibinfo {year}
  {2010})\BibitemShut {NoStop}%
\bibitem [{\citenamefont {Krinner}\ \emph {et~al.}(2022)\citenamefont
  {Krinner}, \citenamefont {Lacroix}, \citenamefont {Remm}, \citenamefont
  {Di~Paolo}, \citenamefont {Genois}, \citenamefont {Leroux}, \citenamefont
  {Hellings}, \citenamefont {Lazar}, \citenamefont {Swiadek}, \citenamefont
  {Herrmann}, \citenamefont {Norris}, \citenamefont {Andersen}, \citenamefont
  {Müller}, \citenamefont {Blais}, \citenamefont {Eichler},\ and\
  \citenamefont {Wallraff}}]{krinner2021}%
  \BibitemOpen
  \bibinfo {author} {S.~Krinner}, \bibinfo {author} {N.~Lacroix}, \bibinfo
  {author} {A.~Remm}, \bibinfo {author} {A.~Di~Paolo}, \bibinfo {author}
  {E.~Genois}, \bibinfo {author} {C.~Leroux}, \bibinfo {author} {C.~Hellings},
  \bibinfo {author} {S.~Lazar}, \bibinfo {author} {F.~Swiadek}, \bibinfo
  {author} {J.~Herrmann}, \bibinfo {author} {G.~J. Norris}, \bibinfo {author}
  {C.~K. Andersen}, \bibinfo {author} {M.~Müller}, \bibinfo {author}
  {A.~Blais}, \bibinfo {author} {C.~Eichler}\ and\ \bibinfo {author}
  {A.~Wallraff},\ \emph {\bibinfo {title} {Realizing repeated quantum error
  correction in a distance-three surface code}},\ \href
  {\doibase10.1038/s41586-022-04566-8} {\bibfield  {journal} {\bibinfo
  {journal} {Nature}\ }\textbf {\bibinfo {volume} {605}},\ \bibinfo {pages}
  {669} (\bibinfo {year} {2022})}\BibitemShut {NoStop}%
\bibitem [{\citenamefont {Andersen}\ \emph {et~al.}(2020)\citenamefont
  {Andersen}, \citenamefont {Remm}, \citenamefont {Lazar}, \citenamefont
  {Krinner}, \citenamefont {Lacroix}, \citenamefont {Norris}, \citenamefont
  {Gabureac}, \citenamefont {Eichler},\ and\ \citenamefont
  {Wallraff}}]{AndersenNatPhys20}%
  \BibitemOpen
  \bibinfo {author} {C.~K. Andersen}, \bibinfo {author} {A.~Remm}, \bibinfo
  {author} {S.~Lazar}, \bibinfo {author} {S.~Krinner}, \bibinfo {author}
  {N.~Lacroix}, \bibinfo {author} {G.~J. Norris}, \bibinfo {author}
  {M.~Gabureac}, \bibinfo {author} {C.~Eichler}\ and\ \bibinfo {author}
  {A.~Wallraff},\ \emph {\bibinfo {title} {Repeated quantum error detection in
  a surface code}},\ \href {\doibase10.1038/s41567-020-0920-y} {\bibfield
  {journal} {\bibinfo  {journal} {Nature Physics}\ }\textbf {\bibinfo {volume}
  {16}},\ \bibinfo {pages} {875} (\bibinfo {year} {2020})}\BibitemShut
  {NoStop}%
\bibitem [{\citenamefont {Marques}\ \emph {et~al.}(2022)\citenamefont
  {Marques}, \citenamefont {Varbanov}, \citenamefont {Moreira}, \citenamefont
  {Ali}, \citenamefont {Muthusubramanian}, \citenamefont {Zachariadis},
  \citenamefont {Battistel}, \citenamefont {Beekman}, \citenamefont {Haider},
  \citenamefont {Vlothuizen}, \citenamefont {Bruno}, \citenamefont {Terhal},\
  and\ \citenamefont {DiCarlo}}]{Marques2021}%
  \BibitemOpen
  \bibinfo {author} {J.~F. Marques}, \bibinfo {author} {B.~M. Varbanov},
  \bibinfo {author} {M.~S. Moreira}, \bibinfo {author} {H.~Ali}, \bibinfo
  {author} {N.~Muthusubramanian}, \bibinfo {author} {C.~Zachariadis}, \bibinfo
  {author} {F.~Battistel}, \bibinfo {author} {M.~Beekman}, \bibinfo {author}
  {N.~Haider}, \bibinfo {author} {W.~Vlothuizen}, \bibinfo {author} {A.~Bruno},
  \bibinfo {author} {B.~M. Terhal}\ and\ \bibinfo {author} {L.~DiCarlo},\ \emph
  {\bibinfo {title} {Logical-qubit operations in an error-detecting surface
  code}},\ \href {\doibase10.1038/s41567-021-01423-9} {\bibfield  {journal}
  {\bibinfo  {journal} {Nature Physics}\ }\textbf {\bibinfo {volume} {18}},\
  \bibinfo {pages} {80} (\bibinfo {year} {2022})}\BibitemShut {NoStop}%
\bibitem [{\citenamefont {Ryan-Anderson}\ \emph {et~al.}(2021)\citenamefont
  {Ryan-Anderson}, \citenamefont {Bohnet}, \citenamefont {Lee}, \citenamefont
  {Gresh}, \citenamefont {Hankin}, \citenamefont {Gaebler}, \citenamefont
  {Francois}, \citenamefont {Chernoguzov}, \citenamefont {Lucchetti},
  \citenamefont {Brown}, \citenamefont {Gatterman}, \citenamefont {Halit},
  \citenamefont {Gilmore}, \citenamefont {Gerber}, \citenamefont {Neyenhuis},
  \citenamefont {Hayes},\ and\ \citenamefont {Stutz}}]{Ryananderson2021}%
  \BibitemOpen
  \bibinfo {author} {C.~Ryan-Anderson}, \bibinfo {author} {J.~G. Bohnet},
  \bibinfo {author} {K.~Lee}, \bibinfo {author} {D.~Gresh}, \bibinfo {author}
  {A.~Hankin}, \bibinfo {author} {J.~P. Gaebler}, \bibinfo {author}
  {D.~Francois}, \bibinfo {author} {A.~Chernoguzov}, \bibinfo {author}
  {D.~Lucchetti}, \bibinfo {author} {N.~C. Brown}, \bibinfo {author} {T.~M.
  Gatterman}, \bibinfo {author} {S.~K. Halit}, \bibinfo {author} {K.~Gilmore},
  \bibinfo {author} {J.~A. Gerber}, \bibinfo {author} {B.~Neyenhuis}, \bibinfo
  {author} {D.~Hayes}\ and\ \bibinfo {author} {R.~P. Stutz},\ \emph {\bibinfo
  {title} {Realization of Real-Time Fault-Tolerant Quantum Error Correction}},\
  \href {\doibase10.1103/PhysRevX.11.041058} {\bibfield  {journal} {\bibinfo
  {journal} {Phys. Rev. X}\ }\textbf {\bibinfo {volume} {11}},\ \bibinfo
  {pages} {041058} (\bibinfo {year} {2021})}\BibitemShut {NoStop}%
\bibitem [{\citenamefont {Chen}\ \emph {et~al.}(2021)\citenamefont {Chen},
  \citenamefont {Satzinger}, \citenamefont {Atalaya}, \citenamefont {Korotkov},
  \citenamefont {Dunsworth}, \citenamefont {Sank}, \citenamefont {Quintana},
  \citenamefont {McEwen}, \citenamefont {Barends}, \citenamefont {Klimov},
  \citenamefont {Hong}, \citenamefont {Jones}, \citenamefont {Petukhov},
  \citenamefont {Kafri}, \citenamefont {Demura}, \citenamefont {Burkett},
  \citenamefont {Gidney}, \citenamefont {Fowler}, \citenamefont {Paler},
  \citenamefont {Putterman}, \citenamefont {Aleiner}, \citenamefont {Arute},
  \citenamefont {Arya}, \citenamefont {Babbush}, \citenamefont {Bardin},
  \citenamefont {Bengtsson}, \citenamefont {Bourassa}, \citenamefont
  {Broughton}, \citenamefont {Buckley}, \citenamefont {Buell}, \citenamefont
  {Bushnell}, \citenamefont {Chiaro}, \citenamefont {Collins}, \citenamefont
  {Courtney}, \citenamefont {Derk}, \citenamefont {Eppens}, \citenamefont
  {Erickson}, \citenamefont {Farhi}, \citenamefont {Foxen}, \citenamefont
  {Giustina}, \citenamefont {Greene}, \citenamefont {Gross}, \citenamefont
  {Harrigan}, \citenamefont {Harrington}, \citenamefont {Hilton}, \citenamefont
  {Ho}, \citenamefont {Huang}, \citenamefont {Huggins}, \citenamefont {Ioffe},
  \citenamefont {Isakov}, \citenamefont {Jeffrey}, \citenamefont {Jiang},
  \citenamefont {Kechedzhi}, \citenamefont {Kim}, \citenamefont {Kitaev},
  \citenamefont {Kostritsa}, \citenamefont {Landhuis}, \citenamefont {Laptev},
  \citenamefont {Lucero}, \citenamefont {Martin}, \citenamefont {McClean},
  \citenamefont {McCourt}, \citenamefont {Mi}, \citenamefont {Miao},
  \citenamefont {Mohseni}, \citenamefont {Montazeri}, \citenamefont
  {Mruczkiewicz}, \citenamefont {Mutus}, \citenamefont {Naaman}, \citenamefont
  {Neeley}, \citenamefont {Neill}, \citenamefont {Newman}, \citenamefont {Niu},
  \citenamefont {O’Brien}, \citenamefont {Opremcak}, \citenamefont {Ostby},
  \citenamefont {Pató}, \citenamefont {Redd}, \citenamefont {Roushan},
  \citenamefont {Rubin}, \citenamefont {Shvarts}, \citenamefont {Strain},
  \citenamefont {Szalay}, \citenamefont {Trevithick}, \citenamefont
  {Villalonga}, \citenamefont {White}, \citenamefont {Yao}, \citenamefont
  {Yeh}, \citenamefont {Yoo}, \citenamefont {Zalcman}, \citenamefont {Neven},
  \citenamefont {Boixo}, \citenamefont {Smelyanskiy}, \citenamefont {Chen},
  \citenamefont {Megrant},\ and\ \citenamefont {Kelly}}]{Chen2021}%
  \BibitemOpen
  \bibinfo {author} {Z.~Chen}, \bibinfo {author} {K.~J. Satzinger}, \bibinfo
  {author} {J.~Atalaya}, \bibinfo {author} {A.~N. Korotkov}, \bibinfo {author}
  {A.~Dunsworth}, \bibinfo {author} {D.~Sank}, \bibinfo {author} {C.~Quintana},
  \bibinfo {author} {M.~McEwen}, \bibinfo {author} {R.~Barends}, \bibinfo
  {author} {P.~V. Klimov} et~al.,\ \emph {\bibinfo {title} {Exponential
  suppression of bit or phase errors with cyclic error correction}},\ \href
  {\doibase10.1038/s41586-021-03588-y} {\bibfield  {journal} {\bibinfo
  {journal} {Nature 2021 595:7867}\ }\textbf {\bibinfo {volume} {595}},\
  \bibinfo {pages} {383} (\bibinfo {year} {2021})}\BibitemShut {NoStop}%
\bibitem [{\citenamefont {Albert}\ \emph {et~al.}(2018)\citenamefont {Albert},
  \citenamefont {Noh}, \citenamefont {Duivenvoorden}, \citenamefont {Young},
  \citenamefont {Brierley}, \citenamefont {Reinhold}, \citenamefont {Vuillot},
  \citenamefont {Li}, \citenamefont {Shen}, \citenamefont {Girvin},
  \citenamefont {Terhal},\ and\ \citenamefont {Jiang}}]{AlbertPRA2018}%
  \BibitemOpen
  \bibinfo {author} {V.~V. Albert}, \bibinfo {author} {K.~Noh}, \bibinfo
  {author} {K.~Duivenvoorden}, \bibinfo {author} {D.~J. Young}, \bibinfo
  {author} {R.~T. Brierley}, \bibinfo {author} {P.~Reinhold}, \bibinfo {author}
  {C.~Vuillot}, \bibinfo {author} {L.~Li}, \bibinfo {author} {C.~Shen},
  \bibinfo {author} {S.~M. Girvin}, \bibinfo {author} {B.~M. Terhal}\ and\
  \bibinfo {author} {L.~Jiang},\ \emph {\bibinfo {title} {Performance and
  structure of single-mode bosonic codes}},\ \href
  {\doibase10.1103/PhysRevA.97.032346} {\bibfield  {journal} {\bibinfo
  {journal} {Phys. Rev. A}\ }\textbf {\bibinfo {volume} {97}},\ \bibinfo
  {pages} {032346} (\bibinfo {year} {2018})}\BibitemShut {NoStop}%
\bibitem [{\citenamefont {Terhal}\ \emph {et~al.}(2020)\citenamefont {Terhal},
  \citenamefont {Conrad},\ and\ \citenamefont
  {Vuillot}}]{TerhalQuantumScienceandTechnology2020}%
  \BibitemOpen
  \bibinfo {author} {B.~M. Terhal}, \bibinfo {author} {J.~Conrad}\ and\
  \bibinfo {author} {C.~Vuillot},\ \emph {\bibinfo {title} {Towards scalable
  bosonic quantum error correction}},\ \href {\doibase10.1088/2058-9565/ab98a5}
  {\bibfield  {journal} {\bibinfo  {journal} {Quantum Science and Technology}\
  }\textbf {\bibinfo {volume} {5}},\ \bibinfo {pages} {043001} (\bibinfo {year}
  {2020})}\BibitemShut {NoStop}%
\bibitem [{\citenamefont {Joshi}\ \emph {et~al.}(2021)\citenamefont {Joshi},
  \citenamefont {Noh},\ and\ \citenamefont {Gao}}]{JoshiQST21}%
  \BibitemOpen
  \bibinfo {author} {A.~Joshi}, \bibinfo {author} {K.~Noh}\ and\ \bibinfo
  {author} {Y.~Y. Gao},\ \emph {\bibinfo {title} {Quantum information
  processing with bosonic qubits in circuit {QED}}},\ \href
  {\doibase10.1088/2058-9565/abe989} {\bibfield  {journal} {\bibinfo  {journal}
  {Quantum Science and Technology}\ }\textbf {\bibinfo {volume} {6}},\ \bibinfo
  {pages} {033001} (\bibinfo {year} {2021})}\BibitemShut {NoStop}%
\bibitem [{\citenamefont {Houck}\ \emph {et~al.}(2012)\citenamefont {Houck},
  \citenamefont {Tureci},\ and\ \citenamefont {Koch}}]{HouckNatPhys12}%
  \BibitemOpen
  \bibinfo {author} {A.~A. Houck}, \bibinfo {author} {H.~E. Tureci}\ and\
  \bibinfo {author} {J.~Koch},\ \emph {\bibinfo {title} {On-chip quantum
  simulation with superconducting circuits}},\ \href
  {http://dx.doi.org/10.1038/nphys2251} {\bibfield  {journal} {\bibinfo
  {journal} {Nat. Phys.}\ }\textbf {\bibinfo {volume} {8}},\ \bibinfo {pages}
  {292} (\bibinfo {year} {2012})}\BibitemShut {NoStop}%
\bibitem [{\citenamefont {Carusotto}\ and\ \citenamefont
  {Ciuti}(2013)}]{Carusotto_RMP_2013_quantum_fluids_light}%
  \BibitemOpen
  \bibinfo {author} {I.~Carusotto}\ and\ \bibinfo {author} {C.~Ciuti},\ \emph
  {\bibinfo {title} {Quantum fluids of light}},\ \href
  {https://link.aps.org/doi/10.1103/RevModPhys.85.299} {\bibfield  {journal}
  {\bibinfo  {journal} {Rev. Mod. Phys.}\ }\textbf {\bibinfo {volume} {85}},\
  \bibinfo {pages} {299} (\bibinfo {year} {2013})}\BibitemShut {NoStop}%
\bibitem [{\citenamefont {Aspelmeyer}\ \emph {et~al.}(2014)\citenamefont
  {Aspelmeyer}, \citenamefont {Kippenberg},\ and\ \citenamefont
  {Marquardt}}]{AspelmeyerRMP2014}%
  \BibitemOpen
  \bibinfo {author} {M.~Aspelmeyer}, \bibinfo {author} {T.~J. Kippenberg}\ and\
  \bibinfo {author} {F.~Marquardt},\ \emph {\bibinfo {title} {Cavity
  optomechanics}},\ \href {\doibase10.1103/RevModPhys.86.1391} {\bibfield
  {journal} {\bibinfo  {journal} {Rev. Mod. Phys.}\ }\textbf {\bibinfo {volume}
  {86}},\ \bibinfo {pages} {1391} (\bibinfo {year} {2014})}\BibitemShut
  {NoStop}%
\bibitem [{\citenamefont {Gu}\ \emph {et~al.}(2017)\citenamefont {Gu},
  \citenamefont {Kockum}, \citenamefont {Miranowicz}, \citenamefont {xi~Liu},\
  and\ \citenamefont {Nori}}]{Gu2017}%
  \BibitemOpen
  \bibinfo {author} {X.~Gu}, \bibinfo {author} {A.~F. Kockum}, \bibinfo
  {author} {A.~Miranowicz}, \bibinfo {author} {Y.~xi~Liu}\ and\ \bibinfo
  {author} {F.~Nori},\ \emph {\bibinfo {title} {Microwave photonics with
  superconducting quantum circuits}},\ \href
  {\doibase10.1016/j.physrep.2017.10.002} {\bibfield  {journal} {\bibinfo
  {journal} {Physics Reports}\ }\textbf {\bibinfo {volume} {718-719}},\
  \bibinfo {pages} {1} (\bibinfo {year} {2017})}\BibitemShut {NoStop}%
\bibitem [{\citenamefont {Lebreuilly}\ \emph {et~al.}(2021)\citenamefont
  {Lebreuilly}, \citenamefont {Noh}, \citenamefont {Wang}, \citenamefont
  {Girvin},\ and\ \citenamefont {Jiang}}]{Lebreuilly2021}%
  \BibitemOpen
  \bibinfo {author} {J.~Lebreuilly}, \bibinfo {author} {K.~Noh}, \bibinfo
  {author} {C.-H. Wang}, \bibinfo {author} {S.~M. Girvin}\ and\ \bibinfo
  {author} {L.~Jiang},\ \href@noop {} {\emph {\bibinfo {title} {Autonomous
  quantum error correction and quantum computation}}} (\bibinfo {year}
  {2021}),\ \Eprint {http://arxiv.org/abs/2103.05007} {arXiv:2103.05007
  [quant-ph]} \BibitemShut {NoStop}%
\bibitem [{\citenamefont {Knill}\ \emph {et~al.}(2001)\citenamefont {Knill},
  \citenamefont {Laflamme},\ and\ \citenamefont {Milburn}}]{Knill2001}%
  \BibitemOpen
  \bibinfo {author} {E.~Knill}, \bibinfo {author} {R.~Laflamme}\ and\ \bibinfo
  {author} {G.~J. Milburn},\ \emph {\bibinfo {title} {A scheme for efficient
  quantum computation with linear optics}},\ \href {\doibase10.1038/35051009}
  {\bibfield  {journal} {\bibinfo  {journal} {Nature 2001 409:6816}\ }\textbf
  {\bibinfo {volume} {409}},\ \bibinfo {pages} {46} (\bibinfo {year}
  {2001})}\BibitemShut {NoStop}%
\bibitem [{\citenamefont {O'Brien}\ \emph {et~al.}(2003)\citenamefont
  {O'Brien}, \citenamefont {Pryde}, \citenamefont {White}, \citenamefont
  {Ralph},\ and\ \citenamefont {Branning}}]{OBrien2003}%
  \BibitemOpen
  \bibinfo {author} {J.~L. O'Brien}, \bibinfo {author} {G.~J. Pryde}, \bibinfo
  {author} {A.~G. White}, \bibinfo {author} {T.~C. Ralph}\ and\ \bibinfo
  {author} {D.~Branning},\ \emph {\bibinfo {title} {Demonstration of an
  all-optical quantum controlled-NOT gate}},\ \href
  {\doibase10.1038/nature02054} {\bibfield  {journal} {\bibinfo  {journal}
  {Nature 2003 426:6964}\ }\textbf {\bibinfo {volume} {426}},\ \bibinfo {pages}
  {264} (\bibinfo {year} {2003})}\BibitemShut {NoStop}%
\bibitem [{\citenamefont {Leung}\ \emph {et~al.}(1997)\citenamefont {Leung},
  \citenamefont {Nielsen}, \citenamefont {Chuang},\ and\ \citenamefont
  {Yamamoto}}]{LeungPRA1997}%
  \BibitemOpen
  \bibinfo {author} {D.~W. Leung}, \bibinfo {author} {M.~A. Nielsen}, \bibinfo
  {author} {I.~L. Chuang}\ and\ \bibinfo {author} {Y.~Yamamoto},\ \emph
  {\bibinfo {title} {Approximate quantum error correction can lead to better
  codes}},\ \href {\doibase10.1103/PhysRevA.56.2567} {\bibfield  {journal}
  {\bibinfo  {journal} {Phys. Rev. A}\ }\textbf {\bibinfo {volume} {56}},\
  \bibinfo {pages} {2567} (\bibinfo {year} {1997})}\BibitemShut {NoStop}%
\bibitem [{\citenamefont {Chuang}\ and\ \citenamefont
  {Yamamoto}(1995)}]{ChuangPRA1995}%
  \BibitemOpen
  \bibinfo {author} {I.~L. Chuang}\ and\ \bibinfo {author} {Y.~Yamamoto},\
  \emph {\bibinfo {title} {Simple quantum computer}},\ \href
  {\doibase10.1103/PhysRevA.52.3489} {\bibfield  {journal} {\bibinfo  {journal}
  {Phys. Rev. A}\ }\textbf {\bibinfo {volume} {52}},\ \bibinfo {pages} {3489}
  (\bibinfo {year} {1995})}\BibitemShut {NoStop}%
\bibitem [{\citenamefont {Gottesman}\ \emph {et~al.}(2001)\citenamefont
  {Gottesman}, \citenamefont {Kitaev},\ and\ \citenamefont
  {Preskill}}]{GottesmanPRA01}%
  \BibitemOpen
  \bibinfo {author} {D.~Gottesman}, \bibinfo {author} {A.~Kitaev}\ and\
  \bibinfo {author} {J.~Preskill},\ \emph {\bibinfo {title} {Encoding a qubit
  in an oscillator}},\ \href {\doibase10.1103/PhysRevA.64.012310} {\bibfield
  {journal} {\bibinfo  {journal} {Phys. Rev. A}\ }\textbf {\bibinfo {volume}
  {64}},\ \bibinfo {pages} {012310} (\bibinfo {year} {2001})}\BibitemShut
  {NoStop}%
\bibitem [{\citenamefont {Grimsmo}\ and\ \citenamefont
  {Puri}(2021)}]{GrimsmoPRXQuantum2021}%
  \BibitemOpen
  \bibinfo {author} {A.~L. Grimsmo}\ and\ \bibinfo {author} {S.~Puri},\ \emph
  {\bibinfo {title} {Quantum Error Correction with the
  Gottesman-Kitaev-Preskill Code}},\ \href
  {\doibase10.1103/PRXQuantum.2.020101} {\bibfield  {journal} {\bibinfo
  {journal} {PRX Quantum}\ }\textbf {\bibinfo {volume} {2}},\ \bibinfo {pages}
  {020101} (\bibinfo {year} {2021})}\BibitemShut {NoStop}%
\bibitem [{Note1()}]{Note1}%
  \BibitemOpen
  \bibinfo {note} {From now on, we will use the terminology ``particle loss''
  instead of ``photon loss'', to refer to more general physical realizations
  such as, for example, phonons in optomechanical systems.}\BibitemShut {Stop}%
\bibitem [{\citenamefont {Flühmann}\ \emph {et~al.}(2019)\citenamefont
  {Flühmann}, \citenamefont {Nguyen}, \citenamefont {Marinelli}, \citenamefont
  {Negnevitsky}, \citenamefont {Mehta},\ and\ \citenamefont
  {Home}}]{FluhmannNature2019}%
  \BibitemOpen
  \bibinfo {author} {C.~Flühmann}, \bibinfo {author} {T.~L. Nguyen}, \bibinfo
  {author} {M.~Marinelli}, \bibinfo {author} {V.~Negnevitsky}, \bibinfo
  {author} {K.~Mehta}\ and\ \bibinfo {author} {J.~P. Home},\ \emph {\bibinfo
  {title} {Encoding a qubit in a trapped-ion mechanical oscillator}},\ \href
  {\doibase10.1038/s41586-019-0960-6} {\bibfield  {journal} {\bibinfo
  {journal} {Nature}\ }\textbf {\bibinfo {volume} {566}},\ \bibinfo {pages}
  {513} (\bibinfo {year} {2019})}\BibitemShut {NoStop}%
\bibitem [{\citenamefont {Campagne-Ibarcq}\ \emph {et~al.}(2020)\citenamefont
  {Campagne-Ibarcq}, \citenamefont {Eickbusch}, \citenamefont {Touzard},
  \citenamefont {Zalys-Geller}, \citenamefont {Frattini}, \citenamefont
  {Sivak}, \citenamefont {Reinhold}, \citenamefont {Puri}, \citenamefont
  {Shankar}, \citenamefont {Schoelkopf}, \citenamefont {Frunzio}, \citenamefont
  {Mirrahimi},\ and\ \citenamefont {Devoret}}]{Campagne2020}%
  \BibitemOpen
  \bibinfo {author} {P.~Campagne-Ibarcq}, \bibinfo {author} {A.~Eickbusch},
  \bibinfo {author} {S.~Touzard}, \bibinfo {author} {E.~Zalys-Geller}, \bibinfo
  {author} {N.~E. Frattini}, \bibinfo {author} {V.~V. Sivak}, \bibinfo {author}
  {P.~Reinhold}, \bibinfo {author} {S.~Puri}, \bibinfo {author} {S.~Shankar},
  \bibinfo {author} {R.~J. Schoelkopf}, \bibinfo {author} {L.~Frunzio},
  \bibinfo {author} {M.~Mirrahimi}\ and\ \bibinfo {author} {M.~H. Devoret},\
  \emph {\bibinfo {title} {Quantum error correction of a qubit encoded in grid
  states of an oscillator}},\ \href {\doibase10.1038/s41586-020-2603-3}
  {\bibfield  {journal} {\bibinfo  {journal} {Nature 2020 584:7821}\ }\textbf
  {\bibinfo {volume} {584}},\ \bibinfo {pages} {368} (\bibinfo {year}
  {2020})}\BibitemShut {NoStop}%
\bibitem [{\citenamefont {Ralph}\ \emph {et~al.}(2003)\citenamefont {Ralph},
  \citenamefont {Gilchrist}, \citenamefont {Milburn}, \citenamefont {Munro},\
  and\ \citenamefont {Glancy}}]{RalphPRA2003}%
  \BibitemOpen
  \bibinfo {author} {T.~C. Ralph}, \bibinfo {author} {A.~Gilchrist}, \bibinfo
  {author} {G.~J. Milburn}, \bibinfo {author} {W.~J. Munro}\ and\ \bibinfo
  {author} {S.~Glancy},\ \emph {\bibinfo {title} {Quantum computation with
  optical coherent states}},\ \href {\doibase10.1103/PhysRevA.68.042319}
  {\bibfield  {journal} {\bibinfo  {journal} {Phys. Rev. A}\ }\textbf {\bibinfo
  {volume} {68}},\ \bibinfo {pages} {042319} (\bibinfo {year}
  {2003})}\BibitemShut {NoStop}%
\bibitem [{\citenamefont {Ofek}\ \emph {et~al.}(2016)\citenamefont {Ofek},
  \citenamefont {Petrenko}, \citenamefont {Heeres}, \citenamefont {Reinhold},
  \citenamefont {Leghtas}, \citenamefont {Vlastakis}, \citenamefont {Liu},
  \citenamefont {Frunzio}, \citenamefont {Girvin}, \citenamefont {Jiang},
  \citenamefont {Mirrahimi}, \citenamefont {Devoret},\ and\ \citenamefont
  {Schoelkopf}}]{OfekNat16}%
  \BibitemOpen
  \bibinfo {author} {N.~Ofek}, \bibinfo {author} {A.~Petrenko}, \bibinfo
  {author} {R.~Heeres}, \bibinfo {author} {P.~Reinhold}, \bibinfo {author}
  {Z.~Leghtas}, \bibinfo {author} {B.~Vlastakis}, \bibinfo {author} {Y.~Liu},
  \bibinfo {author} {L.~Frunzio}, \bibinfo {author} {S.~M. Girvin}, \bibinfo
  {author} {L.~Jiang}, \bibinfo {author} {M.~Mirrahimi}, \bibinfo {author}
  {M.~H. Devoret}\ and\ \bibinfo {author} {R.~J. Schoelkopf},\ \emph {\bibinfo
  {title} {Extending the lifetime of a quantum bit with error correction in
  superconducting circuits}},\ \href {https://doi.org/10.1038/nature18949}
  {\bibfield  {journal} {\bibinfo  {journal} {Nature (London)}\ }\textbf
  {\bibinfo {volume} {536}},\ \bibinfo {pages} {441} (\bibinfo {year}
  {2016})}\BibitemShut {NoStop}%
\bibitem [{\citenamefont {Bergmann}\ and\ \citenamefont {van
  Loock}(2016)}]{BergmannPRA2016}%
  \BibitemOpen
  \bibinfo {author} {M.~Bergmann}\ and\ \bibinfo {author} {P.~van Loock},\
  \emph {\bibinfo {title} {Quantum error correction against photon loss using
  multicomponent cat states}},\ \href {\doibase10.1103/PhysRevA.94.042332}
  {\bibfield  {journal} {\bibinfo  {journal} {Phys. Rev. A}\ }\textbf {\bibinfo
  {volume} {94}},\ \bibinfo {pages} {042332} (\bibinfo {year}
  {2016})}\BibitemShut {NoStop}%
\bibitem [{\citenamefont {Lescanne}\ \emph {et~al.}(2020)\citenamefont
  {Lescanne}, \citenamefont {Villiers}, \citenamefont {Peronnin}, \citenamefont
  {Sarlette}, \citenamefont {Delbecq}, \citenamefont {Huard}, \citenamefont
  {Kontos}, \citenamefont {Mirrahimi},\ and\ \citenamefont
  {Leghtas}}]{LescanneNatPhys2020}%
  \BibitemOpen
  \bibinfo {author} {R.~Lescanne}, \bibinfo {author} {M.~Villiers}, \bibinfo
  {author} {T.~Peronnin}, \bibinfo {author} {A.~Sarlette}, \bibinfo {author}
  {M.~Delbecq}, \bibinfo {author} {B.~Huard}, \bibinfo {author} {T.~Kontos},
  \bibinfo {author} {M.~Mirrahimi}\ and\ \bibinfo {author} {Z.~Leghtas},\ \emph
  {\bibinfo {title} {Exponential suppression of bit-flips in a qubit encoded in
  an oscillator}},\ \href {\doibase10.1038/s41567-020-0824-x} {\bibfield
  {journal} {\bibinfo  {journal} {Nature Physics}\ }\textbf {\bibinfo {volume}
  {16}},\ \bibinfo {pages} {509} (\bibinfo {year} {2020})}\BibitemShut
  {NoStop}%
\bibitem [{\citenamefont {Grimm}\ \emph {et~al.}(2020)\citenamefont {Grimm},
  \citenamefont {Frattini}, \citenamefont {Puri}, \citenamefont {Mundhada},
  \citenamefont {Touzard}, \citenamefont {Mirrahimi}, \citenamefont {Girvin},
  \citenamefont {Shankar},\ and\ \citenamefont {Devoret}}]{GrimmNature2020}%
  \BibitemOpen
  \bibinfo {author} {A.~Grimm}, \bibinfo {author} {N.~E. Frattini}, \bibinfo
  {author} {S.~Puri}, \bibinfo {author} {S.~O. Mundhada}, \bibinfo {author}
  {S.~Touzard}, \bibinfo {author} {M.~Mirrahimi}, \bibinfo {author} {S.~M.
  Girvin}, \bibinfo {author} {S.~Shankar}\ and\ \bibinfo {author} {M.~H.
  Devoret},\ \emph {\bibinfo {title} {Stabilization and operation of a
  {Kerr}-cat qubit}},\ \href {\doibase10.1038/s41586-020-2587-z} {\bibfield
  {journal} {\bibinfo  {journal} {Nature}\ }\textbf {\bibinfo {volume} {584}},\
  \bibinfo {pages} {205} (\bibinfo {year} {2020})}\BibitemShut {NoStop}%
\bibitem [{\citenamefont {Puri}\ \emph {et~al.}(2017)\citenamefont {Puri},
  \citenamefont {Boutin},\ and\ \citenamefont {Blais}}]{PurinpjQI17}%
  \BibitemOpen
  \bibinfo {author} {S.~Puri}, \bibinfo {author} {S.~Boutin}\ and\ \bibinfo
  {author} {A.~Blais},\ \emph {\bibinfo {title} {Engineering the quantum states
  of light in a Kerr-nonlinear resonator by two-photon driving}},\ \href
  {https://doi.org/10.1038/s41534-017-0019-1} {\bibfield  {journal} {\bibinfo
  {journal} {npj Quantum Information}\ }\textbf {\bibinfo {volume} {3}},\
  \bibinfo {pages} {18} (\bibinfo {year} {2017})}\BibitemShut {NoStop}%
\bibitem [{\citenamefont {Puri}\ \emph {et~al.}(2019)\citenamefont {Puri},
  \citenamefont {Grimm}, \citenamefont {Campagne-Ibarcq}, \citenamefont
  {Eickbusch}, \citenamefont {Noh}, \citenamefont {Roberts}, \citenamefont
  {Jiang}, \citenamefont {Mirrahimi}, \citenamefont {Devoret},\ and\
  \citenamefont {Girvin}}]{PuriPRX2019}%
  \BibitemOpen
  \bibinfo {author} {S.~Puri}, \bibinfo {author} {A.~Grimm}, \bibinfo {author}
  {P.~Campagne-Ibarcq}, \bibinfo {author} {A.~Eickbusch}, \bibinfo {author}
  {K.~Noh}, \bibinfo {author} {G.~Roberts}, \bibinfo {author} {L.~Jiang},
  \bibinfo {author} {M.~Mirrahimi}, \bibinfo {author} {M.~H. Devoret}\ and\
  \bibinfo {author} {S.~M. Girvin},\ \emph {\bibinfo {title} {Stabilized Cat in
  a Driven Nonlinear Cavity: A Fault-Tolerant Error Syndrome Detector}},\ \href
  {\doibase10.1103/PhysRevX.9.041009} {\bibfield  {journal} {\bibinfo
  {journal} {Phys. Rev. X}\ }\textbf {\bibinfo {volume} {9}},\ \bibinfo {pages}
  {041009} (\bibinfo {year} {2019})}\BibitemShut {NoStop}%
\bibitem [{\citenamefont {Jeong}\ and\ \citenamefont {Kim}(2002)}]{JeongPRA02}%
  \BibitemOpen
  \bibinfo {author} {H.~Jeong}\ and\ \bibinfo {author} {M.~S. Kim},\ \emph
  {\bibinfo {title} {Efficient quantum computation using coherent states}},\
  \href {\doibase10.1103/PhysRevA.65.042305} {\bibfield  {journal} {\bibinfo
  {journal} {Phys. Rev. A}\ }\textbf {\bibinfo {volume} {65}},\ \bibinfo
  {pages} {042305} (\bibinfo {year} {2002})}\BibitemShut {NoStop}%
\bibitem [{\citenamefont {Leghtas}\ \emph
  {et~al.}(2013{\natexlab{a}})\citenamefont {Leghtas}, \citenamefont
  {Kirchmair}, \citenamefont {Vlastakis}, \citenamefont {Devoret},
  \citenamefont {Schoelkopf},\ and\ \citenamefont
  {Mirrahimi}}]{LeghtasPRA2013}%
  \BibitemOpen
  \bibinfo {author} {Z.~Leghtas}, \bibinfo {author} {G.~Kirchmair}, \bibinfo
  {author} {B.~Vlastakis}, \bibinfo {author} {M.~H. Devoret}, \bibinfo {author}
  {R.~J. Schoelkopf}\ and\ \bibinfo {author} {M.~Mirrahimi},\ \emph {\bibinfo
  {title} {Deterministic protocol for mapping a qubit to coherent state
  superpositions in a cavity}},\ \href {\doibase10.1103/PhysRevA.87.042315}
  {\bibfield  {journal} {\bibinfo  {journal} {Phys. Rev. A}\ }\textbf {\bibinfo
  {volume} {87}},\ \bibinfo {pages} {042315} (\bibinfo {year}
  {2013}{\natexlab{a}})}\BibitemShut {NoStop}%
\bibitem [{\citenamefont {Leghtas}\ \emph
  {et~al.}(2013{\natexlab{b}})\citenamefont {Leghtas}, \citenamefont
  {Kirchmair}, \citenamefont {Vlastakis}, \citenamefont {Schoelkopf},
  \citenamefont {Devoret},\ and\ \citenamefont {Mirrahimi}}]{LeghtasPRL2013}%
  \BibitemOpen
  \bibinfo {author} {Z.~Leghtas}, \bibinfo {author} {G.~Kirchmair}, \bibinfo
  {author} {B.~Vlastakis}, \bibinfo {author} {R.~J. Schoelkopf}, \bibinfo
  {author} {M.~H. Devoret}\ and\ \bibinfo {author} {M.~Mirrahimi},\ \emph
  {\bibinfo {title} {Hardware-Efficient Autonomous Quantum Memory
  Protection}},\ \href {\doibase10.1103/PhysRevLett.111.120501} {\bibfield
  {journal} {\bibinfo  {journal} {Phys. Rev. Lett.}\ }\textbf {\bibinfo
  {volume} {111}},\ \bibinfo {pages} {120501} (\bibinfo {year}
  {2013}{\natexlab{b}})}\BibitemShut {NoStop}%
\bibitem [{\citenamefont {Mirrahimi}\ \emph {et~al.}(2014)\citenamefont
  {Mirrahimi}, \citenamefont {Leghtas}, \citenamefont {Albert}, \citenamefont
  {Touzard}, \citenamefont {Schoelkopf}, \citenamefont {Jiang},\ and\
  \citenamefont {Devoret}}]{MirrahimiNJP14}%
  \BibitemOpen
  \bibinfo {author} {M.~Mirrahimi}, \bibinfo {author} {M.~Leghtas}, \bibinfo
  {author} {V.~Albert}, \bibinfo {author} {S.~Touzard}, \bibinfo {author}
  {R.~Schoelkopf}, \bibinfo {author} {L.~Jiang}\ and\ \bibinfo {author}
  {M.~Devoret},\ \emph {\bibinfo {title} {Dynamically protected cat-qubits: a
  new paradigm for universal quantum computation}},\ \href
  {http://dx.doi.org/10.1088/1367-2630/16/4/045014} {\bibfield  {journal}
  {\bibinfo  {journal} {New Journal of Physics}\ }\textbf {\bibinfo {volume}
  {16}},\ \bibinfo {pages} {045014} (\bibinfo {year} {2014})}\BibitemShut
  {NoStop}%
\bibitem [{\citenamefont {Leghtas}\ \emph {et~al.}(2015)\citenamefont
  {Leghtas}, \citenamefont {Touzard}, \citenamefont {Pop}, \citenamefont {Kou},
  \citenamefont {Vlastakis}, \citenamefont {Petrenko}, \citenamefont {Sliwa},
  \citenamefont {Narla}, \citenamefont {Shankar}, \citenamefont {Hatridge},
  \citenamefont {Reagor}, \citenamefont {Frunzio}, \citenamefont {Schoelkopf},
  \citenamefont {Mirrahimi},\ and\ \citenamefont {Devoret}}]{LeghtasScience15}%
  \BibitemOpen
  \bibinfo {author} {Z.~Leghtas}, \bibinfo {author} {S.~Touzard}, \bibinfo
  {author} {I.~M. Pop}, \bibinfo {author} {A.~Kou}, \bibinfo {author}
  {B.~Vlastakis}, \bibinfo {author} {A.~Petrenko}, \bibinfo {author} {K.~M.
  Sliwa}, \bibinfo {author} {A.~Narla}, \bibinfo {author} {S.~Shankar},
  \bibinfo {author} {M.~J. Hatridge}, \bibinfo {author} {M.~Reagor}, \bibinfo
  {author} {L.~Frunzio}, \bibinfo {author} {R.~J. Schoelkopf}, \bibinfo
  {author} {M.~Mirrahimi}\ and\ \bibinfo {author} {M.~H. Devoret},\ \emph
  {\bibinfo {title} {Confining the state of light to a quantum manifold by
  engineered two-photon loss}},\ \href
  {http://dx.doi.org/10.1126/science.aaa2085} {\bibfield  {journal} {\bibinfo
  {journal} {Science}\ }\textbf {\bibinfo {volume} {347}},\ \bibinfo {pages}
  {853} (\bibinfo {year} {2015})}\BibitemShut {NoStop}%
\bibitem [{\citenamefont {Vlastakis}\ \emph {et~al.}(2013)\citenamefont
  {Vlastakis}, \citenamefont {Kirchmair}, \citenamefont {Leghtas},
  \citenamefont {Nigg}, \citenamefont {Frunzio}, \citenamefont {Girvin},
  \citenamefont {Mirrahimi}, \citenamefont {Devoret},\ and\ \citenamefont
  {Schoelkopf}}]{VlastakisScience2013}%
  \BibitemOpen
  \bibinfo {author} {B.~Vlastakis}, \bibinfo {author} {G.~Kirchmair}, \bibinfo
  {author} {Z.~Leghtas}, \bibinfo {author} {S.~E. Nigg}, \bibinfo {author}
  {L.~Frunzio}, \bibinfo {author} {S.~M. Girvin}, \bibinfo {author}
  {M.~Mirrahimi}, \bibinfo {author} {M.~H. Devoret}\ and\ \bibinfo {author}
  {R.~J. Schoelkopf},\ \emph {\bibinfo {title} {Deterministically encoding
  quantum information using 100-photon Schrödinger cat states}},\ \href
  {\doibase10.1126/SCIENCE.1243289/SUPPL_FILE/VLASTAKIS.SM.PDF} {\bibfield
  {journal} {\bibinfo  {journal} {Science}\ }\textbf {\bibinfo {volume}
  {342}},\ \bibinfo {pages} {607} (\bibinfo {year} {2013})}\BibitemShut
  {NoStop}%
\bibitem [{\citenamefont {Monroe}\ \emph {et~al.}(1996)\citenamefont {Monroe},
  \citenamefont {Meekhof}, \citenamefont {King},\ and\ \citenamefont
  {Wineland}}]{Monroe1996}%
  \BibitemOpen
  \bibinfo {author} {C.~Monroe}, \bibinfo {author} {D.~M. Meekhof}, \bibinfo
  {author} {B.~E. King}\ and\ \bibinfo {author} {D.~J. Wineland},\ \emph
  {\bibinfo {title} {A “Schrödinger Cat” Superposition State of an
  Atom}},\ \href {\doibase10.1126/SCIENCE.272.5265.1131} {\bibfield  {journal}
  {\bibinfo  {journal} {Science}\ }\textbf {\bibinfo {volume} {272}},\ \bibinfo
  {pages} {1131} (\bibinfo {year} {1996})}\BibitemShut {NoStop}%
\bibitem [{\citenamefont {Del{\'e}glise}\ \emph {et~al.}(2008)\citenamefont
  {Del{\'e}glise}, \citenamefont {Dotsenko}, \citenamefont {Sayrin},
  \citenamefont {Bernu}, \citenamefont {Brune}, \citenamefont {Raimond},\ and\
  \citenamefont {Haroche}}]{DelegliseNat08}%
  \BibitemOpen
  \bibinfo {author} {S.~Del{\'e}glise}, \bibinfo {author} {I.~Dotsenko},
  \bibinfo {author} {C.~Sayrin}, \bibinfo {author} {J.~Bernu}, \bibinfo
  {author} {M.~Brune}, \bibinfo {author} {J.-M. Raimond}\ and\ \bibinfo
  {author} {S.~Haroche},\ \emph {\bibinfo {title} {Reconstruction of
  non-classical cavity field states with snapshots of their decoherence}},\
  \href {https://doi.org/10.1038/nature07288} {\bibfield  {journal} {\bibinfo
  {journal} {Nature (London)}\ }\textbf {\bibinfo {volume} {455}},\ \bibinfo
  {pages} {510} (\bibinfo {year} {2008})}\BibitemShut {NoStop}%
\bibitem [{\citenamefont {Hofheinz}\ \emph {et~al.}(2009)\citenamefont
  {Hofheinz}, \citenamefont {Wang}, \citenamefont {Ansmann}, \citenamefont
  {Bialczak}, \citenamefont {Lucero}, \citenamefont {Neeley}, \citenamefont
  {O'Connell}, \citenamefont {Sank}, \citenamefont {Wenner}, \citenamefont
  {Martinis},\ and\ \citenamefont {Cleland}}]{Hofheinz2009}%
  \BibitemOpen
  \bibinfo {author} {M.~Hofheinz}, \bibinfo {author} {H.~Wang}, \bibinfo
  {author} {M.~Ansmann}, \bibinfo {author} {R.~C. Bialczak}, \bibinfo {author}
  {E.~Lucero}, \bibinfo {author} {M.~Neeley}, \bibinfo {author} {A.~D.
  O'Connell}, \bibinfo {author} {D.~Sank}, \bibinfo {author} {J.~Wenner},
  \bibinfo {author} {J.~M. Martinis}\ and\ \bibinfo {author} {A.~N. Cleland},\
  \emph {\bibinfo {title} {Synthesizing arbitrary quantum states in a
  superconducting resonator}},\ \href {\doibase10.1038/nature08005} {\bibfield
  {journal} {\bibinfo  {journal} {Nature 2009 459:7246}\ }\textbf {\bibinfo
  {volume} {459}},\ \bibinfo {pages} {546} (\bibinfo {year}
  {2009})}\BibitemShut {NoStop}%
\bibitem [{\citenamefont {Wang}\ \emph {et~al.}(2016)\citenamefont {Wang},
  \citenamefont {Gao}, \citenamefont {Reinhold}, \citenamefont {Heeres},
  \citenamefont {Ofek}, \citenamefont {Chou}, \citenamefont {Axline},
  \citenamefont {Reagor}, \citenamefont {Blumoff}, \citenamefont {Sliwa},
  \citenamefont {Frunzio}, \citenamefont {Girvin}, \citenamefont {Jiang},
  \citenamefont {Mirrahimi}, \citenamefont {Devoret},\ and\ \citenamefont
  {Schoelkopf}}]{WangScience2016}%
  \BibitemOpen
  \bibinfo {author} {C.~Wang}, \bibinfo {author} {Y.~Y. Gao}, \bibinfo {author}
  {P.~Reinhold}, \bibinfo {author} {R.~W. Heeres}, \bibinfo {author} {N.~Ofek},
  \bibinfo {author} {K.~Chou}, \bibinfo {author} {C.~Axline}, \bibinfo {author}
  {M.~Reagor}, \bibinfo {author} {J.~Blumoff}, \bibinfo {author} {K.~M. Sliwa},
  \bibinfo {author} {L.~Frunzio}, \bibinfo {author} {S.~M. Girvin}, \bibinfo
  {author} {L.~Jiang}, \bibinfo {author} {M.~Mirrahimi}, \bibinfo {author}
  {M.~H. Devoret}\ and\ \bibinfo {author} {R.~J. Schoelkopf},\ \emph {\bibinfo
  {title} {A {Schrödinger} cat living in two boxes}},\ \href
  {\doibase10.1126/science.aaf2941} {\bibfield  {journal} {\bibinfo  {journal}
  {Science}\ }\textbf {\bibinfo {volume} {352}},\ \bibinfo {pages} {1087}
  (\bibinfo {year} {2016})}\BibitemShut {NoStop}%
\bibitem [{\citenamefont {Guillaud}\ and\ \citenamefont
  {Mirrahimi}(2019)}]{GuillaudPRX2019}%
  \BibitemOpen
  \bibinfo {author} {J.~Guillaud}\ and\ \bibinfo {author} {M.~Mirrahimi},\
  \emph {\bibinfo {title} {Repetition Cat Qubits for Fault-Tolerant Quantum
  Computation}},\ \href {\doibase10.1103/PhysRevX.9.041053} {\bibfield
  {journal} {\bibinfo  {journal} {Phys. Rev. X}\ }\textbf {\bibinfo {volume}
  {9}},\ \bibinfo {pages} {041053} (\bibinfo {year} {2019})}\BibitemShut
  {NoStop}%
\bibitem [{\citenamefont {Puri}\ \emph {et~al.}(2020)\citenamefont {Puri},
  \citenamefont {St-Jean}, \citenamefont {Gross}, \citenamefont {Grimm},
  \citenamefont {Frattini}, \citenamefont {Iyer}, \citenamefont {Krishna},
  \citenamefont {Touzard}, \citenamefont {Jiang}, \citenamefont {Blais},
  \citenamefont {Flammia},\ and\ \citenamefont
  {Girvin}}]{PuriScienceAdvances2020}%
  \BibitemOpen
  \bibinfo {author} {S.~Puri}, \bibinfo {author} {L.~St-Jean}, \bibinfo
  {author} {J.~A. Gross}, \bibinfo {author} {A.~Grimm}, \bibinfo {author}
  {N.~E. Frattini}, \bibinfo {author} {P.~S. Iyer}, \bibinfo {author}
  {A.~Krishna}, \bibinfo {author} {S.~Touzard}, \bibinfo {author} {L.~Jiang},
  \bibinfo {author} {A.~Blais}, \bibinfo {author} {S.~T. Flammia}\ and\
  \bibinfo {author} {S.~M. Girvin},\ \emph {\bibinfo {title} {Bias-preserving
  gates with stabilized cat qubits}},\ \href {\doibase10.1126/sciadv.aay5901}
  {\bibfield  {journal} {\bibinfo  {journal} {Science Advances}\ }\textbf
  {\bibinfo {volume} {6}},\ \bibinfo {pages} {eaay5901} (\bibinfo {year}
  {2020})}\BibitemShut {NoStop}%
\bibitem [{\citenamefont {Chamberland}\ \emph {et~al.}(2022)\citenamefont
  {Chamberland}, \citenamefont {Noh}, \citenamefont {Arrangoiz-Arriola},
  \citenamefont {Campbell}, \citenamefont {Hann}, \citenamefont {Iverson},
  \citenamefont {Putterman}, \citenamefont {Bohdanowicz}, \citenamefont
  {Flammia}, \citenamefont {Keller}, \citenamefont {Refael}, \citenamefont
  {Preskill}, \citenamefont {Jiang}, \citenamefont {Safavi-Naeini},
  \citenamefont {Painter},\ and\ \citenamefont {Brand\~ao}}]{Chamberland2022}%
  \BibitemOpen
  \bibinfo {author} {C.~Chamberland}, \bibinfo {author} {K.~Noh}, \bibinfo
  {author} {P.~Arrangoiz-Arriola}, \bibinfo {author} {E.~T. Campbell}, \bibinfo
  {author} {C.~T. Hann}, \bibinfo {author} {J.~Iverson}, \bibinfo {author}
  {H.~Putterman}, \bibinfo {author} {T.~C. Bohdanowicz}, \bibinfo {author}
  {S.~T. Flammia}, \bibinfo {author} {A.~Keller}, \bibinfo {author}
  {G.~Refael}, \bibinfo {author} {J.~Preskill}, \bibinfo {author} {L.~Jiang},
  \bibinfo {author} {A.~H. Safavi-Naeini}, \bibinfo {author} {O.~Painter}\ and\
  \bibinfo {author} {F.~G. Brand\~ao},\ \emph {\bibinfo {title} {Building a
  Fault-Tolerant Quantum Computer Using Concatenated Cat Codes}},\ \href
  {\doibase10.1103/PRXQuantum.3.010329} {\bibfield  {journal} {\bibinfo
  {journal} {PRX Quantum}\ }\textbf {\bibinfo {volume} {3}},\ \bibinfo {pages}
  {010329} (\bibinfo {year} {2022})}\BibitemShut {NoStop}%
\bibitem [{\citenamefont {Guillaud}\ and\ \citenamefont
  {Mirrahimi}(2021)}]{GuillaudPRA2021}%
  \BibitemOpen
  \bibinfo {author} {J.~Guillaud}\ and\ \bibinfo {author} {M.~Mirrahimi},\
  \emph {\bibinfo {title} {Error rates and resource overheads of repetition cat
  qubits}},\ \href {\doibase10.1103/PhysRevA.103.042413} {\bibfield  {journal}
  {\bibinfo  {journal} {Phys. Rev. A}\ }\textbf {\bibinfo {volume} {103}},\
  \bibinfo {pages} {042413} (\bibinfo {year} {2021})}\BibitemShut {NoStop}%
\bibitem [{\citenamefont {Li}\ \emph {et~al.}(2017)\citenamefont {Li},
  \citenamefont {Zou}, \citenamefont {Albert}, \citenamefont {Muralidharan},
  \citenamefont {Girvin},\ and\ \citenamefont {Jiang}}]{LiPRL2017}%
  \BibitemOpen
  \bibinfo {author} {L.~Li}, \bibinfo {author} {C.-L. Zou}, \bibinfo {author}
  {V.~V. Albert}, \bibinfo {author} {S.~Muralidharan}, \bibinfo {author} {S.~M.
  Girvin}\ and\ \bibinfo {author} {L.~Jiang},\ \emph {\bibinfo {title} {Cat
  Codes with Optimal Decoherence Suppression for a Lossy Bosonic Channel}},\
  \href {\doibase10.1103/PhysRevLett.119.030502} {\bibfield  {journal}
  {\bibinfo  {journal} {Phys. Rev. Lett.}\ }\textbf {\bibinfo {volume} {119}},\
  \bibinfo {pages} {030502} (\bibinfo {year} {2017})}\BibitemShut {NoStop}%
\bibitem [{\citenamefont {Gertler}\ \emph {et~al.}(2021)\citenamefont
  {Gertler}, \citenamefont {Baker}, \citenamefont {Li}, \citenamefont {Shirol},
  \citenamefont {Koch},\ and\ \citenamefont {Wang}}]{GertlerNature2021}%
  \BibitemOpen
  \bibinfo {author} {J.~M. Gertler}, \bibinfo {author} {B.~Baker}, \bibinfo
  {author} {J.~Li}, \bibinfo {author} {S.~Shirol}, \bibinfo {author} {J.~Koch}\
  and\ \bibinfo {author} {C.~Wang},\ \emph {\bibinfo {title} {Protecting a
  bosonic qubit with autonomous quantum error correction}},\ \href
  {\doibase10.1038/s41586-021-03257-0} {\bibfield  {journal} {\bibinfo
  {journal} {Nature}\ }\textbf {\bibinfo {volume} {590}},\ \bibinfo {pages}
  {243} (\bibinfo {year} {2021})}\BibitemShut {NoStop}%
\bibitem [{\citenamefont {Albert}\ \emph {et~al.}(2019)\citenamefont {Albert},
  \citenamefont {Mundhada}, \citenamefont {Grimm}, \citenamefont {Touzard},
  \citenamefont {Devoret},\ and\ \citenamefont {Jiang}}]{AlbertIOP2019}%
  \BibitemOpen
  \bibinfo {author} {V.~V. Albert}, \bibinfo {author} {S.~O. Mundhada},
  \bibinfo {author} {A.~Grimm}, \bibinfo {author} {S.~Touzard}, \bibinfo
  {author} {M.~H. Devoret}\ and\ \bibinfo {author} {L.~Jiang},\ \emph {\bibinfo
  {title} {Pair-cat codes: Autonomous error-correction with low-order
  nonlinearity}},\ \href {\doibase10.1088/2058-9565/ab1e69} {\bibfield
  {journal} {\bibinfo  {journal} {Quantum Science and Technology}\ }\textbf
  {\bibinfo {volume} {4}},\ \bibinfo {pages} {35007} (\bibinfo {year}
  {2019})},\ \Eprint {http://arxiv.org/abs/1801.05897} {1801.05897}
  \BibitemShut {NoStop}%
\bibitem [{\citenamefont {Michael}\ \emph {et~al.}(2016)\citenamefont
  {Michael}, \citenamefont {Silveri}, \citenamefont {Brierley}, \citenamefont
  {Albert}, \citenamefont {Salmilehto}, \citenamefont {Jiang},\ and\
  \citenamefont {Girvin}}]{MariosPRX16}%
  \BibitemOpen
  \bibinfo {author} {M.~H. Michael}, \bibinfo {author} {M.~Silveri}, \bibinfo
  {author} {R.~T. Brierley}, \bibinfo {author} {V.~V. Albert}, \bibinfo
  {author} {J.~Salmilehto}, \bibinfo {author} {L.~Jiang}\ and\ \bibinfo
  {author} {S.~M. Girvin},\ \emph {\bibinfo {title} {New Class of Quantum
  Error-Correcting Codes for a Bosonic Mode}},\ \href
  {\doibase10.1103/PhysRevX.6.031006} {\bibfield  {journal} {\bibinfo
  {journal} {Phys. Rev. X}\ }\textbf {\bibinfo {volume} {6}},\ \bibinfo {pages}
  {031006} (\bibinfo {year} {2016})}\BibitemShut {NoStop}%
\bibitem [{\citenamefont {Hu}\ \emph {et~al.}(2019)\citenamefont {Hu},
  \citenamefont {Ma}, \citenamefont {Cai}, \citenamefont {Mu}, \citenamefont
  {Xu}, \citenamefont {Wang}, \citenamefont {Wu}, \citenamefont {Wang},
  \citenamefont {Song}, \citenamefont {Zou}, \citenamefont {Girvin},
  \citenamefont {Duan},\ and\ \citenamefont {Sun}}]{HuNaturePhys2019}%
  \BibitemOpen
  \bibinfo {author} {L.~Hu}, \bibinfo {author} {Y.~Ma}, \bibinfo {author}
  {W.~Cai}, \bibinfo {author} {X.~Mu}, \bibinfo {author} {Y.~Xu}, \bibinfo
  {author} {W.~Wang}, \bibinfo {author} {Y.~Wu}, \bibinfo {author} {H.~Wang},
  \bibinfo {author} {Y.~P. Song}, \bibinfo {author} {C.-L. Zou}, \bibinfo
  {author} {S.~M. Girvin}, \bibinfo {author} {L.-M. Duan}\ and\ \bibinfo
  {author} {L.~Sun},\ \emph {\bibinfo {title} {Quantum error correction and
  universal gate set operation on a binomial bosonic logical qubit}},\ \href
  {\doibase10.1038/s41567-018-0414-3} {\bibfield  {journal} {\bibinfo
  {journal} {Nature Physics}\ }\textbf {\bibinfo {volume} {15}},\ \bibinfo
  {pages} {503} (\bibinfo {year} {2019})}\BibitemShut {NoStop}%
\bibitem [{\citenamefont {Grimsmo}\ \emph {et~al.}(2020)\citenamefont
  {Grimsmo}, \citenamefont {Combes},\ and\ \citenamefont
  {Baragiola}}]{GrimsmoPRX2020}%
  \BibitemOpen
  \bibinfo {author} {A.~L. Grimsmo}, \bibinfo {author} {J.~Combes}\ and\
  \bibinfo {author} {B.~Q. Baragiola},\ \emph {\bibinfo {title} {Quantum
  Computing with Rotation-Symmetric Bosonic Codes}},\ \href
  {\doibase10.1103/PhysRevX.10.011058} {\bibfield  {journal} {\bibinfo
  {journal} {Phys. Rev. X}\ }\textbf {\bibinfo {volume} {10}},\ \bibinfo
  {pages} {011058} (\bibinfo {year} {2020})}\BibitemShut {NoStop}%
\bibitem [{\citenamefont {Li}\ \emph {et~al.}(2021)\citenamefont {Li},
  \citenamefont {Young}, \citenamefont {Albert}, \citenamefont {Noh},
  \citenamefont {Zou},\ and\ \citenamefont {Jiang}}]{LiPRA2021}%
  \BibitemOpen
  \bibinfo {author} {L.~Li}, \bibinfo {author} {D.~J. Young}, \bibinfo {author}
  {V.~V. Albert}, \bibinfo {author} {K.~Noh}, \bibinfo {author} {C.-L. Zou}\
  and\ \bibinfo {author} {L.~Jiang},\ \emph {\bibinfo {title} {Phase-engineered
  bosonic quantum codes}},\ \href {\doibase10.1103/PhysRevA.103.062427}
  {\bibfield  {journal} {\bibinfo  {journal} {Phys. Rev. A}\ }\textbf {\bibinfo
  {volume} {103}},\ \bibinfo {pages} {062427} (\bibinfo {year}
  {2021})}\BibitemShut {NoStop}%
\bibitem [{\citenamefont {Wang}\ \emph {et~al.}(2022)\citenamefont {Wang},
  \citenamefont {Rajabzadeh}, \citenamefont {Lee},\ and\ \citenamefont
  {Safavi-Naeini}}]{Wang2021}%
  \BibitemOpen
  \bibinfo {author} {Z.~Wang}, \bibinfo {author} {T.~Rajabzadeh}, \bibinfo
  {author} {N.~Lee}\ and\ \bibinfo {author} {A.~H. Safavi-Naeini},\ \emph
  {\bibinfo {title} {Automated Discovery of Autonomous Quantum Error Correction
  Schemes}},\ \href {\doibase10.1103/PRXQuantum.3.020302} {\bibfield  {journal}
  {\bibinfo  {journal} {PRX Quantum}\ }\textbf {\bibinfo {volume} {3}},\
  \bibinfo {pages} {020302} (\bibinfo {year} {2022})}\BibitemShut {NoStop}%
\bibitem [{\citenamefont {Shukla}\ \emph {et~al.}(2021)\citenamefont {Shukla},
  \citenamefont {Nimmrichter},\ and\ \citenamefont {Sanders}}]{ShuklaPRA2021}%
  \BibitemOpen
  \bibinfo {author} {N.~Shukla}, \bibinfo {author} {S.~Nimmrichter}\ and\
  \bibinfo {author} {B.~C. Sanders},\ \emph {\bibinfo {title} {Squeezed comb
  states}},\ \href {\doibase10.1103/PhysRevA.103.012408} {\bibfield  {journal}
  {\bibinfo  {journal} {Phys. Rev. A}\ }\textbf {\bibinfo {volume} {103}},\
  \bibinfo {pages} {012408} (\bibinfo {year} {2021})}\BibitemShut {NoStop}%
\bibitem [{\citenamefont {Nicacio}\ \emph {et~al.}(2010)\citenamefont
  {Nicacio}, \citenamefont {Maia}, \citenamefont {Toscano},\ and\ \citenamefont
  {Vallejos}}]{NicacioPRA2010}%
  \BibitemOpen
  \bibinfo {author} {F.~Nicacio}, \bibinfo {author} {R.~N.~P. Maia}, \bibinfo
  {author} {F.~Toscano}\ and\ \bibinfo {author} {R.~O. Vallejos},\ \emph
  {\bibinfo {title} {Phase space structure of generalized {Gaussian} cat
  states}},\ \href {\doibase10.1016/j.physleta.2010.08.076} {\bibfield
  {journal} {\bibinfo  {journal} {Physics Letters A}\ }\textbf {\bibinfo
  {volume} {374}},\ \bibinfo {pages} {4385} (\bibinfo {year}
  {2010})}\BibitemShut {NoStop}%
\bibitem [{\citenamefont {Chao}\ \emph {et~al.}(2019)\citenamefont {Chao},
  \citenamefont {Xiong},\ and\ \citenamefont
  {Zhou}}]{ChaoAnnalenderPhysik2019}%
  \BibitemOpen
  \bibinfo {author} {S.-L. Chao}, \bibinfo {author} {B.~Xiong}\ and\ \bibinfo
  {author} {L.~Zhou},\ \emph {\bibinfo {title} {Generating a
  {Squeezed}-{Coherent}-{Cat} {State} in a {Double}-{Cavity} {Optomechanical}
  {System}}},\ \href {\doibase10.1002/andp.201900196} {\bibfield  {journal}
  {\bibinfo  {journal} {Annalen der Physik}\ }\textbf {\bibinfo {volume}
  {531}},\ \bibinfo {pages} {1900196} (\bibinfo {year} {2019})}\BibitemShut
  {NoStop}%
\bibitem [{\citenamefont {Teh}\ \emph {et~al.}(2020)\citenamefont {Teh},
  \citenamefont {Drummond},\ and\ \citenamefont {Reid}}]{TehPRR2020}%
  \BibitemOpen
  \bibinfo {author} {R.~Y. Teh}, \bibinfo {author} {P.~D. Drummond}\ and\
  \bibinfo {author} {M.~D. Reid},\ \emph {\bibinfo {title} {Overcoming
  decoherence of Schr\"odinger cat states formed in a cavity using
  squeezed-state inputs}},\ \href {\doibase10.1103/PhysRevResearch.2.043387}
  {\bibfield  {journal} {\bibinfo  {journal} {Phys. Rev. Research}\ }\textbf
  {\bibinfo {volume} {2}},\ \bibinfo {pages} {043387} (\bibinfo {year}
  {2020})}\BibitemShut {NoStop}%
\bibitem [{\citenamefont {Liu}\ \emph {et~al.}(2005)\citenamefont {Liu},
  \citenamefont {Wei},\ and\ \citenamefont {Nori}}]{LiuPRA2005}%
  \BibitemOpen
  \bibinfo {author} {Y.-x. Liu}, \bibinfo {author} {L.~F. Wei}\ and\ \bibinfo
  {author} {F.~Nori},\ \emph {\bibinfo {title} {Preparation of macroscopic
  quantum superposition states of a cavity field via coupling to a
  superconducting charge qubit}},\ \href {\doibase10.1103/PhysRevA.71.063820}
  {\bibfield  {journal} {\bibinfo  {journal} {Phys. Rev. A}\ }\textbf {\bibinfo
  {volume} {71}},\ \bibinfo {pages} {063820} (\bibinfo {year}
  {2005})}\BibitemShut {NoStop}%
\bibitem [{\citenamefont {Ourjoumtsev}\ \emph {et~al.}(2007)\citenamefont
  {Ourjoumtsev}, \citenamefont {Jeong}, \citenamefont {Tualle-Brouri},\ and\
  \citenamefont {Grangier}}]{Ourjoumtsev2007}%
  \BibitemOpen
  \bibinfo {author} {A.~Ourjoumtsev}, \bibinfo {author} {H.~Jeong}, \bibinfo
  {author} {R.~Tualle-Brouri}\ and\ \bibinfo {author} {P.~Grangier},\ \emph
  {\bibinfo {title} {Generation of optical ‘Schrödinger cats’ from photon
  number states}},\ \href {\doibase10.1038/nature06054} {\bibfield  {journal}
  {\bibinfo  {journal} {Nature 2007 448:7155}\ }\textbf {\bibinfo {volume}
  {448}},\ \bibinfo {pages} {784} (\bibinfo {year} {2007})}\BibitemShut
  {NoStop}%
\bibitem [{\citenamefont {Lo}\ \emph {et~al.}(2015)\citenamefont {Lo},
  \citenamefont {Kienzler}, \citenamefont {Clercq}, \citenamefont {Marinelli},
  \citenamefont {Negnevitsky}, \citenamefont {Keitch},\ and\ \citenamefont
  {Home}}]{Lo2015}%
  \BibitemOpen
  \bibinfo {author} {H.~Y. Lo}, \bibinfo {author} {D.~Kienzler}, \bibinfo
  {author} {L.~D. Clercq}, \bibinfo {author} {M.~Marinelli}, \bibinfo {author}
  {V.~Negnevitsky}, \bibinfo {author} {B.~C. Keitch}\ and\ \bibinfo {author}
  {J.~P. Home},\ \emph {\bibinfo {title} {Spin–motion entanglement and state
  diagnosis with squeezed oscillator wavepackets}},\ \href
  {\doibase10.1038/nature14458} {\bibfield  {journal} {\bibinfo  {journal}
  {Nature 2015 521:7552}\ }\textbf {\bibinfo {volume} {521}},\ \bibinfo {pages}
  {336} (\bibinfo {year} {2015})}\BibitemShut {NoStop}%
\bibitem [{\citenamefont {Le~Jeannic}\ \emph {et~al.}(2018)\citenamefont
  {Le~Jeannic}, \citenamefont {Cavaill\`es}, \citenamefont {Huang},
  \citenamefont {Filip},\ and\ \citenamefont {Laurat}}]{JeannicPRL2018}%
  \BibitemOpen
  \bibinfo {author} {H.~Le~Jeannic}, \bibinfo {author} {A.~Cavaill\`es},
  \bibinfo {author} {K.~Huang}, \bibinfo {author} {R.~Filip}\ and\ \bibinfo
  {author} {J.~Laurat},\ \emph {\bibinfo {title} {Slowing Quantum Decoherence
  by Squeezing in Phase Space}},\ \href
  {\doibase10.1103/PhysRevLett.120.073603} {\bibfield  {journal} {\bibinfo
  {journal} {Phys. Rev. Lett.}\ }\textbf {\bibinfo {volume} {120}},\ \bibinfo
  {pages} {073603} (\bibinfo {year} {2018})}\BibitemShut {NoStop}%
\bibitem [{\citenamefont {Gorini}\ \emph {et~al.}(1976)\citenamefont {Gorini},
  \citenamefont {Kossakowski},\ and\ \citenamefont {Sudarshan}}]{GoriniJMP76}%
  \BibitemOpen
  \bibinfo {author} {V.~Gorini}, \bibinfo {author} {A.~Kossakowski}\ and\
  \bibinfo {author} {E.~C.~G. Sudarshan},\ \emph {\bibinfo {title} {Completely
  positive dynamical semigroups of N-level systems}},\ \href
  {https://aip.scitation.org/doi/abs/10.1063/1.522979} {\bibfield  {journal}
  {\bibinfo  {journal} {Journal of Mathematical Physics}\ }\textbf {\bibinfo
  {volume} {17}},\ \bibinfo {pages} {821} (\bibinfo {year} {1976})}\BibitemShut
  {NoStop}%
\bibitem [{\citenamefont {Lindblad}(1976)}]{LindbladCMP76}%
  \BibitemOpen
  \bibinfo {author} {G.~Lindblad},\ \emph {\bibinfo {title} {On the generators
  of quantum dynamical semigroups}},\ \href
  {https://doi.org/10.1007/BF01608499} {\bibfield  {journal} {\bibinfo
  {journal} {Communications in Mathematical Physics}\ }\textbf {\bibinfo
  {volume} {48}},\ \bibinfo {pages} {119} (\bibinfo {year} {1976})}\BibitemShut
  {NoStop}%
\bibitem [{\citenamefont {Breuer}\ and\ \citenamefont
  {Petruccione}(2007)}]{BreuerBookOpen}%
  \BibitemOpen
  \bibinfo {author} {H.~Breuer}\ and\ \bibinfo {author} {F.~Petruccione},\
  \href@noop {} {\emph {\bibinfo {title} {The Theory of Open Quantum
  Systems}}}\ (\bibinfo  {publisher} {Oxford University Press},\ \bibinfo
  {address} {Oxford},\ \bibinfo {year} {2007})\BibitemShut {NoStop}%
\bibitem [{\citenamefont {Dassonneville}\ \emph {et~al.}(2021)\citenamefont
  {Dassonneville}, \citenamefont {Assouly}, \citenamefont {Peronnin},
  \citenamefont {Clerk}, \citenamefont {Bienfait},\ and\ \citenamefont
  {Huard}}]{DassonnevillePRX2021}%
  \BibitemOpen
  \bibinfo {author} {R.~Dassonneville}, \bibinfo {author} {R.~Assouly},
  \bibinfo {author} {T.~Peronnin}, \bibinfo {author} {A.~Clerk}, \bibinfo
  {author} {A.~Bienfait}\ and\ \bibinfo {author} {B.~Huard},\ \emph {\bibinfo
  {title} {Dissipative Stabilization of Squeezing Beyond 3 dB in a Microwave
  Mode}},\ \href {\doibase10.1103/PRXQuantum.2.020323} {\bibfield  {journal}
  {\bibinfo  {journal} {PRX Quantum}\ }\textbf {\bibinfo {volume} {2}},\
  \bibinfo {pages} {020323} (\bibinfo {year} {2021})}\BibitemShut {NoStop}%
\bibitem [{\citenamefont {Eickbusch}\ \emph {et~al.}(2021)\citenamefont
  {Eickbusch}, \citenamefont {Sivak}, \citenamefont {Ding}, \citenamefont
  {Elder}, \citenamefont {Jha}, \citenamefont {Venkatraman}, \citenamefont
  {Royer}, \citenamefont {Girvin}, \citenamefont {Schoelkopf},\ and\
  \citenamefont {Devoret}}]{Eickbusch2021}%
  \BibitemOpen
  \bibinfo {author} {A.~Eickbusch}, \bibinfo {author} {V.~Sivak}, \bibinfo
  {author} {A.~Z. Ding}, \bibinfo {author} {S.~S. Elder}, \bibinfo {author}
  {S.~R. Jha}, \bibinfo {author} {J.~Venkatraman}, \bibinfo {author}
  {B.~Royer}, \bibinfo {author} {S.~M. Girvin}, \bibinfo {author} {R.~J.
  Schoelkopf}\ and\ \bibinfo {author} {M.~H. Devoret},\ \href@noop {} {\emph
  {\bibinfo {title} {Fast Universal Control of an Oscillator with Weak
  Dispersive Coupling to a Qubit}}} (\bibinfo {year} {2021}),\ \Eprint
  {http://arxiv.org/abs/2111.06414} {arXiv:2111.06414 [quant-ph]} \BibitemShut
  {NoStop}%
\bibitem [{\citenamefont {Han}\ \emph {et~al.}(2021)\citenamefont {Han},
  \citenamefont {Wang},\ and\ \citenamefont {Zhang}}]{Han:21}%
  \BibitemOpen
  \bibinfo {author} {K.~Han}, \bibinfo {author} {Y.~Wang}\ and\ \bibinfo
  {author} {G.-Q. Zhang},\ \emph {\bibinfo {title} {Enhancement of microwave
  squeezing via parametric down-conversion in a superconducting quantum
  circuit}},\ \href {\doibase10.1364/OE.423373} {\bibfield  {journal} {\bibinfo
   {journal} {Opt. Express}\ }\textbf {\bibinfo {volume} {29}},\ \bibinfo
  {pages} {13451} (\bibinfo {year} {2021})}\BibitemShut {NoStop}%
\bibitem [{\citenamefont {Xie}\ \emph {et~al.}(2020)\citenamefont {Xie},
  \citenamefont {Ma}, \citenamefont {Ren}, \citenamefont {Li},\ and\
  \citenamefont {Li}}]{XiePRA2020}%
  \BibitemOpen
  \bibinfo {author} {J.-k. Xie}, \bibinfo {author} {S.-l. Ma}, \bibinfo
  {author} {Y.-l. Ren}, \bibinfo {author} {X.-k. Li}\ and\ \bibinfo {author}
  {F.-l. Li},\ \emph {\bibinfo {title} {Dissipative generation of steady-state
  squeezing of superconducting resonators via parametric driving}},\ \href
  {\doibase10.1103/PhysRevA.101.012348} {\bibfield  {journal} {\bibinfo
  {journal} {Phys. Rev. A}\ }\textbf {\bibinfo {volume} {101}},\ \bibinfo
  {pages} {012348} (\bibinfo {year} {2020})}\BibitemShut {NoStop}%
\bibitem [{\citenamefont {Kraus}(1971)}]{KrausAnnalsofPhysics1971}%
  \BibitemOpen
  \bibinfo {author} {K.~Kraus},\ \emph {\bibinfo {title} {General state changes
  in quantum theory}},\ \href {\doibase10.1016/0003-4916(71)90108-4} {\bibfield
   {journal} {\bibinfo  {journal} {Annals of Physics}\ }\textbf {\bibinfo
  {volume} {64}},\ \bibinfo {pages} {311} (\bibinfo {year} {1971})}\BibitemShut
  {NoStop}%
\bibitem [{\citenamefont {Bennett}\ \emph {et~al.}(1996)\citenamefont
  {Bennett}, \citenamefont {DiVincenzo}, \citenamefont {Smolin},\ and\
  \citenamefont {Wootters}}]{BennetPRA1996}%
  \BibitemOpen
  \bibinfo {author} {C.~H. Bennett}, \bibinfo {author} {D.~P. DiVincenzo},
  \bibinfo {author} {J.~A. Smolin}\ and\ \bibinfo {author} {W.~K. Wootters},\
  \emph {\bibinfo {title} {Mixed-state entanglement and quantum error
  correction}},\ \href {\doibase10.1103/PhysRevA.54.3824} {\bibfield  {journal}
  {\bibinfo  {journal} {Phys. Rev. A}\ }\textbf {\bibinfo {volume} {54}},\
  \bibinfo {pages} {3824} (\bibinfo {year} {1996})}\BibitemShut {NoStop}%
\bibitem [{\citenamefont {Knill}\ and\ \citenamefont
  {Laflamme}(1997)}]{KnillPRA97}%
  \BibitemOpen
  \bibinfo {author} {E.~Knill}\ and\ \bibinfo {author} {R.~Laflamme},\ \emph
  {\bibinfo {title} {Theory of quantum error-correcting codes}},\ \href
  {\doibase10.1103/PhysRevA.55.900} {\bibfield  {journal} {\bibinfo  {journal}
  {Phys. Rev. A}\ }\textbf {\bibinfo {volume} {55}},\ \bibinfo {pages} {900}
  (\bibinfo {year} {1997})}\BibitemShut {NoStop}%
\bibitem [{\citenamefont {Schumacher}\ and\ \citenamefont
  {Westmoreland}(2002)}]{SchumacherQuInfoProcessing2002}%
  \BibitemOpen
  \bibinfo {author} {B.~Schumacher}\ and\ \bibinfo {author} {M.~D.
  Westmoreland},\ \emph {\bibinfo {title} {Approximate {Quantum} {Error}
  {Correction}}},\ \href {\doibase10.1023/A:1019653202562} {\bibfield
  {journal} {\bibinfo  {journal} {Quantum Information Processing}\ }\textbf
  {\bibinfo {volume} {1}},\ \bibinfo {pages} {5} (\bibinfo {year}
  {2002})}\BibitemShut {NoStop}%
\bibitem [{\citenamefont {Ng}\ and\ \citenamefont
  {Mandayam}(2010)}]{NgPRA2010}%
  \BibitemOpen
  \bibinfo {author} {H.~K. Ng}\ and\ \bibinfo {author} {P.~Mandayam},\ \emph
  {\bibinfo {title} {Simple approach to approximate quantum error correction
  based on the transpose channel}},\ \href {\doibase10.1103/PhysRevA.81.062342}
  {\bibfield  {journal} {\bibinfo  {journal} {Phys. Rev. A}\ }\textbf {\bibinfo
  {volume} {81}},\ \bibinfo {pages} {062342} (\bibinfo {year}
  {2010})}\BibitemShut {NoStop}%
\bibitem [{\citenamefont {B\'eny}\ and\ \citenamefont
  {Oreshkov}(2010)}]{BenyPRL2010}%
  \BibitemOpen
  \bibinfo {author} {C.~B\'eny}\ and\ \bibinfo {author} {O.~Oreshkov},\ \emph
  {\bibinfo {title} {General Conditions for Approximate Quantum Error
  Correction and Near-Optimal Recovery Channels}},\ \href
  {\doibase10.1103/PhysRevLett.104.120501} {\bibfield  {journal} {\bibinfo
  {journal} {Phys. Rev. Lett.}\ }\textbf {\bibinfo {volume} {104}},\ \bibinfo
  {pages} {120501} (\bibinfo {year} {2010})}\BibitemShut {NoStop}%
\bibitem [{\citenamefont {Mandayam}\ and\ \citenamefont
  {Ng}(2012)}]{MandayamPRA2012}%
  \BibitemOpen
  \bibinfo {author} {P.~Mandayam}\ and\ \bibinfo {author} {H.~K. Ng},\ \emph
  {\bibinfo {title} {Towards a unified framework for approximate quantum error
  correction}},\ \href {\doibase10.1103/PhysRevA.86.012335} {\bibfield
  {journal} {\bibinfo  {journal} {Phys. Rev. A}\ }\textbf {\bibinfo {volume}
  {86}},\ \bibinfo {pages} {012335} (\bibinfo {year} {2012})}\BibitemShut
  {NoStop}%
\bibitem [{\citenamefont {Schumacher}(1996)}]{SchuhmacherPRA1996}%
  \BibitemOpen
  \bibinfo {author} {B.~Schumacher},\ \emph {\bibinfo {title} {Sending
  entanglement through noisy quantum channels}},\ \href
  {\doibase10.1103/PhysRevA.54.2614} {\bibfield  {journal} {\bibinfo  {journal}
  {Phys. Rev. A}\ }\textbf {\bibinfo {volume} {54}},\ \bibinfo {pages} {2614}
  (\bibinfo {year} {1996})}\BibitemShut {NoStop}%
\bibitem [{Note2()}]{Note2}%
  \BibitemOpen
  \bibinfo {note} {As discussed in~\cite {AlbertPRA2018}, the average channel
  fidelity of a single logical qubit, with codewords $\ket *{0}_{\protect \rm
  L}$ and $\ket *{1}_{\protect \rm L}$, can be equivalently expressed in terms
  of the Pauli operators in the code space $\{\protect \hat {\protect \mathds
  {1}}, \protect \hat {\sigma }_x, \protect \hat {\sigma }_y, \protect \hat
  {\sigma }_z \}$ as \begin {equation} \protect \mathcal {F} = \protect \frac
  {1}{4}\DOTSB \sum@ \slimits@ _i{\protect \mathcal {E}_{ii}}\protect \,, \end
  {equation} where $\protect \mathcal {E}_{ij} = \protect \genfrac
  {}{}{}1{1}{2}\Tr {\sigma _i \protect \mathcal {E}(\sigma _j)}$}\BibitemShut
  {NoStop}%
\bibitem [{\citenamefont {Yamamoto}\ \emph {et~al.}(2005)\citenamefont
  {Yamamoto}, \citenamefont {Hara},\ and\ \citenamefont
  {Tsumura}}]{YamamotoPRA2005}%
  \BibitemOpen
  \bibinfo {author} {N.~Yamamoto}, \bibinfo {author} {S.~Hara}\ and\ \bibinfo
  {author} {K.~Tsumura},\ \emph {\bibinfo {title} {Suboptimal
  quantum-error-correcting procedure based on semidefinite programming}},\
  \href {\doibase10.1103/PhysRevA.71.022322} {\bibfield  {journal} {\bibinfo
  {journal} {Phys. Rev. A}\ }\textbf {\bibinfo {volume} {71}},\ \bibinfo
  {pages} {022322} (\bibinfo {year} {2005})}\BibitemShut {NoStop}%
\bibitem [{\citenamefont {Fletcher}\ \emph {et~al.}(2007)\citenamefont
  {Fletcher}, \citenamefont {Shor},\ and\ \citenamefont
  {Win}}]{FletcherPRA2007}%
  \BibitemOpen
  \bibinfo {author} {A.~S. Fletcher}, \bibinfo {author} {P.~W. Shor}\ and\
  \bibinfo {author} {M.~Z. Win},\ \emph {\bibinfo {title} {Optimum quantum
  error recovery using semidefinite programming}},\ \href
  {\doibase10.1103/PhysRevA.75.012338} {\bibfield  {journal} {\bibinfo
  {journal} {Phys. Rev. A}\ }\textbf {\bibinfo {volume} {75}},\ \bibinfo
  {pages} {012338} (\bibinfo {year} {2007})}\BibitemShut {NoStop}%
\bibitem [{\citenamefont {Kosut}\ and\ \citenamefont
  {Lidar}(2009)}]{KosutQuantum2009}%
  \BibitemOpen
  \bibinfo {author} {R.~L. Kosut}\ and\ \bibinfo {author} {D.~A. Lidar},\ \emph
  {\bibinfo {title} {Quantum error correction via convex optimization}},\ \href
  {\doibase10.1007/s11128-009-0120-2} {\bibfield  {journal} {\bibinfo
  {journal} {Quantum Information Processing}\ }\textbf {\bibinfo {volume}
  {8}},\ \bibinfo {pages} {443} (\bibinfo {year} {2009})}\BibitemShut {NoStop}%
\bibitem [{\citenamefont {Haroche}\ and\ \citenamefont
  {Raimond}(2006)}]{Haroche_BOOK_Quantum}%
  \BibitemOpen
  \bibinfo {author} {S.~Haroche}\ and\ \bibinfo {author} {J.~M. Raimond},\
  \href@noop {} {\emph {\bibinfo {title} {Exploring the Quantum: Atoms,
  Cavities, and Photons}}}\ (\bibinfo  {publisher} {Oxford University Press},\
  \bibinfo {address} {Oxford},\ \bibinfo {year} {2006})\BibitemShut {NoStop}%
\bibitem [{\citenamefont {Minganti}\ \emph {et~al.}(2016)\citenamefont
  {Minganti}, \citenamefont {Bartolo}, \citenamefont {Lolli}, \citenamefont
  {Casteels},\ and\ \citenamefont {Ciuti}}]{MingantiSciRep16}%
  \BibitemOpen
  \bibinfo {author} {F.~Minganti}, \bibinfo {author} {N.~Bartolo}, \bibinfo
  {author} {J.~Lolli}, \bibinfo {author} {W.~Casteels}\ and\ \bibinfo {author}
  {C.~Ciuti},\ \emph {\bibinfo {title} {Exact results for Schr{\"o}dinger cats
  in driven-dissipative systems and their feedback control}},\ \href
  {http://dx.doi.org/10.1038/srep26987} {\bibfield  {journal} {\bibinfo
  {journal} {Scientific Reports}\ }\textbf {\bibinfo {volume} {6}},\ \bibinfo
  {pages} {26987} (\bibinfo {year} {2016})}\BibitemShut {NoStop}%
\bibitem [{\citenamefont {Bartolo}\ \emph {et~al.}(2016)\citenamefont
  {Bartolo}, \citenamefont {Minganti}, \citenamefont {Casteels},\ and\
  \citenamefont {Ciuti}}]{BartoloPRA16}%
  \BibitemOpen
  \bibinfo {author} {N.~Bartolo}, \bibinfo {author} {F.~Minganti}, \bibinfo
  {author} {W.~Casteels}\ and\ \bibinfo {author} {C.~Ciuti},\ \emph {\bibinfo
  {title} {Exact steady state of a Kerr resonator with one- and two-photon
  driving and dissipation: Controllable Wigner-function multimodality and
  dissipative phase transitions}},\ \href
  {https://link.aps.org/doi/10.1103/PhysRevA.94.033841} {\bibfield  {journal}
  {\bibinfo  {journal} {Phys. Rev. A}\ }\textbf {\bibinfo {volume} {94}},\
  \bibinfo {pages} {033841} (\bibinfo {year} {2016})}\BibitemShut {NoStop}%
\bibitem [{\citenamefont {Yurke}(1987)}]{Yurke:87}%
  \BibitemOpen
  \bibinfo {author} {B.~Yurke},\ \emph {\bibinfo {title} {Squeezed-state
  generation using a Josephson parametric amplifier}},\ \href
  {\doibase10.1364/JOSAB.4.001551} {\bibfield  {journal} {\bibinfo  {journal}
  {J. Opt. Soc. Am. B}\ }\textbf {\bibinfo {volume} {4}},\ \bibinfo {pages}
  {1551} (\bibinfo {year} {1987})}\BibitemShut {NoStop}%
\bibitem [{\citenamefont {Göppl}\ \emph {et~al.}(2008)\citenamefont {Göppl},
  \citenamefont {Fragner}, \citenamefont {Baur}, \citenamefont {Bianchetti},
  \citenamefont {Filipp}, \citenamefont {Fink}, \citenamefont {Leek},
  \citenamefont {Puebla}, \citenamefont {Steffen},\ and\ \citenamefont
  {Wallraff}}]{GopplJAP2008}%
  \BibitemOpen
  \bibinfo {author} {M.~Göppl}, \bibinfo {author} {A.~Fragner}, \bibinfo
  {author} {M.~Baur}, \bibinfo {author} {R.~Bianchetti}, \bibinfo {author}
  {S.~Filipp}, \bibinfo {author} {J.~M. Fink}, \bibinfo {author} {P.~J. Leek},
  \bibinfo {author} {G.~Puebla}, \bibinfo {author} {L.~Steffen}\ and\ \bibinfo
  {author} {A.~Wallraff},\ \emph {\bibinfo {title} {Coplanar waveguide
  resonators for circuit quantum electrodynamics}},\ \href
  {\doibase10.1063/1.3010859} {\bibfield  {journal} {\bibinfo  {journal}
  {Journal of Applied Physics}\ }\textbf {\bibinfo {volume} {104}},\ \bibinfo
  {pages} {113904} (\bibinfo {year} {2008})}\BibitemShut {NoStop}%
\bibitem [{\citenamefont {Reagor}\ \emph {et~al.}(2013)\citenamefont {Reagor},
  \citenamefont {Paik}, \citenamefont {Catelani}, \citenamefont {Sun},
  \citenamefont {Axline}, \citenamefont {Holland}, \citenamefont {Pop},
  \citenamefont {Masluk}, \citenamefont {Brecht}, \citenamefont {Frunzio},
  \citenamefont {Devoret}, \citenamefont {Glazman},\ and\ \citenamefont
  {Schoelkopf}}]{ReagorAPL2013}%
  \BibitemOpen
  \bibinfo {author} {M.~Reagor}, \bibinfo {author} {H.~Paik}, \bibinfo {author}
  {G.~Catelani}, \bibinfo {author} {L.~Sun}, \bibinfo {author} {C.~Axline},
  \bibinfo {author} {E.~Holland}, \bibinfo {author} {I.~M. Pop}, \bibinfo
  {author} {N.~A. Masluk}, \bibinfo {author} {T.~Brecht}, \bibinfo {author}
  {L.~Frunzio}, \bibinfo {author} {M.~H. Devoret}, \bibinfo {author}
  {L.~Glazman}\ and\ \bibinfo {author} {R.~J. Schoelkopf},\ \emph {\bibinfo
  {title} {Reaching 10 ms single photon lifetimes for superconducting
  aluminum cavities}},\ \href {\doibase10.1063/1.4807015} {\bibfield  {journal}
  {\bibinfo  {journal} {Applied Physics Letters}\ }\textbf {\bibinfo {volume}
  {102}},\ \bibinfo {pages} {192604} (\bibinfo {year} {2013})}\BibitemShut
  {NoStop}%
\bibitem [{\citenamefont {Myatt}\ \emph {et~al.}(2000)\citenamefont {Myatt},
  \citenamefont {King}, \citenamefont {Turchette}, \citenamefont {Sackett},
  \citenamefont {Kielpinski}, \citenamefont {Itano}, \citenamefont {Monroe},\
  and\ \citenamefont {Wineland}}]{MyattNature2000}%
  \BibitemOpen
  \bibinfo {author} {C.~J. Myatt}, \bibinfo {author} {B.~E. King}, \bibinfo
  {author} {Q.~A. Turchette}, \bibinfo {author} {C.~A. Sackett}, \bibinfo
  {author} {D.~Kielpinski}, \bibinfo {author} {W.~M. Itano}, \bibinfo {author}
  {C.~Monroe}\ and\ \bibinfo {author} {D.~J. Wineland},\ \emph {\bibinfo
  {title} {Decoherence of quantum superpositions through coupling to engineered
  reservoirs}},\ \href {\doibase10.1038/35002001} {\bibfield  {journal}
  {\bibinfo  {journal} {Nature}\ }\textbf {\bibinfo {volume} {403}},\ \bibinfo
  {pages} {269} (\bibinfo {year} {2000})}\BibitemShut {NoStop}%
\bibitem [{\citenamefont {Ourjoumtsev}\ \emph {et~al.}(2006)\citenamefont
  {Ourjoumtsev}, \citenamefont {Tualle-Brouri}, \citenamefont {Laurat},\ and\
  \citenamefont {Grangier}}]{OurjoumtsevScience06}%
  \BibitemOpen
  \bibinfo {author} {A.~Ourjoumtsev}, \bibinfo {author} {R.~Tualle-Brouri},
  \bibinfo {author} {J.~Laurat}\ and\ \bibinfo {author} {P.~Grangier},\ \emph
  {\bibinfo {title} {Generating Optical Schr{\"o}dinger Kittens for Quantum
  Information Processing}},\ \href {http://dx.doi.org/10.1126/science.1122858}
  {\bibfield  {journal} {\bibinfo  {journal} {Science}\ }\textbf {\bibinfo
  {volume} {312}},\ \bibinfo {pages} {83} (\bibinfo {year} {2006})}\BibitemShut
  {NoStop}%
\bibitem [{\citenamefont {Neergaard-Nielsen}\ \emph {et~al.}(2006)\citenamefont
  {Neergaard-Nielsen}, \citenamefont {Nielsen}, \citenamefont {Hettich},
  \citenamefont {Mølmer},\ and\ \citenamefont
  {Polzik}}]{Neergaard-NielsenPRL2006}%
  \BibitemOpen
  \bibinfo {author} {J.~S. Neergaard-Nielsen}, \bibinfo {author} {B.~M.
  Nielsen}, \bibinfo {author} {C.~Hettich}, \bibinfo {author} {K.~Mølmer}\
  and\ \bibinfo {author} {E.~S. Polzik},\ \emph {\bibinfo {title} {Generation
  of a Superposition of Odd Photon Number States for Quantum Information
  Networks}},\ \href {\doibase10.1103/PhysRevLett.97.083604} {\bibfield
  {journal} {\bibinfo  {journal} {Phys. Rev. Lett.}\ }\textbf {\bibinfo
  {volume} {97}},\ \bibinfo {pages} {083604} (\bibinfo {year}
  {2006})}\BibitemShut {NoStop}%
\bibitem [{\citenamefont {Takahashi}\ \emph {et~al.}(2008)\citenamefont
  {Takahashi}, \citenamefont {Wakui}, \citenamefont {Suzuki}, \citenamefont
  {Takeoka}, \citenamefont {Hayasaka}, \citenamefont {Furusawa},\ and\
  \citenamefont {Sasaki}}]{TakahashiPRL2008}%
  \BibitemOpen
  \bibinfo {author} {H.~Takahashi}, \bibinfo {author} {K.~Wakui}, \bibinfo
  {author} {S.~Suzuki}, \bibinfo {author} {M.~Takeoka}, \bibinfo {author}
  {K.~Hayasaka}, \bibinfo {author} {A.~Furusawa}\ and\ \bibinfo {author}
  {M.~Sasaki},\ \emph {\bibinfo {title} {Generation of {Large}-{Amplitude}
  {Coherent}-{State} {Superposition} via {Ancilla}-{Assisted} {Photon}
  {Subtraction}}},\ \href {\doibase10.1103/PhysRevLett.101.233605} {\bibfield
  {journal} {\bibinfo  {journal} {Phys. Rev. Lett.}\ }\textbf {\bibinfo
  {volume} {101}},\ \bibinfo {pages} {233605} (\bibinfo {year}
  {2008})}\BibitemShut {NoStop}%
\bibitem [{\citenamefont {Etesse}\ \emph {et~al.}(2015)\citenamefont {Etesse},
  \citenamefont {Bouillard}, \citenamefont {Kanseri},\ and\ \citenamefont
  {Tualle-Brouri}}]{EtessePRL2015}%
  \BibitemOpen
  \bibinfo {author} {J.~Etesse}, \bibinfo {author} {M.~Bouillard}, \bibinfo
  {author} {B.~Kanseri}\ and\ \bibinfo {author} {R.~Tualle-Brouri},\ \emph
  {\bibinfo {title} {Experimental {Generation} of {Squeezed} {Cat} {States}
  with an {Operation} {Allowing} {Iterative} {Growth}}},\ \href
  {\doibase10.1103/PhysRevLett.114.193602} {\bibfield  {journal} {\bibinfo
  {journal} {Phys. Rev. Lett.}\ }\textbf {\bibinfo {volume} {114}},\ \bibinfo
  {pages} {193602} (\bibinfo {year} {2015})}\BibitemShut {NoStop}%
\bibitem [{\citenamefont {Gilles}\ \emph {et~al.}(1994)\citenamefont {Gilles},
  \citenamefont {Garraway},\ and\ \citenamefont {Knight}}]{GillesPRA94}%
  \BibitemOpen
  \bibinfo {author} {L.~Gilles}, \bibinfo {author} {B.~M. Garraway}\ and\
  \bibinfo {author} {P.~L. Knight},\ \emph {\bibinfo {title} {Generation of
  nonclassical light by dissipative two-photon processes}},\ \href
  {http://dx.doi.org/10.1103/PhysRevA.49.2785} {\bibfield  {journal} {\bibinfo
  {journal} {Phys. Rev. A}\ }\textbf {\bibinfo {volume} {49}},\ \bibinfo
  {pages} {2785} (\bibinfo {year} {1994})}\BibitemShut {NoStop}%
\bibitem [{\citenamefont {S\'anchez Mu\~noz}\ and\ \citenamefont
  {Jaksch}(2021)}]{Munozarxiv20}%
  \BibitemOpen
  \bibinfo {author} {C.~S\'anchez Mu\~noz}\ and\ \bibinfo {author}
  {D.~Jaksch},\ \emph {\bibinfo {title} {Squeezed Lasing}},\ \href
  {\doibase10.1103/PhysRevLett.127.183603} {\bibfield  {journal} {\bibinfo
  {journal} {Phys. Rev. Lett.}\ }\textbf {\bibinfo {volume} {127}},\ \bibinfo
  {pages} {183603} (\bibinfo {year} {2021})}\BibitemShut {NoStop}%
\bibitem [{\citenamefont {Khaneja}\ \emph {et~al.}(2005)\citenamefont
  {Khaneja}, \citenamefont {Reiss}, \citenamefont {Kehlet}, \citenamefont
  {Schulte-Herbrüggen},\ and\ \citenamefont
  {Glaser}}]{KhanejaJournalMagneticResonance2005}%
  \BibitemOpen
  \bibinfo {author} {N.~Khaneja}, \bibinfo {author} {T.~Reiss}, \bibinfo
  {author} {C.~Kehlet}, \bibinfo {author} {T.~Schulte-Herbrüggen}\ and\
  \bibinfo {author} {S.~J. Glaser},\ \emph {\bibinfo {title} {Optimal control
  of coupled spin dynamics: design of {NMR} pulse sequences by gradient ascent
  algorithms}},\ \href {\doibase10.1016/j.jmr.2004.11.004} {\bibfield
  {journal} {\bibinfo  {journal} {Journal of Magnetic Resonance}\ }\textbf
  {\bibinfo {volume} {172}},\ \bibinfo {pages} {296} (\bibinfo {year}
  {2005})}\BibitemShut {NoStop}%
\bibitem [{\citenamefont {Glaser}\ \emph {et~al.}(2015)\citenamefont {Glaser},
  \citenamefont {Boscain}, \citenamefont {Calarco}, \citenamefont {Koch},
  \citenamefont {Köckenberger}, \citenamefont {Kosloff}, \citenamefont
  {Kuprov}, \citenamefont {Luy}, \citenamefont {Schirmer}, \citenamefont
  {Schulte-Herbrüggen}, \citenamefont {Sugny},\ and\ \citenamefont
  {Wilhelm}}]{GlaserEuropeanPhysJournalD2015}%
  \BibitemOpen
  \bibinfo {author} {S.~J. Glaser}, \bibinfo {author} {U.~Boscain}, \bibinfo
  {author} {T.~Calarco}, \bibinfo {author} {C.~P. Koch}, \bibinfo {author}
  {W.~Köckenberger}, \bibinfo {author} {R.~Kosloff}, \bibinfo {author}
  {I.~Kuprov}, \bibinfo {author} {B.~Luy}, \bibinfo {author} {S.~Schirmer},
  \bibinfo {author} {T.~Schulte-Herbrüggen}, \bibinfo {author} {D.~Sugny}\
  and\ \bibinfo {author} {F.~K. Wilhelm},\ \emph {\bibinfo {title} {Training
  {Schrödinger}’s cat: quantum optimal control}},\ \href
  {\doibase10.1140/epjd/e2015-60464-1} {\bibfield  {journal} {\bibinfo
  {journal} {The European Physical Journal D}\ }\textbf {\bibinfo {volume}
  {69}},\ \bibinfo {pages} {279} (\bibinfo {year} {2015})}\BibitemShut
  {NoStop}%
\bibitem [{\citenamefont {M{\o}lmer}\ \emph {et~al.}(1993)\citenamefont
  {M{\o}lmer}, \citenamefont {Castin},\ and\ \citenamefont
  {Dalibard}}]{Molmer1993}%
  \BibitemOpen
  \bibinfo {author} {K.~M{\o}lmer}, \bibinfo {author} {Y.~Castin}\ and\
  \bibinfo {author} {J.~Dalibard},\ \emph {\bibinfo {title} {Monte Carlo
  wave-function method in quantum optics}},\ \href
  {http://josab.osa.org/abstract.cfm?URI=josab-10-3-524} {\bibfield  {journal}
  {\bibinfo  {journal} {J. Opt. Soc. Am. B}\ }\textbf {\bibinfo {volume}
  {10}},\ \bibinfo {pages} {524} (\bibinfo {year} {1993})}\BibitemShut
  {NoStop}%
\bibitem [{\citenamefont {Dalibard}\ \emph {et~al.}(1992)\citenamefont
  {Dalibard}, \citenamefont {Castin},\ and\ \citenamefont
  {M\o{}lmer}}]{DalibardPRL92}%
  \BibitemOpen
  \bibinfo {author} {J.~Dalibard}, \bibinfo {author} {Y.~Castin}\ and\ \bibinfo
  {author} {K.~M\o{}lmer},\ \emph {\bibinfo {title} {Wave-function approach to
  dissipative processes in quantum optics}},\ \href
  {https://link.aps.org/doi/10.1103/PhysRevLett.68.580} {\bibfield  {journal}
  {\bibinfo  {journal} {Phys. Rev. Lett.}\ }\textbf {\bibinfo {volume} {68}},\
  \bibinfo {pages} {580} (\bibinfo {year} {1992})}\BibitemShut {NoStop}%
\bibitem [{\citenamefont {Gardiner}\ and\ \citenamefont
  {Zoller}(2004)}]{Gardiner_BOOK_Quantum}%
  \BibitemOpen
  \bibinfo {author} {C.~Gardiner}\ and\ \bibinfo {author} {P.~Zoller},\
  \href@noop {} {\emph {\bibinfo {title} {Quantum Noise: A Handbook of
  Markovian and Non-Markovian Quantum Stochastic Methods with Applications to
  Quantum Optics}}}\ (\bibinfo  {publisher} {Springer},\ \bibinfo {address}
  {Berlin},\ \bibinfo {year} {2004})\BibitemShut {NoStop}%
\bibitem [{\citenamefont {Nakata}(2010)}]{NakataIEEE2010}%
  \BibitemOpen
  \bibinfo {author} {M.~Nakata},\ in\ \href
  {\doibase10.1109/CACSD.2010.5612693} {\emph {\bibinfo {booktitle} {2010
  {IEEE} {International} {Symposium} on {Computer}-{Aided} {Control} {System}
  {Design}}}}\ (\bibinfo {year} {2010})\ pp.\ \bibinfo {pages}
  {29--34}\BibitemShut {NoStop}%
\bibitem [{\citenamefont {Udell}\ \emph {et~al.}(2014)\citenamefont {Udell},
  \citenamefont {Mohan}, \citenamefont {Zeng}, \citenamefont {Hong},
  \citenamefont {Diamond},\ and\ \citenamefont {Boyd}}]{UdellSC142014}%
  \BibitemOpen
  \bibinfo {author} {M.~Udell}, \bibinfo {author} {K.~Mohan}, \bibinfo {author}
  {D.~Zeng}, \bibinfo {author} {J.~Hong}, \bibinfo {author} {S.~Diamond}\ and\
  \bibinfo {author} {S.~Boyd},\ in\ \href {\doibase10.1109/HPTCDL.2014.5}
  {\emph {\bibinfo {booktitle} {2014 First Workshop for High Performance
  Technical Computing in Dynamic Languages}}}\ (\bibinfo {year} {2014})\ pp.\
  \bibinfo {pages} {18--28}\BibitemShut {NoStop}%
\bibitem [{\citenamefont {Bezanson}\ \emph {et~al.}(2017)\citenamefont
  {Bezanson}, \citenamefont {Edelman}, \citenamefont {Karpinski},\ and\
  \citenamefont {Shah}}]{BezansonSIAM2017}%
  \BibitemOpen
  \bibinfo {author} {J.~Bezanson}, \bibinfo {author} {A.~Edelman}, \bibinfo
  {author} {S.~Karpinski}\ and\ \bibinfo {author} {V.~B. Shah},\ \emph
  {\bibinfo {title} {Julia: A fresh approach to numerical computing}},\ \href
  {https://doi.org/10.1137/141000671} {\bibfield  {journal} {\bibinfo
  {journal} {SIAM review}\ }\textbf {\bibinfo {volume} {59}},\ \bibinfo {pages}
  {65} (\bibinfo {year} {2017})}\BibitemShut {NoStop}%
\bibitem [{\citenamefont {Kr{\"a}mer}\ \emph {et~al.}(2018)\citenamefont
  {Kr{\"a}mer}, \citenamefont {Plankensteiner}, \citenamefont {Ostermann},\
  and\ \citenamefont {Ritsch}}]{KramerComputerPhysicsCommunications2018}%
  \BibitemOpen
  \bibinfo {author} {S.~Kr{\"a}mer}, \bibinfo {author} {D.~Plankensteiner},
  \bibinfo {author} {L.~Ostermann}\ and\ \bibinfo {author} {H.~Ritsch},\ \emph
  {\bibinfo {title} {QuantumOptics. jl: A Julia framework for simulating open
  quantum systems}},\ \href {\doibase10.1016/j.cpc.2018.02.004} {\bibfield
  {journal} {\bibinfo  {journal} {Computer Physics Communications}\ }\textbf
  {\bibinfo {volume} {227}},\ \bibinfo {pages} {109} (\bibinfo {year}
  {2018})}\BibitemShut {NoStop}%
\bibitem [{\citenamefont {Reinhold}(2019)}]{Reinhold_PhD_Thesis}%
  \BibitemOpen
  \bibinfo {author} {P.~Reinhold},\ \emph {\bibinfo {title} {Controlling
  Error-Correctable Bosonic Qubits}},\ \href@noop {} {\bibinfo {type} {{Ph.D.
  Thesis}}},\ \bibinfo  {school} {Yale University} (\bibinfo {year}
  {2019})\BibitemShut {NoStop}%
\end{thebibliography}%

\clearpage
\title{Supplementary Material:\\
Quantum error correction using squeezed Schrödinger cat states}
\maketitle

\onecolumngrid

\setcounter{page}{1}
\pagestyle{fancy}

\fancyhf{}

\rhead{Supplementary Material \thepage}

\newcounter{appsection}
\newcounter{appsubsection}[appsection]

\newcommand{\appsection}[1]{\refstepcounter{appsection}
$ \ $ \\
$ \ $\\
\begin{center}
    \textbf{ \uppercase{}~S.\Alph{appsection}:~\uppercase{#1}}
\end{center} 
}

\newcommand{\appsubsection}[1]{\refstepcounter{appsubsection}
$ \ $ \\
\begin{center}
    \textbf{S.\Alph{appsection}.\Roman{appsubsection}:~#1}
\end{center} 
}

\appsection{KL conditions for $\bra*{\mathcal{C}_{\alpha, \xi}^\pm} \hat{E}_{l}^\dagger \hat{E}_{l'} \ket*{\mathcal{C}_{\alpha, \xi}^\pm}$}

\appsubsection
{$\bra*{\mathcal{C}_{\alpha, \xi}^\pm} \hat{\mathds{1}} \ket*{\mathcal{C}_{\alpha, \xi}^\pm}$}

\begin{equation}
    \bra*{\mathcal{C}_{\alpha, \xi}^\pm} \hat{\mathds{1}} \ket*{\mathcal{C}_{\alpha, \xi}^\pm} =1
\end{equation}

\appsubsection
{$\bra*{\mathcal{C}_{\alpha, \xi}^\pm} \hat{a} \ket*{\mathcal{C}_{\alpha, \xi}^\pm}$}

\begin{equation}
    \bra*{\mathcal{C}_{\alpha, \xi}^\pm} \hat{a} \ket*{\mathcal{C}_{\alpha, \xi}^\pm} =0
\end{equation}

\appsubsection{
$\bra*{\mathcal{C}_{\alpha, \xi}^\pm} \hat{a}^\dagger \hat{a} \ket*{\mathcal{C}_{\alpha, \xi}^\pm}$}

\begin{equation}
    \bra*{\mathcal{C}_{\alpha, \xi}^\pm} \hat{a}^\dagger \hat{a} \ket*{\mathcal{C}_{\alpha, \xi}^\pm} =\sinh (\xi ) \left(\sinh (\xi )-2
   \gamma ^2 \cosh (\xi
   )\right)+\frac{\gamma ^2 \cosh (2
   \xi ) \left(N_{\alpha \xi }^{\mp
   }\right){}^2}{\left(N_{\alpha \xi
   }^{\pm }\right){}^2}
\end{equation}

\appsubsection{
$\bra*{\mathcal{C}_{\alpha, \xi}^\pm} (\hat{a}^\dagger \hat{a})^2 \ket*{\mathcal{C}_{\alpha, \xi}^\pm}$}

\begin{equation}\begin{split}
    \bra*{\mathcal{C}_{\alpha, \xi}^\pm} (\hat{a}^\dagger \hat{a})^2 \ket*{\mathcal{C}_{\alpha, \xi}^\pm} &= \frac{1}{4} \Bigg(\left(4 \gamma
   ^4+1\right) \cosh (4 \xi )+4 \gamma
   ^2 \sinh (2 \xi )-6 \gamma ^2 \sinh
   (4 \xi )-2 \cosh (2 \xi )  \\ &  \qquad +\frac{2
   \gamma ^2 \left(-2 \gamma ^2 \sinh
   (4 \xi )-2 \cosh (2 \xi )+3 \cosh
   (4 \xi )+1\right) \left(N_{\alpha
   \xi }^{\mp
   }\right){}^2}{\left(N_{\alpha \xi
   }^{\pm }\right){}^2}+1\Bigg)
   \end{split}
\end{equation}

\appsubsection{
$\bra*{\mathcal{C}_{\alpha, \xi}^\pm} \hat{a}^\dagger  \ket*{\mathcal{C}_{\alpha, \xi}^\pm}$}

\begin{equation}
    \bra*{\mathcal{C}_{\alpha, \xi}^\pm} \hat{a}^\dagger  \ket*{\mathcal{C}_{\alpha, \xi}^\pm}=0
\end{equation}

\appsubsection{
$\bra*{\mathcal{C}_{\alpha, \xi}^\pm} (\hat{a}^\dagger )^2 \hat{a}  \ket*{\mathcal{C}_{\alpha, \xi}^\pm}$}

\begin{equation}
    \bra*{\mathcal{C}_{\alpha, \xi}^\pm} (\hat{a}^\dagger )^2 \hat{a}  \ket*{\mathcal{C}_{\alpha, \xi}^\pm}=0
\end{equation}

\appsubsection{
$\bra*{\mathcal{C}_{\alpha, \xi}^\pm} (\hat{a}^\dagger )^2 \hat{a} \hat{a}^\dagger \hat{a}  \ket*{\mathcal{C}_{\alpha, \xi}^\pm}$}

\begin{equation}
    \bra*{\mathcal{C}_{\alpha, \xi}^\pm} (\hat{a}^\dagger )^2 \hat{a} \hat{a}^\dagger \hat{a}  \ket*{\mathcal{C}_{\alpha, \xi}^\pm}=0
\end{equation}

\appsubsection{
$\bra*{\mathcal{C}_{\alpha, \xi}^\pm} \hat{a}^\dagger  \hat{a}^2 \ket*{\mathcal{C}_{\alpha, \xi}^\pm}$}

\begin{equation}
\bra*{\mathcal{C}_{\alpha, \xi}^\pm} \hat{a}^\dagger  \hat{a}^2 \ket*{\mathcal{C}_{\alpha, \xi}^\pm}=0
\end{equation}

\appsubsection{
$\bra*{\mathcal{C}_{\alpha, \xi}^\pm} (\hat{a}^\dagger \hat{a})^3 \ket*{\mathcal{C}_{\alpha, \xi}^\pm}$}

\begin{equation}\begin{split}
    \bra*{\mathcal{C}_{\alpha, \xi}^\pm} (\hat{a}^\dagger \hat{a})^3 \ket*{\mathcal{C}_{\alpha, \xi}^\pm} &= 
    \frac{1}{32} \Bigg[\left(24 \gamma
   ^4+11\right) \cosh (2 \xi )- \\ &  2
   \left(24 \gamma ^4+5\right) \cosh
   (4 \xi )-2 \gamma ^2 \left(\left(16
   \gamma ^4+45\right) \sinh (6 \xi
   )-60 \gamma ^2 \cosh (6 \xi )+\sinh
   (2 \xi )-36 \sinh (4 \xi )\right) \\ &   +5
   \cosh (6 \xi ) -6  +2 \gamma^2 \left( 
   \frac{   N_{\alpha \xi }^{\mp
   } }{N_{\alpha \xi
   }^{\pm }}
 \right)^2  \Big( 
  \left(16 \gamma ^4+45\right)
   \cosh (6 \xi )  \\ & -12 \left(\gamma ^2
   (\sinh (2 \xi )-2 \sinh (4 \xi )+5
   \sinh (6 \xi ))+1\right)+19 \cosh
   (2 \xi )-36 \cosh (4 \xi ) \Big) 
    \Bigg]
   \end{split}
\end{equation}

\appsubsection{
$\bra*{\mathcal{C}_{\alpha, \xi}^\pm} \hat{a}^\dagger \hat{a} \hat{a}^\dagger \hat{a}^2  \ket*{\mathcal{C}_{\alpha, \xi}^\pm}$}

\begin{equation}\begin{split}
    \bra*{\mathcal{C}_{\alpha, \xi}^\pm} \hat{a}^\dagger \hat{a} \hat{a}^\dagger \hat{a}^2  \ket*{\mathcal{C}_{\alpha, \xi}^\pm}=0 
   \end{split}
\end{equation}

\appsubsection{
$\bra*{\mathcal{C}_{\alpha, \xi}^\pm} (\hat{a}^\dagger \hat{a})^4 \ket*{\mathcal{C}_{\alpha, \xi}^\pm}$}

\begin{equation}\begin{split}
    \bra*{\mathcal{C}_{\alpha, \xi}^\pm} (\hat{a}^\dagger \hat{a})^4 \ket*{\mathcal{C}_{\alpha, \xi}^\pm} &= 
\frac{1}{64} \Bigg[-6 \left(16 \gamma
   ^4+3\right) \cosh (2 \xi )   +16
   \left(7 \gamma ^4+1\right) \cosh (4
   \xi )-2 \left(240 \gamma
   ^4+7\right) \cosh (6 \xi )  \\ & +72
   \gamma ^4-56 \gamma ^2 \sinh (2 \xi
   ) +\left(64 \gamma ^8+840 \gamma
   ^4+7\right) \cosh (8 \xi ) 
   +8
   \left(9-8 \gamma ^4\right) \gamma
   ^2 \sinh (4 \xi )\\ & +8 \left(16 \gamma
   ^4+45\right) \gamma ^2 \sinh (6 \xi
   )  -28 \left(16 \gamma ^4+15\right)
   \gamma ^2 \sinh (8 \xi ) +9  \\ &  
   + 4
   \gamma ^2    
   \left(\frac{N_{\alpha \xi }^{\mp
   }}{N_{\alpha \xi
   }^{\pm }} \right)^2  \Big(4 \left(4 \gamma
   ^4+3\right) \cosh (4 \xi )-2
   \left(16 \gamma ^4+45\right) \cosh
   (6 \xi ) \\ & -22 \cosh (2
   \xi ) +105 \cosh (8 \xi )+11 
   \\ & + 2 \gamma ^2 \left( 56 \gamma ^2 \cosh (8 \xi ) -\left(8
   \gamma ^4+105\right) \sinh (8 \xi
   )+12
   \sinh (2 \xi )-14 \sinh (4 \xi )+60
   \sinh (6 \xi )\right)
   \Big)
   \Bigg]
   \end{split}
\end{equation}

\appsection{KL conditions for $\bra*{\mathcal{C}_{\alpha, \xi}^\pm} \hat{E}_{l}^\dagger \hat{E}_{l'} \ket*{\mathcal{C}_{\alpha, \xi}^\mp}$}

\appsubsection
{$\bra*{\mathcal{C}_{\alpha, \xi}^\pm} \hat{\mathds{1}} \ket*{\mathcal{C}_{\alpha, \xi}^\mp}$}

\begin{equation}
    \bra*{\mathcal{C}_{\alpha, \xi}^\pm} \hat{\mathds{1}} \ket*{\mathcal{C}_{\alpha, \xi}^\mp} =0
\end{equation}

\appsubsection
{$\bra*{\mathcal{C}_{\alpha, \xi}^\pm} \hat{a} \ket*{\mathcal{C}_{\alpha, \xi}^\mp}$}

\begin{equation}
    \bra*{\mathcal{C}_{\alpha, \xi}^\pm} \hat{a} \ket*{\mathcal{C}_{\alpha, \xi}^\mp} =\frac{\gamma  \cosh (\xi ) N_{\alpha
   \xi }^{\pm }}{N_{\alpha \xi }^{\mp
   }}-\frac{\gamma  \sinh (\xi )
   N_{\alpha \xi }^{\mp }}{N_{\alpha
   \xi }^{\pm }}
\end{equation}

\appsubsection{
$\bra*{\mathcal{C}_{\alpha, \xi}^\pm} \hat{a}^\dagger \hat{a} \ket*{\mathcal{C}_{\alpha, \xi}^\mp}$}

\begin{equation}
    \bra*{\mathcal{C}_{\alpha, \xi}^\pm} \hat{a}^\dagger \hat{a} \ket*{\mathcal{C}_{\alpha, \xi}^\mp} =0
\end{equation}

\appsubsection{
$\bra*{\mathcal{C}_{\alpha, \xi}^\pm} (\hat{a}^\dagger \hat{a})^2 \ket*{\mathcal{C}_{\alpha, \xi}^\mp}$}

\begin{equation}\begin{split}
    \bra*{\mathcal{C}_{\alpha, \xi}^\pm} (\hat{a}^\dagger \hat{a})^2 \ket*{\mathcal{C}_{\alpha, \xi}^\mp} &= 0
   \end{split}
\end{equation}

\appsubsection{
$\bra*{\mathcal{C}_{\alpha, \xi}^\pm} \hat{a}^\dagger  \ket*{\mathcal{C}_{\alpha, \xi}^\mp}$}

\begin{equation}
    \bra*{\mathcal{C}_{\alpha, \xi}^\pm} \hat{a}^\dagger  \ket*{\mathcal{C}_{\alpha, \xi}^\mp}=\frac{\gamma  \cosh (\xi ) N_{\alpha
   \xi }^{\mp }}{N_{\alpha \xi }^{\pm
   }}-\frac{\gamma  \sinh (\xi )
   N_{\alpha \xi }^{\pm }}{N_{\alpha
   \xi }^{\mp }}
\end{equation}

\appsubsection{
$\bra*{\mathcal{C}_{\alpha, \xi}^\pm} (\hat{a}^\dagger )^2 \hat{a}  \ket*{\mathcal{C}_{\alpha, \xi}^\mp}$}

\begin{equation}\begin{split}
    \bra*{\mathcal{C}_{\alpha, \xi}^\pm} (\hat{a}^\dagger )^2 \hat{a}  \ket*{\mathcal{C}_{\alpha, \xi}^\mp}&=\frac{\gamma  \left(4 \gamma ^2 \cosh
   (3 \xi )+5 \sinh (\xi )-3 \sinh (3
   \xi )\right) N_{\alpha \xi }^{\pm
   }}{4 N_{\alpha \xi }^{\mp
   }} \\ & \quad -\frac{\gamma  \left(4 \gamma ^2
   \sinh (3 \xi )+3 \cosh (\xi )-3
   \cosh (3 \xi )\right) N_{\alpha \xi
   }^{\mp }}{4 N_{\alpha \xi }^{\pm }}
   \end{split}
\end{equation}

\appsubsection{
$\bra*{\mathcal{C}_{\alpha, \xi}^\pm} (\hat{a}^\dagger )^2 \hat{a} \hat{a}^\dagger \hat{a}  \ket*{\mathcal{C}_{\alpha, \xi}^\mp}$}

\begin{equation}\begin{split}
    \bra*{\mathcal{C}_{\alpha, \xi}^\pm} (\hat{a}^\dagger )^2 \hat{a} \hat{a}^\dagger \hat{a}  \ket*{\mathcal{C}_{\alpha, \xi}^\mp} &=\frac{ \gamma N_{\alpha \xi }^{\pm }}{16
   N_{\alpha \xi }^{\mp
   }}
    \Big[ 
      8 \gamma ^2
   \left(-2 \gamma ^2 \sinh (5 \xi
   )+\cosh (\xi )-4 \cosh (3 \xi )+5
   \cosh (5 \xi )\right) \\ & -22 \sinh (\xi
   )+27 \sinh (3 \xi )-15 \sinh (5 \xi
   )\Big]      
   \\ &
   +\frac{\gamma N_{\alpha \xi }^{\mp }}{16
   N_{\alpha \xi }^{\pm }}  \Big[\left(16
   \gamma ^4+15\right) \cosh (5 \xi
   )-8 \gamma ^2 (\sinh (\xi ) \\ & -4 \sinh
   (3 \xi )+5 \sinh (5 \xi ))+6 \cosh
   (\xi )-21 \cosh (3 \xi )\Big]
   \end{split}
\end{equation}

\appsubsection{
$\bra*{\mathcal{C}_{\alpha, \xi}^\pm} \hat{a}^\dagger  \hat{a}^2\ket*{\mathcal{C}_{\alpha, \xi}^\mp} $}

\begin{equation}
\begin{split}
\bra*{\mathcal{C}_{\alpha, \xi}^\pm} \hat{a}^\dagger  \hat{a}^2 \ket*{\mathcal{C}_{\alpha, \xi}^\mp} = & \frac{\gamma  \left(4 \gamma ^2 \cosh
   (3 \xi )+5 \sinh (\xi )-3 \sinh (3
   \xi )\right) N_{\alpha \xi }^{\mp
   }}{4 N_{\alpha \xi }^{\pm
   }} \\ & -\frac{\gamma  \left(4 \gamma ^2
   \sinh (3 \xi )+3 \cosh (\xi )-3
   \cosh (3 \xi )\right) N_{\alpha \xi
   }^{\pm }}{4 N_{\alpha \xi }^{\mp }}
\end{split}   
\end{equation}

\appsubsection{
$\bra*{\mathcal{C}_{\alpha, \xi}^\pm} (\hat{a}^\dagger \hat{a})^3 \ket*{\mathcal{C}_{\alpha, \xi}^\mp}$}

\begin{equation}\begin{split}
    \bra*{\mathcal{C}_{\alpha, \xi}^\pm} (\hat{a}^\dagger \hat{a})^3 \ket*{\mathcal{C}_{\alpha, \xi}^\mp} &= 0
   \end{split}
\end{equation}


\appsubsection{
$\bra*{\mathcal{C}_{\alpha, \xi}^\pm} \hat{a}^\dagger \hat{a} \hat{a}^\dagger \hat{a}^2  \ket*{\mathcal{C}_{\alpha, \xi}^\mp}$}

\begin{equation}\begin{split}
    \bra*{\mathcal{C}_{\alpha, \xi}^\pm} \hat{a}^\dagger \hat{a} \hat{a}^\dagger \hat{a}^2  \ket*{\mathcal{C}_{\alpha, \xi}^\mp} &
    =\frac{\gamma N_{\alpha \xi }^{\mp }}{16
   N_{\alpha \xi }^{\pm
   }}  \Big[ 8 \gamma ^2
   \left(-2 \gamma ^2 \sinh (5 \xi
   )+\cosh (\xi )-4 \cosh (3 \xi )+5
   \cosh (5 \xi )\right) \\ & -22 \sinh (\xi
   )+27 \sinh (3 \xi )-15 \sinh (5 \xi
   )\Big]  \\ &+\frac{\gamma   N_{\alpha \xi }^{\pm }}{16
   N_{\alpha \xi }^{\mp }}   \Big[\left(16
   \gamma ^4+15\right) \cosh (5 \xi
   )-8 \gamma ^2 (\sinh (\xi ) \\ & -4 \sinh
   (3 \xi )+5 \sinh (5 \xi ))+6 \cosh
   (\xi )-21 \cosh (3 \xi )\Big]
   \end{split}
\end{equation}

\appsubsection{
$\bra*{\mathcal{C}_{\alpha, \xi}^\pm} (\hat{a}^\dagger \hat{a})^4 \ket*{\mathcal{C}_{\alpha, \xi}^\mp}$}

\begin{equation}\begin{split}
    \bra*{\mathcal{C}_{\alpha, \xi}^\pm} (\hat{a}^\dagger \hat{a})^4 \ket*{\mathcal{C}_{\alpha, \xi}^\mp} &= 0
   \end{split}
\end{equation}

\end{document}